\documentclass[fleqn,usenatbib]{mnras}

\usepackage{newtxtext,newtxmath}
\usepackage[T1]{fontenc}
\usepackage{ae,aecompl}

\usepackage{graphicx}	% Including figure files
\usepackage{amsmath}	% Advanced maths commands
\usepackage[flushleft]{threeparttable}

\title[ASAS-SN Catalog of Variable Stars II]{The ASAS-SN Catalog of Variable Stars II: \textit{Uniform Classification of 412,000 Known Variables}}

\author[T. Jayasinghe et al.]{T. Jayasinghe$^{1,2}$\thanks{E-mail: jayasinghearachchilage.1@osu.edu},
K. Z. Stanek$^{1,2}$,
C. S. Kochanek$^{1,2}$,
B. J. Shappee$^{3}$,
\newauthor 
T. W. -S. Holoien$^{4}$,
Todd A. Thompson$^{1,2}$,
J. L. Prieto$^{5,6}$,
Subo Dong$^{7}$,
M. Pawlak$^{8}$,
\newauthor 
O. Pejcha$^{8}$,
J. V. Shields$^{1}$,
G. Pojmanski$^{9}$,
S. Otero$^{10}$,
C. A. Britt$^{11}$,
D. Will$^{1,11}$
\\
$^{1}$Department of Astronomy, The Ohio State University, 140 West 18th Avenue, Columbus, OH 43210, USA\\
$^{2}$Center for Cosmology and Astroparticle Physics, The Ohio State University, 191 W. Woodruff Avenue, Columbus, OH 43210, USA\\
$^{3}$Institute for Astronomy, University of Hawaii, 2680 Woodlawn Drive, Honolulu, HI 96822,USA\\
$^{4}$Carnegie Observatories, 813 Santa Barbara Street, Pasadena, CA 91101, USA\\
$^{5}$N\'ucleo de Astronom\'ia de la Facultad de Ingenier\'ia y Ciencias, Universidad Diego Portales, Av. Ej\'ercito 441, Santiago, Chile\\
$^{6}$Millennium Institute of Astrophysics, Santiago, Chile\\
$^{7}$Kavli Institute for Astronomy and Astrophysics, Peking University, Yi He Yuan Road 5, Hai Dian District, China\\
$^{8}$Institute of Theoretical Physics, Faculty of Mathematics and Physics, Charles University in Prague, Czech Republic\\
$^{9}$Warsaw University Observatory, Al Ujazdowskie 4, 00-478 Warsaw, Poland\\
$^{10}$The American Association of Variable Star Observers, 49 Bay State Road, Cambridge, MA 02138, USA\\
$^{11}$ASC Technology Services, 433 Mendenhall Laboratory 125 South Oval Mall Columbus OH, 43210, USA\\
}

\date{Accepted XXX. Received YYY; in original form ZZZ}

\pubyear{2018}

\begin{document}
\label{firstpage}
\pagerange{\pageref{firstpage}--\pageref{lastpage}}
\maketitle

\begin{abstract}
The variable stars in the VSX catalog are derived from a multitude of inhomogeneous data sources and classification tools. This inhomogeneity complicates our understanding of variable star types, statistics, and properties, and it directly affects attempts to build training sets for current (and next) generation all-sky, time-domain surveys. We homogeneously analyze the ASAS-SN V-band light curves of ${\sim}412,000$ variables from the VSX catalog. The variables are classified using an updated random forest classifier with an $F_1$ score of 99.4\% and refinement criteria for individual classifications. We have derived periods for ${\sim}52,000$ variables in the VSX catalog that lack a period, and have reclassified ${\sim} 17,000$ sources into new broad variability groups with high confidence. We have also reclassified ${\sim} 94,000$ known variables with miscellaneous/generic classifications. The light curves, classifications, and a range of properties of the variables are all available through the ASAS-SN variable stars database (\url{https://asas-sn.osu.edu/variables}). We also include the V-band light curves for a set of ${\sim}4,000$ rare variables and transient sources, including cataclysmic variables, symbiotic binaries and flare stars. 
\end{abstract}


%%%%%%%%%%%%%%%%%%%%%%%%%%%%%%%%%%%%%%%%%%%%%%%%%%

%%%%%%%%%%%%%%%%% BODY OF PAPER %%%%%%%%%%%%%%%%%%

\section{Introduction}
It is difficult to pinpoint the exact origin of the study of variable stars. Astronomy is the oldest natural science, and many ancient cultures carefully observed celestial objects. Records of transient sources such as supernovae date back to 185 AD, but variable sources are different in the sense that they persist over human lifetimes, rather than disappearing after a short while. It has been suggested that the ancient Egyptians first noted the variability of Algol, an eclipsing binary, over 3000 years ago \citep{2013ApJ...773....1J}. Recently, it has even been claimed that aboriginal Australians observed the variability of pulsating red giants long ago and incorporated this discovery into their culture and lore \citep{2018AuJAn..29...89H} . This is not difficult to believe, as the large variability amplitudes of pulsating red giants make them easy targets for visual, naked-eye observation \citep{2018JAHH...21....7S}. It is possible that many other ancient cultures noted the variability of bright sources without any record of it surviving to the modern day.

The first modern discovery of a periodic variable was made in 1638, when Johannes Holwarda recorded the periodicity of the Mira variable Omicron Ceti. The number of known variable stars gradually increased to ${\sim}12$ by 1786, ${\sim}175$ by 1890, ${\sim}4000$ by 1912 and ${\sim}28450$ by 1983  \citep{1994cpav.book.....A}. In the modern era, surveys such as the All-Sky Automated Survey for SuperNovae (ASAS-SN, \citealt{2014ApJ...788...48S, 2017PASP..129j4502K,2018MNRAS.477.3145J,2018RNAAS...2a..18J}), the All-Sky Automated Survey (ASAS; \citealt{2002AcA....52..397P}), the Optical Gravitational Lensing Experiment (OGLE; \citealt{2003AcA....53..291U}), the Northern Sky Variability Survey (NSVS; \citealt{2004AJ....127.2436W}), MACHO \citep{1997ApJ...486..697A}, EROS \citep{2002A&A...389..149D}, the Catalina Real-Time Transient Survey (CRTS; \citealt{2014ApJS..213....9D}), the Asteroid Terrestrial-impact Last Alert System (ATLAS; \citealt{2018PASP..130f4505T,2018arXiv180402132H}), and Gaia \citep{2018arXiv180409365G,2018arXiv180409373H,gdr2var} have collectively discovered $\gtrsim 10^6$ variables in the span of ${\sim}20$ years. 

ASAS-SN monitors the entire visible sky to a depth of $V\lesssim17$ mag. The ASAS-SN V-band data used in this study were obtained on a cadence of 2-3 days, but with the addition of 3 new g-band units ASAS-SN has significantly improved its cadence to $\lesssim1$ day. The ASAS-SN telescopes are hosted by the Las Cumbres Observatory (LCO; \citealt{2013PASP..125.1031B}) in Hawaii, Chile, Texas and South Africa. While ASAS-SN focuses on the detection of bright supernovae (e.g., \citealt{2017MNRAS.471.4966H}), and other transients (e.g.,  \citealt{2018arXiv180807875T,2018RNAAS...2b...8R}), we simultaneously build up well-sampled light curves for $\gtrsim50$ million bright ($V<17$ mag) sources across the whole sky. In Paper I \citep{2018MNRAS.477.3145J}, we identified ${\sim}66,000$ new variables, most of which are located in regions close to the Galactic plane or Celestial poles which were not well-sampled by previous surveys.

Existing catalogs of variable stars have been derived from a multitude of inhomogeneous data sources and classification tools. This inhomogeneity proves to be a challenge when analyzing populations of variable stars and directly affects the current (and next generation) of all-sky time domain surveys. The OGLE survey provides an excellent catalog of homogeneously classified variable stars in the Magellanic clouds and the Galactic bulge, but an analogous, all-sky catalog of variable stars does not exist. In order to tackle this need, we homogeneously analyze a sample of known variables in this work. While a more complete, all-sky search for variable stars in ASAS-SN is underway, known variables can provide an excellent training set for current and upcoming variability surveys if it can be homogenized.

Missions like the Transiting Exoplanet Survey Satellite (TESS; \citealt{2015JATIS...1a4003R}) are expected to produce a large numbers of high-quality light curves. For example, the TESS input catalog (TIC; \citealt{2018AJ....156..102S}) contains ${\sim} 470$ million sources. Outside of the $400,000$ selected targets observed at a 2 min cadence, the remaining sources are observed with a cadence of 30 min. Each TESS sector will be observed for at least 27 days, thus one can expect each TESS light curve to consist of $\gtrsim1300$ epochs. With such a short cadence, TESS light curves are expected to probe short period variability to great detail. On the other hand, it will be difficult or impossible for TESS to probe the long period variability of most sources. ASAS-SN provides complementary light curves, sampled at a cadence of ${\sim} 1-3$ days, to study variability at longer periods. For large scale time domain surveys like TESS, the need to automatically classify sources becomes critical. Crucial to most classification methods are a well-sampled, diverse and homogeneously analyzed \textit{training set} of variables that can be used to `teach' a classifier to identify new variable sources. 

Random Forest classifiers have been widely used as a means to classify variable stars. \citet{2011ApJ...733...10R} noted the high classification accuracy and speed of the random forest classifier for the classification of variable stars using sparse and noisy time series data. Most wide field surveys have successfully used random forest classifiers for the discovery of variable stars (see for e.g., \citealt{2018arXiv180402132H} for the classification of variable stars by ATLAS and \citealt{gdr2var} for the classification of variable stars in Gaia DR2). Other surveys, like CRTS \citep{2014ApJS..213....9D}, largely used visual inspection of the light curves for classification. In addition to random forests, unsupervised clustering techniques have also been used for the identification of anomalous periodic variables (see for e.g., \citealt{2009arXiv0905.3428R}). 

The variable sources in the Magellanic clouds have provided an excellent training set for random forest classifiers which have been subsequently used to discover numerous variable stars in the LMC and SMC (see for e.g., \citealt{2014A&A...566A..43K,upsilon}). \citet{2012ApJS..203...32R} used a training set consisting of \textit{Hipparcos} and OGLE LMC sources to train a random forest classifier and homogeneously classified a set of ${\sim}50,000$ variable sources in the ASAS catalog of variable stars. This work resulted in the Machine-learned ASAS Classification Catalog (MACC), which is an all-sky, calibrated, probabilistic catalog of variable sources.

We extracted the ASAS-SN light curves of ${\sim}412,000$ variable stars previously discovered by other surveys and in the VSX catalog. In this work, we homogeneously classify the entire sample. In Section $\S2$, we discuss the ASAS-SN observations and data reduction procedure. Section $\S3$ discusses the variability classification pipeline. It is based on random forest classification models with variability type refinements. We derive a training sample with sources from Paper I and a subset of the known VSX variables. In Section $\S4$, we apply our variability classification pipeline to the sources that were not included in our training sample and discuss our approach to estimating variability amplitudes in Section $\S5$. We discuss the overall catalog in Section $\S6$ and present a summary of our work in Section $\S7$.

\section{Observations and Data reduction}
\label{data}
We started with list of previously discovered variables from the VSX \citep{2006SASS...25...47W} database available in April 2018. The VSX database contains over ${\sim} 500,000$ variables discovered by a large number of surveys and individuals and is the most complete all-sky catalog of known variables. At this time, variables from ATLAS \citep{2018arXiv180402132H} and Gaia DR2 \citep{2018arXiv180409373H,gdr2var} have not been included in the VSX catalog. We include the list of variables discovered by the Kilodegree Extremely Little Telescope (KELT; \citealt{2018AJ....155...39O,2007PASP..119..923P}) and the ASAS-SN variables that are in the process of being included in the VSX catalog into our working list of variables. For now, we also remove transients (cataclysmic variables, flaring sources, etc.), sources hosting planets and very-low amplitude rotational variables ($<50$ mmag) to arrive at a list of ${\sim}450,000$ variable sources. Table \ref{tab:vars} describes the different samples of variables referenced in this paper and the section where we discuss each sample.

The V-band observations made by the ``Brutus" (Haleakala, Hawaii) and ``Cassius" (CTIO, Chile) quadruple telescopes between 2013 and 2018 were used to produce the light curves for these sources. Each ASAS-SN field in the V-band has ${\sim}$ 200-600 epochs of observation to a depth of $V\lesssim17$ mag. Each camera has a field of view of 4.5 deg$^2$, the pixel scale is 8\farcs0 and the FWHM is ${\sim}$ 2 pixels. ASAS-SN saturates at ${\sim} 10-11$ mag \citep{2017PASP..129j4502K}.

The light curves used in this work were produced as described in \citet{2018MNRAS.477.3145J}. ASAS-SN data are analyzed using image subtraction \citep{1998ApJ...503..325A,2000A&AS..144..363A}.
The light curves were extracted using aperture photometry on the subtracted images using an aperture with a 2 pixel radius and the IRAF \verb apphot \space package. Calibration was done using the AAVSO Photometric All-Sky Survey (APASS; \citealt{2015AAS...22533616H}). Roughly, ${\sim}30,000$ sources with $<30$ V-band detections were removed from the list to leave us with ${\sim} 420,000$ variables. The light curve uncertainties were rescaled using a rescaling function and zero point offsets between the different cameras were corrected as described in \citet{2018MNRAS.477.3145J}.

\begin{table*}
	\centering
	\caption{Descriptions of the different samples of variables used in this work.}
	\label{tab:vars}
	\begin{tabular}{lrc}
		\hline
		Description & Number of Variables & Section\\
		\hline
        Starting VSX Catalog & ${\sim} 500,000$ & $\S2$\\
        Selecting by the variability type and amplitude & ${\sim} 450,000$ & $\S2$\\
        Selecting by the number of detections in the V-band ($>30$ V-band detections) & ${\sim} 420,000$ & $\S2$\\
        \hline        
        Training set for the V1 classifier (ASAS-SN variables from Paper I) & ${\sim} 66,000$ & $\S3.3$\\
        Initial training set for the V2 classifier & ${\sim} 177,000$ & $\S3.3$\\
        Final training set for the V2 classifier after classification refinement & ${\sim} 166,000$ & $\S3.4,\S3.5, \S3.6$\\
        \hline        
        Variables with $11<V<17$ mag & ${\sim} 356,000$ & $\S6.1$\\
        Variables with miscellaneous classifications & ${\sim} 94,000$ & $\S4.1$\\
        OGLE variables in the VSX catalog & ${\sim} 52,000$ & $\S4.2$\\
        Variables with low V1 classification probabilities & ${\sim} 47,000$ & $\S4.3$\\ 
        \hline
        Variables with definite classifications & ${\sim} 278,000$ & $\S6.1$\\         	 
        Variables with uncertain classifications & ${\sim} 78,000$ & $\S6.2$\\   
        Variables matched to Gaia DR2 variables & ${\sim} 23,000$ & $\S6.5$\\           
        Variables with Gaia DR2 $A_G$ extinction and $E(BP-RP)$ reddening estimates & ${\sim} 169,000$ & $\S6.6$\\   
        Saturated/Faint VSX Variables & ${\sim} 56,000$ & Appendix A\\
        \hline
        Variables classified by ASAS-SN & ${\sim} 412,000$ & $\S6.1$\\    
        Rare and transient variables & ${\sim} 4,000$ & $\S6.3$\\         
        Final catalog on the ASAS-SN Variable Stars Database & ${\sim} 416,000$ & $\S6, \S7$\\        
    
	\end{tabular}

\end{table*}   

\begin{figure}
	\includegraphics[width=0.5\textwidth]{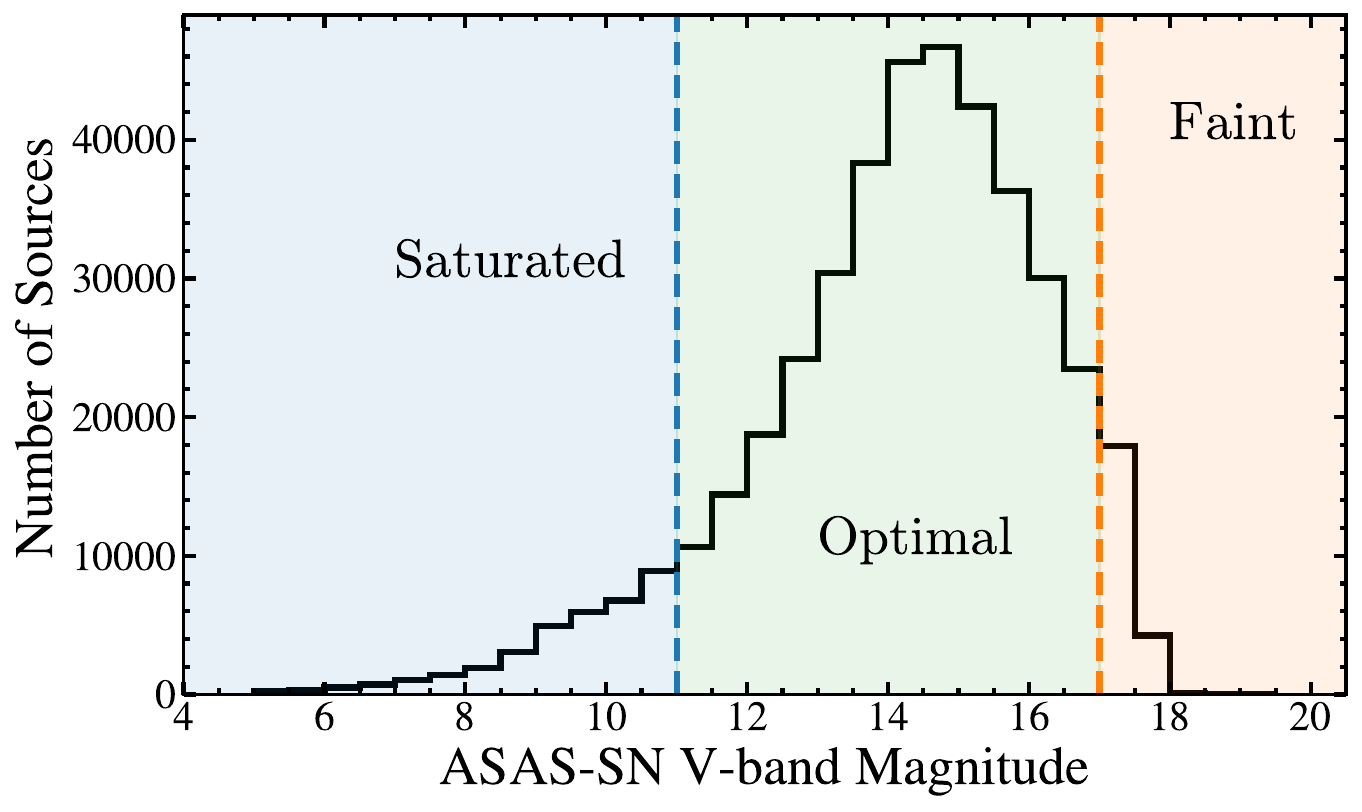}
    \caption{Distribution of variables in their average ASAS-SN V-band magnitude.}
    \label{fig:fig1}
\end{figure}

Figure \ref{fig:fig1} illustrates the distribution of variables by their average V-band magnitude. In this work, we consider a variable to be saturated if $V\leq11$ mag and faint if $V\geq17$ mag. The optimal magnitude range for ASAS-SN is $11< V <17$ mag. We do not take into consideration blending/crowding effects when computing the average magnitudes. The main body of the paper focuses on the ${\sim} 356,000$ variables with $11< V <17$ mag. In the Appendix we discuss the variables with $V\leq11$ mag and $V\geq17$ mag.

%2MASS \citep{2006AJ....131.1163S}
%the Phase Dispersion Minimization (PDM, \citealt{1978ApJ...224..953S})

\section{Constructing a robust variability classifier}
\label{varclass}
Here we describe the procedure we used to construct a training set and develop a variability classification pipeline. In Section $\S3.1$, we describe the cross-matches made to external catalogs. In Section $\S3.2$, we describe the procedure we took to derive periods for these sources. In Section $\S3.3$, we discuss the initial V1 random forest classifier model from Paper I and the steps we took to select variables for our training set. In Section $\S3.4$, we describe our refinement criteria for the initial RF classifications and discuss our enhanced training set in section $\S3.5$. We discuss our final V2 random forest classifier model in Section $\S3.6$.

\subsection{Cross-matches to external catalogs}
We cross-matched the variables with Gaia DR2 \citep{2018arXiv180409365G} using a matching radius of 5\farcs0. When multiple cross-matches are returned, we select the nearest cross-match as the best source from the external catalog. This is reasonable because ASAS-SN astrometry is generally better than 1\farcs0. A significant fraction of these sources have large parallax errors and/or negative parallaxes, hence reliable distances cannot be derived. We utilize the probabilistic distance estimates from \citet{2018AJ....156...58B} in our pipeline. Even poor distance estimates aid in classification because they generally indicate that the source has to be a distant giant rather than a nearby dwarf.

We also crossmatch the variables with 2MASS \citep{2006AJ....131.1163S} and AllWISE \citep{2013yCat.2328....0C,2010AJ....140.1868W} using a matching radius of 10\farcs0. This provides near-infrared, mid-infrared and optical colors for use in classification. We used \verb"TOPCAT" \citep{2005ASPC..347...29T} both to query the Gaia DR2 database, and to cross-match our sources with the 2MASS and AllWISE catalogs .

\subsection{Period Determination}
Periods were derived for the ${\sim}420,000$ sources following the procedure described in \citet{2018MNRAS.477.3145J}.
The Generalized Lomb-Scargle (GLS, \citealt{2009A&A...496..577Z,1982ApJ...263..835S}), the Multi-Harmonic Analysis Of Variance (MHAOV, \citealt{1996ApJ...460L.107S}), and the Box Least Squares (BLS, \citealt{2002A&A...391..369K}) periodograms were used to search for periodicity. We use the \verb"astrobase"  implementation of the GLS, BLS and MHAOV periodograms \citep{astrob}. The \verb"astropy" implementation of the GLS algorithm \citep{2013A&A...558A..33A} was used to derive the window function for each light curve.

In order to minimize the effect of outliers during the period search, we clip the light curve data to select only the epochs with magnitudes between the 1$^{\rm st}$ and 99$^{\rm th}$ percentiles. Periods were searched over the range $0.05\leq P \leq1000$ days. We initialized the MHAOV periodogram with $N_{harm}=5$ harmonics to provide better sensitivity to complex variability signals, while the BLS periodogram was initialized with 200 phase bins and a minimum (maximum) transit duration of $0.1$ $(0.3)$ in phase. The 5 best periods from each periodogram were saved. BLS periods were only selected if the BLS power was $<0.3$.

The observed periodogram for a given light curve is the convolution of a window function that depends on the survey and the `true' periodogram (see the discussion in \citealt{2018ApJS..236...16V}). Ideally, one would want to retrieve the true periodogram by deconvolution, but this is difficult to achieve in practice. A good method to identify potentially false periodicity is to examine the window function. The Lomb-Scargle window function \citep{2013A&A...558A..33A} is found using a light curve with the same temporal sampling as the data but with the magnitude measurements set to a constant. The 10 strongest peaks in this window function are saved and are considered to be the possible aliased periods for the light curve. 
\begin{figure*}
	\includegraphics[width=\textwidth]{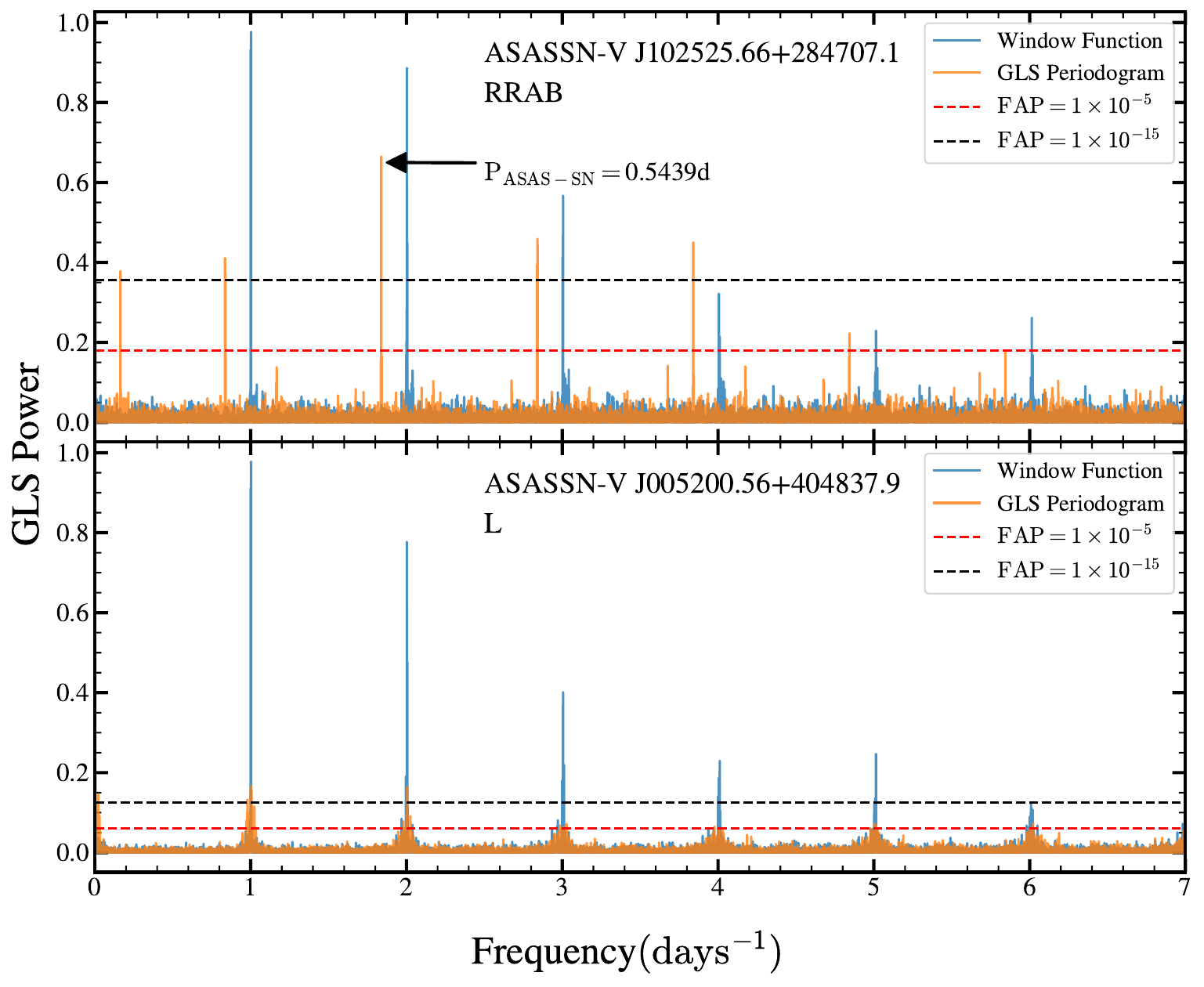}
    \caption{The GLS periodogram (gold) and the associated window function (blue) for the RRAB variable ASASSN-V J102525.66+284707.1 (top) and the irregular variable ASASSN-V J005200.56+404837.9 (bottom). The best period of $\rm P=0.5439$ d for ASASSN-V J102525.66+284707.1 is marked, along with the false alarm probability levels corresponding to $1\times10^{-5}$ (red) and $1\times10^{-15}$ (black).  }
    \label{fig:fig2}
\end{figure*}

Comparisons of the window function and the GLS periodogram for the RRAB variable ASASSN-V J102525.66+284707.1 (top panel) and the irregular variable ASASSN-V J005200.56+404837.9 (bottom panel) are shown in Figure \ref{fig:fig2}. The false alarm probability (FAP) levels corresponding to $1\times10^{-5}$ and $1\times10^{-15}$ are also shown. The false alarm probability is an estimate of the probability that a light curve with no periodicity results in a periodogram peak of a given power and it is a good indicator of the significance of a periodogram peak. In the case of strongly periodic variables like ASASSN-V J102525.66+284707.1, we see that the underlying window function is suppressed and the true period of $\rm P=0.5439$ d dominates the periodogram peaks. For irregular sources and sources with weak periodicity like ASASSN-V J005200.56+404837.9, the window function can dominate the observed periodogram and result in aliases being retrieved as periods. 

After running the GLS, BLS and MHAOV periodograms, we have 15 possible periods. To select the best period from these 15 periods, we used the \verb"gatspy" implementation of the Supersmoother algorithm \citep{gatspy,1994PhDT........20R}. Supersmoother is an algorithm that performs nonparametric regression based on local linear regression with adaptive bandwidths \citep{1994PhDT........20R}. Each of the retrieved periods are compared with the aliases derived from the associated window function, and those periods within $10^{-3}$ of an alias and with a $\rm FAP <1\times10^{-15}$ are removed from consideration. We generate a Supersmoother fit to the light curve and calculate the Supersmoother model score from \verb"gatspy" for each of the remaining periods. The period with the highest Supersmoother score is chosen as the best period. We use the Supersmoother model to further refine this period within a range $0.995 \rm \,P_{\rm best}\leq P \leq1.005 \,\rm P_{\rm best}$. We find that this approach generally works well, except in cases where the Supersmoother model prefers a multiple of the true period over the true period. We will identify and correct most of these cases in $\S3.4$.

We refined this process of period selection using the Lafler-Kinmann string length statistic \citep{1965ApJS...11..216L,2002A&A...386..763C}. We use the definitions of \begin{equation}
    \Theta(P)=\frac{\sum_{i=1}^{\rm N} (m_{i+1}-m_i)^2}{\sum_{i=1}^{\rm N} (m_{i}-\overline m)^2},
	\label{eq:thetp}
\end{equation} 
and
\begin{equation}
    T(\phi | P)=\Theta(P)  \times \frac{(N-1)}{2N}
	\label{eq:tp}
\end{equation} from \citet{2002A&A...386..763C} where the $m_i$ are the magnitudes sorted by phase and $\overline m$ is the mean magnitude. The original Lafler-Kinmann string length statistic ($\Theta(P)$) is derived using the sum of the squares of the vector lengths required to connect re-ordered measurements in phase sequence \citep{2002A&A...386..763C}. This is scaled by a factor of ${(N-1)}/{2N}$ to normalize the result and remove sample-size bias \citep{2002A&A...386..763C}, resulting in the $T(\phi | P)$ statistic used in this work. We use $T(\phi | P)$ to assign a score to each period after sorting the light curve by ascending phase. 

Aliases are removed in a similar fashion to the Supersmoother approach, but the tolerance was made absolute and allowed to vary based on the period being tested. For $0.05\leq P \leq10$ d, $\rm tol=0.005$ d, for $10\leq P \leq300$ d, $\rm tol=0.05$ d, and for $P>300$ d, $\rm tol=1$ d. The best period in this case is the period with the smallest $T(\phi | P)$.

\begin{figure*}
	\includegraphics[width=\textwidth]{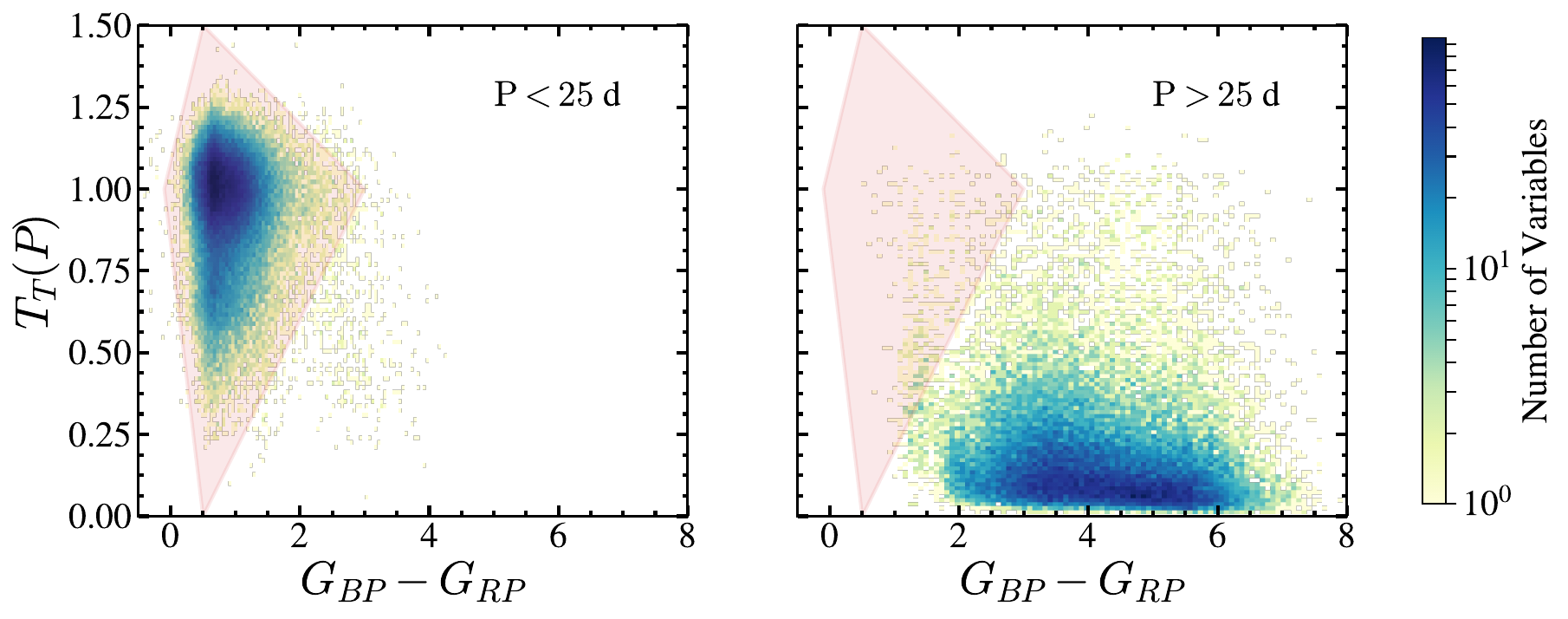}
    \caption{$T(t)$ for short period variables ($\rm P<5$ d, left) and long period variables (right) as a function of the Gaia DR2 $G_{BP}-G_{RP}$ color. The selection region for short period variables is shaded in red.}
    \label{fig:fig3}
\end{figure*}

We also use the Gaia DR2 $G_{BP}-G_{RP}$ color and $T(t)$ calculated on the light curves as in equations \ref{eq:thetp} and \ref{eq:tp} but with the data ordered by time instead of phase to rule out shorter periods for redder, long period variables (LPVs). $T(t)$ is sensitive to the structure and clustering of sorted data points. For the unphased light curves of variables with short periods, $T(t)$ is larger (worse) than for the unphased light curves of LPVs. We can rule out short periods ($P<25$ d) for LPVs by using their location on the $T(t)$ vs. $G_{BP}-G_{RP}$ diagram (Figure \ref{fig:fig3}). Periods shorter than 25 d are only allowed for the variables falling into the red shaded region. This process effectively eliminates the common aliases of a sidereal day that are frequently found for the light curves of red, irregular variable sources.  

\subsection{Initial Random Forest Classification}
We built a variability classifier in Paper I based on a random forest model using \verb"scikit-learn"  \citep{2012arXiv1201.0490P,breiman}. We will refer to this as the V1 classifier. The set of variables used to train this classifier consisted of ${\sim} 56,000$ variables that passed visual inspection of their light curves. We augmented this classifier with the remaining ${\sim} 10,000$ variables in Paper I to increase the diversity of our training set from Paper I.
The goal was to provide general classifications into broad groups: CEPH (Cepheids), RRAB (RR Lyrae, type ab), RRC/RRD (RR Lyrae, types c/d), ROT (Rotational variables), SR/IRR (Semi-regular/irregular variables), M (Mira variables), DSCT (DSCT/HADS variables) and ECL (Eclipsing Binaries). These broad classes were selected to reduce the complexity of the classifier, and to provide an accurate initial classification prior to refining them into subclasses in the next stages.

We choose a sample of known variables with optimal ASAS-SN magnitudes ($11\leq V \leq 17$) for classification with the V1 random forest classifier. Sources with an OGLE identifier in their name were not selected at this point, and are classified later. To this sample of variables, we fit a Fourier model of variable order,
\begin{equation}
    M(\phi)=m_0 + \sum_{i=1}^{N} \left(a_i\sin(2\pi i\phi)+b_i\cos(2\pi i\phi)\right)
	\label{eq:ffit}
\end{equation} where $m_0$ is fixed to be the median magnitude for each source, $4 \leq N \leq 16$ is the order of the Fourier model and $\phi$ is the phase at a given epoch. The amplitude of the $i^{th}$ harmonic is $A_i={(a^2_i+ b^2_i)^{{1}/{2}}}$ and the phase angle is $\Phi_i =\tan^{-1}\space(-{b_i}/{a_i})$. We define the amplitude ratios of the harmonics as $R_{ij}=A_i/A_j$ and the difference in phase angle between two harmonics as $\Phi_{ij} =\Phi_{i}-i\Phi_{j}$. The best fit Fourier model of order $N_{\rm best}$ is the Fourier model that minimizes $\chi^2_{\rm DOF}$. Through this process, we find that a Fourier model of order 6 is sufficient to describe most periodic variables and provide features for classification. During this initial round of classification, we derived distances by simply inverting the Gaia DR2 parallax. In the cases with very large parallax uncertainties, and/or negative parallaxes, an arbitrarily large integer was assigned to the distance to signify that they are very distant. Distance estimates from \citet{2018AJ....156...58B} are used in the V2 classifier (Section $\S3.6$). The complete list of features and their importances in the random forest classifier are summarized in Table \ref{tab:features}. Feature importances are calculated using the mean decrease impurity algorithm (Gini importances) implemented in the \verb"scikit-learn" version of the Random Forest Classifier \citep{2012arXiv1201.0490P}.  Compared to the RF classifier from Paper I, we introduced 12 new Fourier features ($a_2$, $a_4$, $b_2$, $b_4$, $R_{41}$, $R_{32}$, $R_{42}$, $R_{43}$, $\Phi_{41}$, $\Phi_{21}$, $N_{\rm best}$ and $\chi^2_{\rm DOF}$) and 11 other features ($1/\eta$, $\sigma$, $\rm MAD$, $1/\eta$, $m_{\rm slope}$, $J-K_s$, $G_{BP}-G_{RP}$, $W1-W2$, $M_G$, $W_{\rm RP}$ and $W_{JK}$) in this work.
We set the number of decision trees in the forest as \verb"n_estimators=800", pruned the trees at a maximum depth of \verb"max_depth=16" to prevent over-fitting, set the number of samples needed to split a node as \verb"min_samples_split=10" and set the number of samples at a leaf node as \verb"min_samples_leaf=5". To further reduce over-fitting, weights were assigned to each class by initializing \verb"class_weight=`balanced_subsample'". These parameters were optimized using cross-validation to maximize the overall $F_1$ score of the classifier.

\begin{table*}
	\centering
    \setlength\tabcolsep{2pt}
	\caption{Variability features and their importances across the two versions of the ASAS-SN random forest variability classifier}
	\label{tab:features}
	\begin{tabular}{||p{2cm}| p{5cm}| c| c| p{4cm}||} % four columns, alignment for each
		\hline
		Feature & Description & V1 Importance & V2 Importance & Reference \\
		\hline
        logP & Base 10 logarithm of the period & 18\% & 25\%  &  -  \\
        $J-K_s$ & 2MASS $J-K_s$ color  & 9\% & 5\%  &\citet{2006AJ....131.1163S} \\
        $J-H$ & 2MASS $J-H$ color  & -& 7\%  &\citet{2006AJ....131.1163S} \\        
        $G_{BP}-G_{RP}$ & Gaia DR2 $G_{BP}-G_{RP}$ color  & 8\% & 5\% & \citet{2018arXiv180409365G} \\ 
        $W1-W2$ & WISE $W1-W2$ color  & -& 0\%  & \citet{2013yCat.2328....0C,2010AJ....140.1868W}\\ 
        $M_G$ & Absolute Gaia DR2 G-band magnitude  & 7\% & -  & \citet{2018AJ....156...58B,2018arXiv180409365G} \\  
        $W_{RP}$ & Absolute Wesenheit Gaia DR2 $G_{RP}$-band magnitude  & -& 8\%  & \citet{2018AJ....156...58B,2006AJ....131.1163S,1982ApJ...253..575M} \\     
        $W_{JK}$ & Absolute Wesenheit 2MASS $K_s$-band magnitude & -& 6\%  & \citet{2018AJ....156...58B,2006AJ....131.1163S,1982ApJ...253..575M} \\       
        $a_{2}$ & Fourier component $a_{2}$ & 1\% & - &-\\ 
        $b_{2}$ & Fourier component $b_{2}$ & 1\% & - &- \\  
        $a_{4}$ & Fourier component $a_{4}$ & 1\% & - &-\\ 
        $b_{4}$ & Fourier component $b_{4}$ & 0\% & - &- \\   
        $R_{41}$ & Ratio between the amplitudes of the $4^{\rm th}$ and $1^{\rm st}$ harmonics & 2\% & 4\%  & - \\         
        $R_{31}$ & Ratio between the amplitudes of the $3^{\rm rd}$ and $1^{\rm st}$ harmonics & 2\% & 3\%  & - \\  
        $R_{21}$ & Ratio between the amplitudes of the $2^{\rm nd}$ and $1^{\rm st}$ harmonics & 3\% & 9\%  & - \\ 
        $R_{32}$ & Ratio between the amplitudes of the $3^{\rm rd}$ and $2^{\rm nd}$ harmonics & 1\% & - & - \\ 
        $R_{42}$ & Ratio between the amplitudes of the $4^{\rm th}$ and $2^{\rm nd}$ harmonics & 1\% & - & - \\ 
        $R_{43}$ & Ratio between the amplitudes of the $4^{\rm th}$ and $3^{\rm rd}$ harmonics & 1\% & - & - \\         
        $\Phi_{41}$ & Difference between the phase angle of the $4^{\rm th}$ and $1^{\rm st}$ harmonics & 1\% & - & -\\
        $\Phi_{31}$ & Difference between the phase angle of the $3^{\rm rd}$ and $1^{\rm st}$ harmonics & 1\% & - & -\\        
        $\Phi_{21}$ & Difference between the phase angle of the $2^{\rm nd}$ and $1^{\rm st}$ harmonics & 2\% & - & -\\ 
        $N_{\rm best}$ & Best fit order of the Fourier model & 0\%  & - & -\\  
        $\chi^2_{\rm DOF}$ & Reduced $\chi^2$ statistic of the best fit Fourier model & 2\% & - & -\\   
        $A$ & Amplitude of the light curve (between the 5$^{\rm th}$ and 95$^{\rm th}$ percentiles)  & 7\% & 6\%  &- \\ 
        $\rm Skew$ & Skewness of the magnitude distribution & 5\% & 3\%  &-\\ 
        $\rm Kurt$ & Kurtosis of the magnitude distribution & 3\% & - &-\\          
        $T_m$ & M-test statistic & 3\% & - & \citet{2006AJ....132.1202K} \\   
        IQR & Difference between the 75$^{\rm th}$ and 25$^{\rm th}$ percentiles in magnitude & 6\% & 3\%  &-\\         
        $A_{\rm HL}$ & Ratio of magnitudes brighter or fainter than the average & 6\% & 3\%  & \citet{upsilon} \\
        rms & Root-mean-square statistic of the light curve  & 7\% & 5\%  & -  \\ 
        $\sigma$ & Standard deviation of the light curve  & -& 4\%  & -  \\  
        $\rm MAD$ & Median absolute deviation of the light curve  & -& 4\%  & -  \\    
        $1/\eta$ & Inverse of the $\eta$ (Von Neumann index) value for the light curve  & 4\% & - & \citet{vn}  \\   
        $m_{\rm slope}$ & Slope of a linear fit to the light curve  & 1\% & - & -  \\          
		\hline
	\end{tabular}
\end{table*} 

The overall results can be evaluated based on the \begin{equation}
    \rm precision=\frac{\alpha}{\alpha + \beta} \,,
	\label{eq:prec}
\end{equation} \begin{equation}
    \rm recall=\frac{\alpha}{\alpha + \gamma}\,,
	\label{eq:rec}
\end{equation} and the harmonic mean of the two, \begin{equation}
    F_1=2 \bigg( \frac{\rm precision\times \rm recall}{\rm precision+\rm recall}\,\bigg),
	\label{eq:f1}  \end{equation} where $\alpha$, $\beta$ and $\gamma$ are the number of true positives, false positives and false negatives respectively.
The sample of ASAS-SN variables from paper I were split for training ($80\%$) and testing ($20\%$) in order to evaluate the performance of the RF classifier. For any given variable, the RF classifier assigns a classification probability $\rm Prob(C)$ to each of the 8 variable classes where $\sum_{i=1}^{8}\rm Prob(C) =1$. The output classification of the RF classifier is the class with the highest probability. The performance of the classifier is summarized in Table \ref{tab:perf}. We noted that this version of the classifier needed improvement, particularly for the CEPH, ROT and RRC/RRD classes. We suspect that this is due to the small number of these variables in our Paper I training set. The classification of sources into the remaining classes is accurate enough for our purposes. The overall $F_1$ score for version V1 of the classifier is 93.3\%. 

\begin{table*}
	\centering
	\caption{Performance of the ASAS-SN random forest classifiers}
	\label{tab:perf}
	\begin{tabular}{lrrrrrrr}
		\hline
		Class & V1 Precision & V1 Recall & V1 $F_1$ score & V2 Precision & V2 Recall & V2 $F_1$ score& V2 Sources\\
		\hline
        $\delta$ Scuti & 94$\%$ & 94$\%$ & 94$\%$ & 100$\%$ & 100$\%$ & 100$\%$ & 1500\\
		RR Lyrae (Type ab) & 97$\%$ & 99$\%$ & 98$\%$ & 100$\%$ & 100$\%$ & 100$\%$ & 26289 \\
		RR Lyrae (Types c and d) & 90$\%$ & 92$\%$ & 91$\%$ & 99$\%$ & 99$\%$ & 99$\%$ & 5999 \\        
		Cepheids & 100$\%$& 70$\%$& 82$\%$ &  99$\%$ & 100$\%$ & 99$\%$ & 977\\
		Eclipsing Binaries & 98$\%$ & 98$\%$ & 98$\%$ & 100$\%$ & 100$\%$ & 100$\%$ & 52912\\
		Semi-regular/Irregular variables & 100$\%$ & 100$\%$ & 100$\%$ & 100$\%$ & 100$\%$ & 100$\%$ & 66416\\ 
		Rotational variables & 82$\%$  & 95$\%$ & 88$\%$ & 97$\%$ & 100$\%$ & 98$\%$ & 3597\\        
		Mira & 95$\%$ & 100$\%$ & 97$\%$ & 98$\%$ & 100$\%$ & 99$\%$ & 7832\\        
		\hline
	\end{tabular}

\end{table*}   
We want to apply this version of the classifier to the ${\sim}150,000$ other VSX variables with $11< V <17$  and VSX classifications that could be explicitly matched with the outputs of our classifier. For example, we do not include objects with MISC/VAR classifications. We first select ${\sim} 110,000$ variables where the VSX and V1 classifications agree. Next, we compare the ASAS-SN periods and the VSX periods. If any of the periods retrieved from the ASAS-SN light curve are comparable to the VSX period to a tolerance $\Delta P/P_{\rm VSX}$ given in Table \ref{tab:tols}, and depending on the variable class, we assign that ASAS-SN period as the best period for the source. The tolerance was made to vary by variable class due to the differences in periodicity amongst the different variable types (see \citealt{2013MNRAS.434.3423G} for a discussion on the accuracy of period estimation for different variable classes and periodograms). Certain semi-regular variables, rotational variables and Mira variables tend to have poorly defined periods, and in some cases display changes in their period over time. These classes were given a higher matching tolerance to account for this behavior.

In the case of eclipsing binaries, we also check to see if a multiple of the VSX period is retrieved, and in such cases, we correct the ASAS-SN period by multiplying it by the appropriate constant. We report the results of the period matching as well as the number of sources in this sample of variables without an associated VSX period in Table \ref{tab:tols}. The ${\sim} 94,000$ variables that have matching classes and matching periods are assumed to be good. The ${\sim} 11,000$ variables with matching classes and discrepant periods were visually reviewed to verify or correct the new periods before including them in the preliminary V2 training sample.

There were ${\sim} 39,000$ variables with discrepant classifications. For these objects, we kept those with V1 classification probabilities of $\rm Prob>0.5$ for $|b|>25$ deg and $\rm Prob>0.7$ for $|b|<25$ deg. The Galactic latitude is included because crowding and blending will be an increasing problem for ASAS-SN at low latitudes. This left ${\sim} 9,000$ variables which we visually inspected and corrected (46\%), remained unchanged (19\%) or dropped (35\%). This added ${\sim} 5,000$ additional variables to the preliminary V2 training sample.

These procedures provide us with a first pass at a new training set of ${\sim} 177,000$ sources, including ${\sim} 66,000$ ASAS-SN variables. To build V2 of the classifier, we will first refine these classifications using light curve characteristics, distance measurements from \citet{2018AJ....156...58B} and multi-band photometry from Gaia DR2, 2MASS and WISE \citep{2018arXiv180409365G,2006AJ....131.1163S,2013yCat.2328....0C,2010AJ....140.1868W}.
\begin{table*}
	\centering
	\caption{Tolerance when matching periods. The percentage of matching periods and discrepancies given the corresponding tolerance are reported. The number of sources that lack a VSX period in each variability group are also listed.}
	\label{tab:tols}
	\begin{tabular}{lrrrr}
		\hline
		Class & Tolerance ($\Delta P/P_{\rm VSX}$) & Matches & Discrepancies & Unreported VSX periods\\
		\hline
        $\delta$ Scuti &0.001\% & 55\% & 45\% & 26\\
		RR Lyrae (Type ab) & 5\% & 99\% & 1\% & 760  \\
		RR Lyrae (Types c and d) & 5\% & 99\% & 1\% & 56 \\        
		Cepheids & 5\% & 100\% & 0\% & 9\\
		Eclipsing Binaries & 0.1\% & 93\% & 7\% & 906 \\
		Semi-regular/Irregular variables & 25\% & 79\%  & 21\% & 9374\\ 
		Rotational variables & 20\% & 79\% & 21\% & 74\\        
		Mira & 25\% & 91\% & 9\% & 406\\        
		\hline
	\end{tabular}

\end{table*}  

\subsection{Refining Classifications}
While the initial classifications assigned by the RF classifier are generally accurate, we are able to correct some misclassifications and refine some of the broad groups into sub-types using the period, light curve statistics, colors, photometry from external catalogs and distance measurements. We apply these both to the updated training set and to the final sample. 
Distance estimates from \citet{2018AJ....156...58B} were used to derive absolute magnitudes during the refinement process. To account for interstellar extinction in the absolute Gaia DR2 $G_{RP}$ and 2MASS $K_s$ band magnitudes, we use the reddening-free Wesenheit magnitudes \citep{1982ApJ...253..575M,2018arXiv180803659L} \begin{equation}
    W_{RP}=M_{\rm G_{RP}}-1.3(G_{BP}-G_{RP}) \,, 
	\label{eq:wrp}
\end{equation}
and
\begin{equation}
    W_{JK}=M_{\rm K_s}-0.686(J-K_s) \,
	\label{eq:wk}
\end{equation}
instead of the uncorrected absolute magnitudes in V2 of the RF classifier and the classification refinement process. A similar reddening-free formulation of color requires at least 3 different photometric measurements. This is difficult to do with just the two Gaia DR2 bands ($G_{BP}$ and $G_{RP}$), thus we do not use reddening-free colors in this work. However, we discuss the effects of extinction on color and the ASAS-SN classifications in Section $\S6.5$.

Here we discuss the refinement criteria used on the different broad groups that are assigned to the variables classified with the RF classifier.

\subsubsection{Delta Scuti}
$\delta$ Scuti stars pulsate at high frequencies (P$<0.3$ d) and are located towards the lower end of the instability strip \citep{1979PASP...91....5B}. $\delta$ Scuti variables are also known to follow a period-luminosity relationship \citep{1990A&AS...83...51L}. High amplitude $\delta$ Scuti variables (HADS) are a sub-type of the $\delta$ Scuti stars that have amplitudes $A>0.15$ mag. HADS variables are more commonly discovered by sky surveys like ASAS-SN than the lower amplitude DSCT variables. 

\begin{figure*}
	\includegraphics[width=\textwidth]{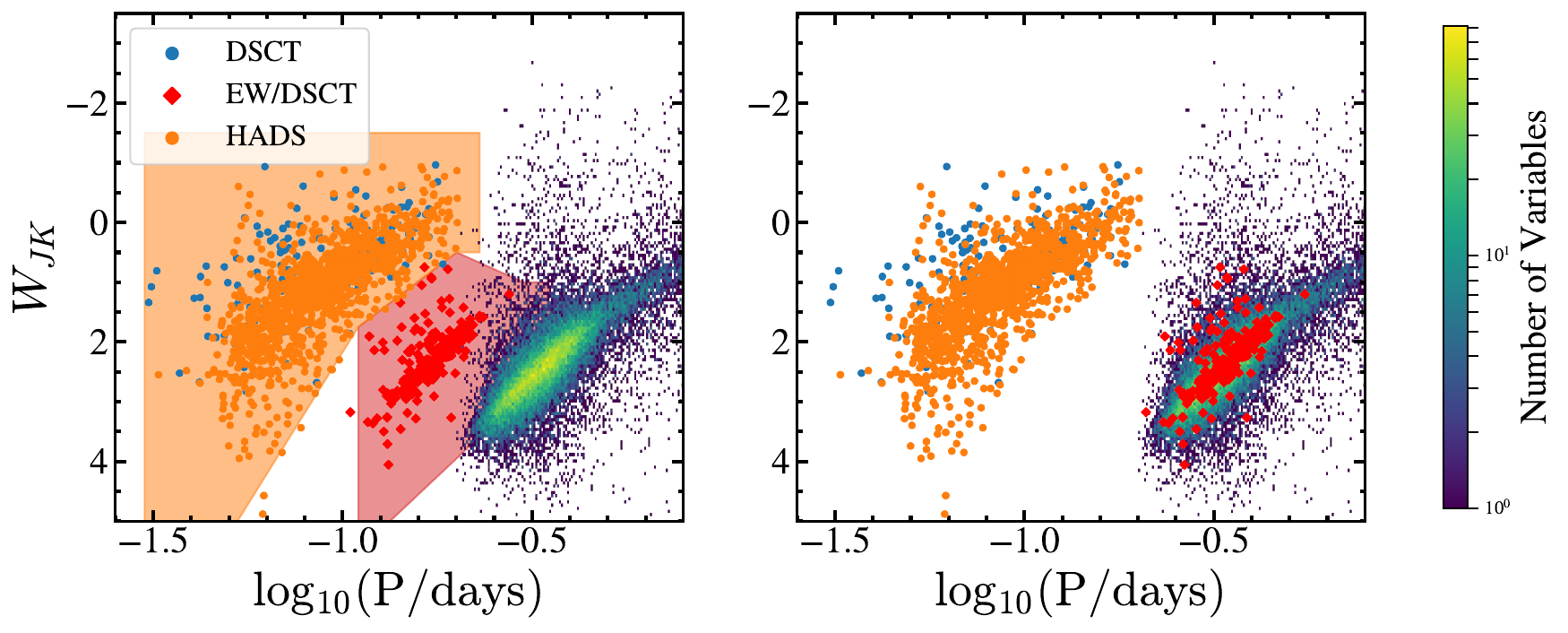}
    \caption{The PLR for $\delta$ Scuti variables (HADS as orange, DSCT as blue) and EW eclipsing binaries (blue to yellow density shading) in the absolute $W_{JK}$ Wesenheit magnitude. The red points in the left panel are EW binaries given half of their correct period and then misidentified as HADS variables. We take objects in the red shaded region and double their periods, which moves them into the locus of EW binaries (right panel). }
    \label{fig:fig4}
\end{figure*}

To refine the HADS/DSCT classifications, we use the period-luminosity relation (PLR) in absolute Wesenheit $W_{JK}$ magnitude (Figure \ref{fig:fig4}). A common classification error that reduces the purity of the HADS/DSCT sample is when contact eclipsing binaries at half their true period are mistaken for HADS/DSCT variables. This commonly happens in the period range $0.1 \leq \rm P \leq 0.25$ d. In the period-luminosity diagram, it is easy to distinguish the two variable types. Contact eclipsing binaries (EW) have a tight PLR \citep{2018ApJ...859..140C} and are fainter than the HADS/DSCT at any given period. Therefore, contact eclipsing binaries with half their true periods form a fainter PLR locus than the HADS/DSCT PLR (left panel in Figure \ref{fig:fig4}). We empirically isolate the misclassified EW variables in $W_{JK}$ vs. $\log \rm P$ space. We double their periods and change their classifications. After doubling the periods, these variables fall along the well-defined EW PLR (Right panel in Figure \ref{fig:fig4}). We identified 1664 misclassified EW variables, which constitute 4\% of all the EW variables in the complete sample of variables classified in this paper.

In a few cases, the RF classifier assigns longer period RR Lyrae to the DSCT class. We reclassify longer period variables ($P>0.25$ d) in the DSCT class into one of the RR Lyrae sub-types (RRAB/RRC) depending on their period. Sources that do not lie on an empirically defined locus for valid HADS/DSCT sources are assigned an uncertainty flag (i.e., DSCT:). 

\subsubsection{RR Lyrae}
RR Lyrae stars are sub-divided as RRAB, RRC and RRD variables. RRAB variables are fundamental mode, high amplitude pulsators with typical periods in the range $0.3 \leq \rm P \leq 1.2$ d. RRC variables are overtone pulsators with nearly symmetric light curves and have typical periods in the range $0.2 \leq \rm P \leq 0.5$ d. RRD variables are double-mode radial pulsators with fundamental periods $P_{FO}{\sim}0.5$ d and period ratios in the range $0.742<{P_{1O}}/{P_{FO}}<0.748$ \citep{2014PASP..126..509P}. The light curves of RRD variables are less regular due to the presence of both the fundamental and overtone modes of pulsation.

\begin{figure*}
	\includegraphics[width=\textwidth]{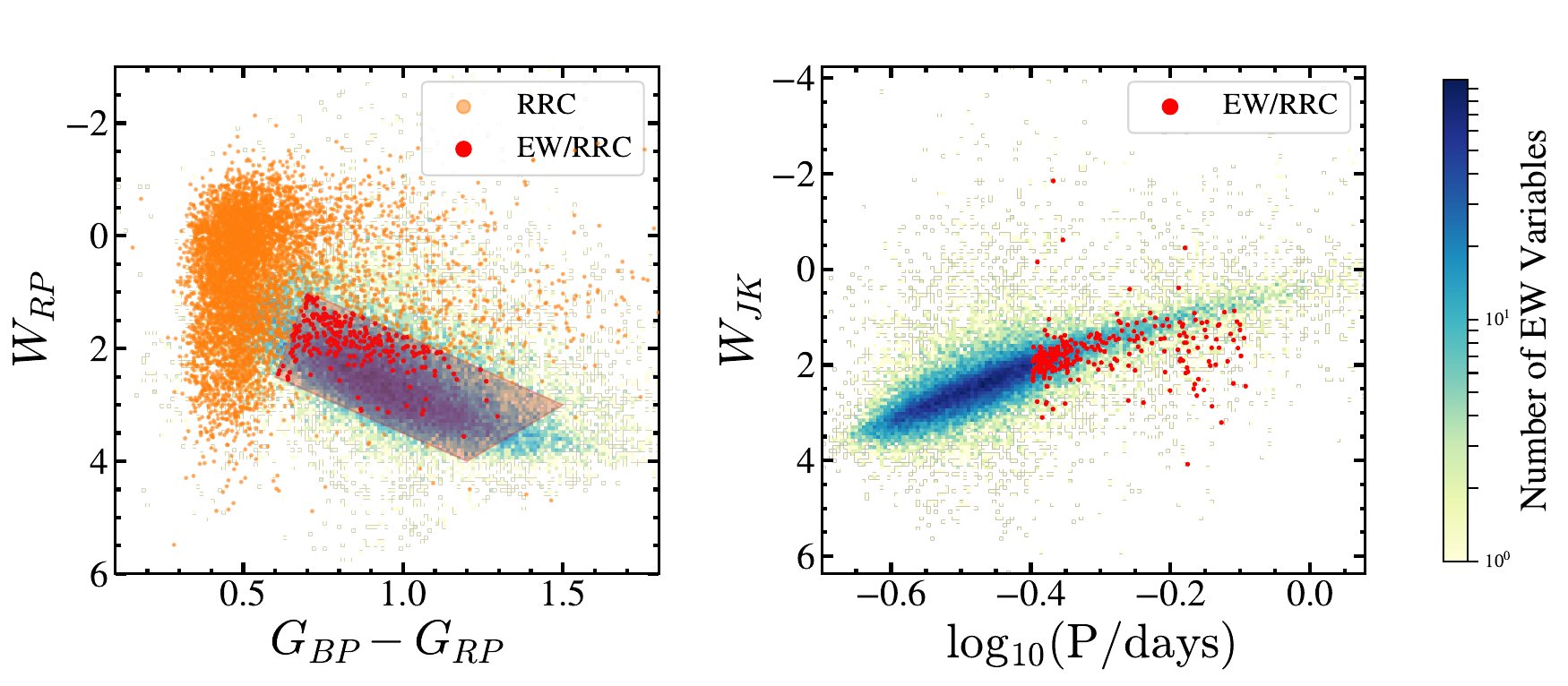}
    \caption{ \textit{Left}: The Wesenheit $W_{RP}$ vs. $G_{BP}$ - $G_{RP}$ color-magnitude diagram for the RRC and EW variables. The locus of misclassified RRC variables is highlighted by the red shaded region. RRC variables whose classifications were changed to that of a contact eclipsing binary are shown in red. The color-magnitude relationship for the EW variables is plotted as a 2-d histogram (blue). \textit{Right}: The $W_{JK}$-band PLR for the reclassified RRC variables after doubling the period. The distribution of the EW variables is shown as a 2-d histogram (blue).}
    \label{fig:fig5}
\end{figure*}

We discussed the issue of cross-contamination between RRC variables and short period contact binaries (EW) in Paper I. Using the period-luminosity space to distinguish between RRC variables and EW variables is difficult since the probabilistic distances to RR Lyrae stars from \citet{2018AJ....156...58B} have such a broad distribution in $W_{JK}$ that many RR Lyrae overlap with the PLR for EW variables. Instead we look at the $W_{RP}$ vs. $G_{BP}-G_{RP}$ color-magnitude space to separate these variables (Figure \ref{fig:fig5}). The EW variables have a tight relationship in the $W_{RP}$ vs. $G_{BP}-G_{RP}$ color-magnitude space which is distinct from the distribution of RRC variables beyond $G_{BP}-G_{RP} {{\sim}} 0.5$ mag. We empirically isolate the locus of EW binaries in this color-magnitude space to identify misclassified RRC variables. We change the classification of the RRC variables that fall in this region to EW and double their periods. These misclassified sources then lie on the contact binary PLR (right panel in Figure \ref{fig:fig5}). From the complete sample, we identify 674 variables that were initially classified as RRC but were reclassified as EW through this procedure. This constitutes 2\% of all EW variables. However, the degree of overlap between the RRC variables and the EW variables suggests that some RRC variables with significant interstellar extinction might also populate this empirically identified locus of EW variables, and these sources are then misclassified as EW variables in our pipeline.

To identify RRD candidates, we searched all the sources classified as RRAB/RRC for secondary periods, and those with period ratios $0.72<{P_{1O}}/{P_{F}}<0.78$ were classified as RRD variables. We also require that the fundamental period lie in the range $0.45<P_{F}<0.60$ d. As in Paper I, we chose this range in ${P_{1O}}/{P_{F}}$ to account for variations observed in RRD populations \citep{2009A&A...494L..17O} and dispersion due to noise. We report the fundamental period for all the RRD candidates.

There is likely some cross-contamination between RRAB and RRC variables in the period range $0.3 \leq \rm P \leq 0.5$ d. RRAB variables in crowded regions will tend to have lower amplitudes as result of blending, and therefore have a non-negligible chance of being classified as RRC variables owing to their smaller amplitudes.

We also check all the RRAB variables with $P>1$ ($P>0.9$) d to see if they follow the PLR for Classical (Type II Cepheids) respectively. RRAB variables falling into these PLR locii have their classifications changed to reflect their correct type. From the training sample, 95 variables initially classified as RRAB were reclassified as Cepheids. We considered the initial classification of these sources as RRAB variables as the correct classification, and calculated the RRAB contamination in the complete sample of Cepheids to be ${\sim}5\%$.

To further improve the purity of the RRAB and RRC samples, we require $-3<W_{JK}<5$ mag in order to account for the large observed dispersion in absolute Wesenheit magnitudes, and that the RRAB/RRC variables follow the period constraints mentioned above. Uncertain classifications are followed by a `:' (i.e., RRAB: and RRC:). RR Lyrae with uncertain classifications may include sources without Gaia DR2 data. RRAB/RRC sources with $P<0.2$ d are reclassified as HADS/DSCT/EW through the DSCT classification routine.

\subsubsection{Cepheids}
Cepheids are pulsators that form an important rung in the distance ladder. The first Cepheids discovered in the Magellanic clouds numbered in the few hundreds \citep{1908AnHar..60...87L,1912HarCi.173....1L}, but owing to the extensive modern sky surveys toward the Magellanic Clouds, close to 10,000 Cepheids \citep{2017AcA....67..103S} have now been identified. In the Milky Way, only ${\sim}1000$ Cepheids were discovered by 2011 \citep{2011A&A...530A..76W}, but thanks to recent variability studies, there are now ${\sim} 2100$ \citep{2018arXiv180502079C}. 

Fundamental mode Cepheids (DCEP) obey a period-luminosity relation \citep{1912HarCi.173....1L} and their light curve morphologies depend on the pulsation period (see for e.g., \citealt{2012ApJ...748..107P} for physically motivated multi-band light curve models for Cepheids). First overtone Cepheids (DCEPS) have periods $P<7$ d and tend to have smaller variability amplitudes than a DCEP variable of the same period. 

Type II Cepheids are intrinsically less luminous than classical Cepheids and commonly have light curves that are morphologically different from classical Cepheids at similar periods (e.g., \citealt{2006MNRAS.370.1979M}). The type II Cepheids are categorized based on their periods: BL Herculis (CWB) variables have periods $P<8$ d while W Virginis variables (CWA) have periods $P>8$ d. RV Tauri variables (RVA/RVB) tend to have periods $16<P<180$ d and have alternating primary/secondary minima of different depths. 

\begin{figure}
	\includegraphics[width=0.5\textwidth]{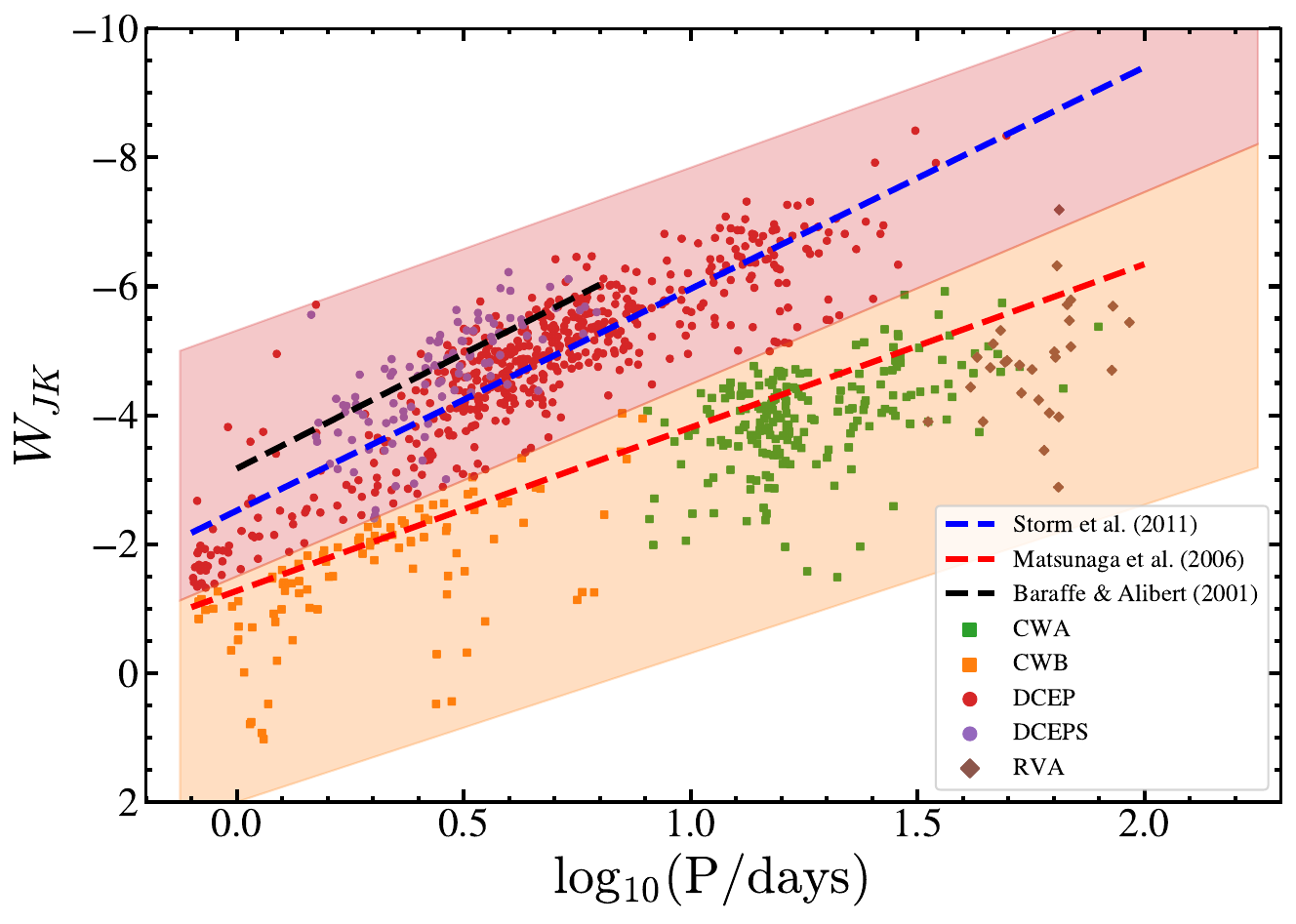}
    \caption{ The Wesenheit $W_{JK}$ magnitude PLR for Cepheids. The locus of classical Cepheids (both fundamental and overtone) is highlighted by the region shaded in red. The locus of Type II Cepheids is highlighted by the region shaded in orange. The Wesenheit $W_{JK}$ magnitude PLR is plotted in black for first overtone Cepheids \citep{2001A&A...371..592B}, in blue for classical fundamental mode Cepheids \citep{2011A&A...534A..94S} and in red for Type II Cepheids \citep{2006MNRAS.370.1979M}.}
    
    \label{fig:fig6}
\end{figure}

To classify Cepheids, we empirically group them into regions for Classical (red) and Type II (orange) pulsators (Figure \ref{fig:fig6}). Previously derived PLRs for first overtone Cepheids \citep{2001A&A...371..592B}, fundamental mode Cepheids \citep{2011A&A...534A..94S} and Type II Cepheids \citep{2006MNRAS.370.1979M} are shown for reference. Cepheids are broadly classified into one of these two classes. Those variables classified as Cepheids that are less luminous than the Type II locus are classified as rotational variables (ROT).

\begin{figure}
	\includegraphics[width=0.5\textwidth]{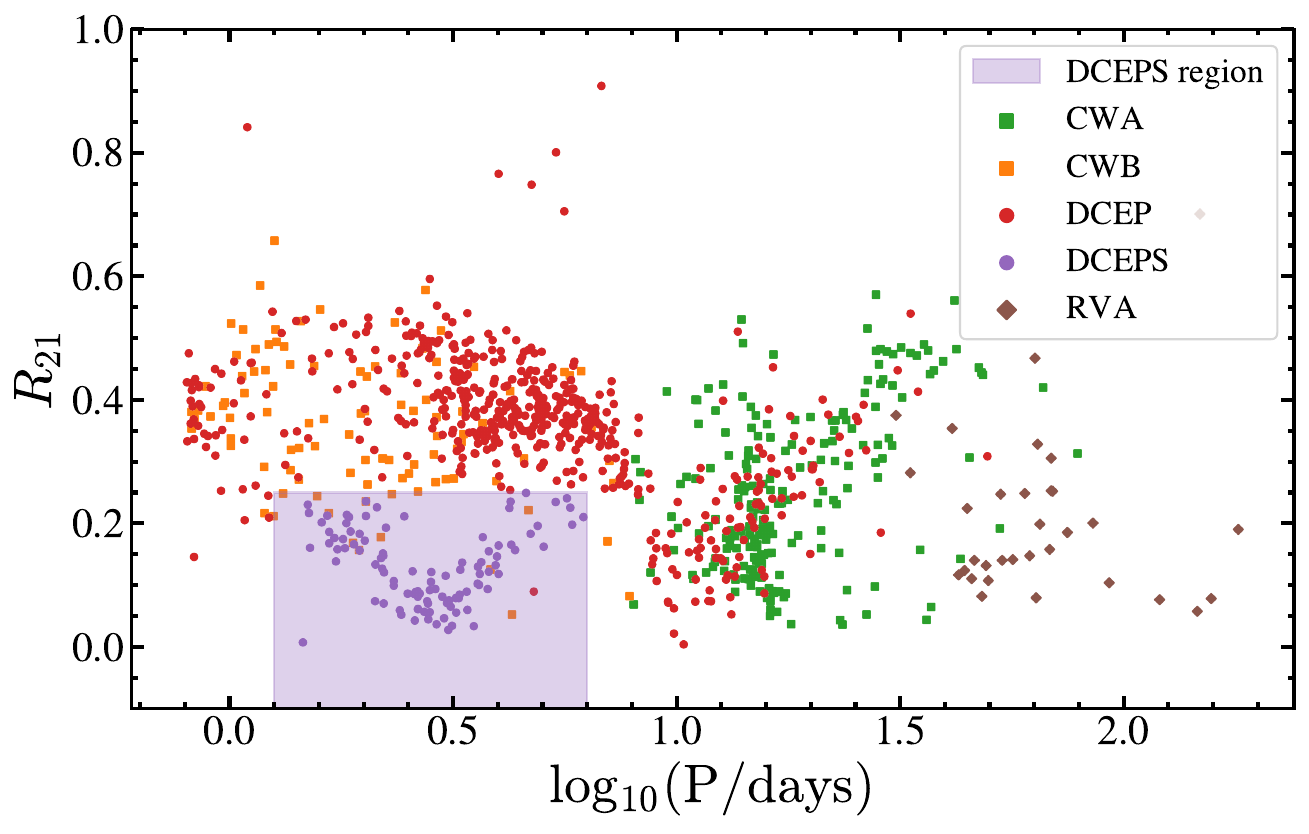}
    \caption{$R_{21}$ vs $\log \rm P$ for Cepheids. The points are colored according to the final classifications. The region occupied by first overtone Cepheids is shaded in purple.}
    
    \label{fig:fig7}
\end{figure}

In order to identify overtone pulsators, we look at the position of the Cepheids in the Fourier amplitude ratio $R_{21}$ vs $\log \rm P$ space (Figure \ref{fig:fig7}). In this space, the DCEPS variables are well-separated from the fundamental mode DCEP variables. The overlapping CWB variables cause no problems because the PLRs distinguish the two classes. 

We also looked more carefully for RV Tauri variables among the $P>16$ d CWA and $P>30$ d DCEP variables. We look for a minimum at phase ${\sim} 0.5$ in the light curve. If a minimum does not exist, we double the period. We then calculate the ratio between the secondary minimum at phase ${\sim} 0.5$ and the primary minimum at phase ${\sim} 0$ ($R_{\rm peaks}$). We call the source an RV Tauri variable if $R_{\rm peaks}<0.9$, $A>0.5$ mag and $16 \leq P \leq 180$ d. Cepheids that do not meet this criteria are reclassified into the CWA and DCEP groups. Periods are updated or halved based on the results of the automated period doubling.  

\begin{figure*}
	\includegraphics[width=\textwidth]{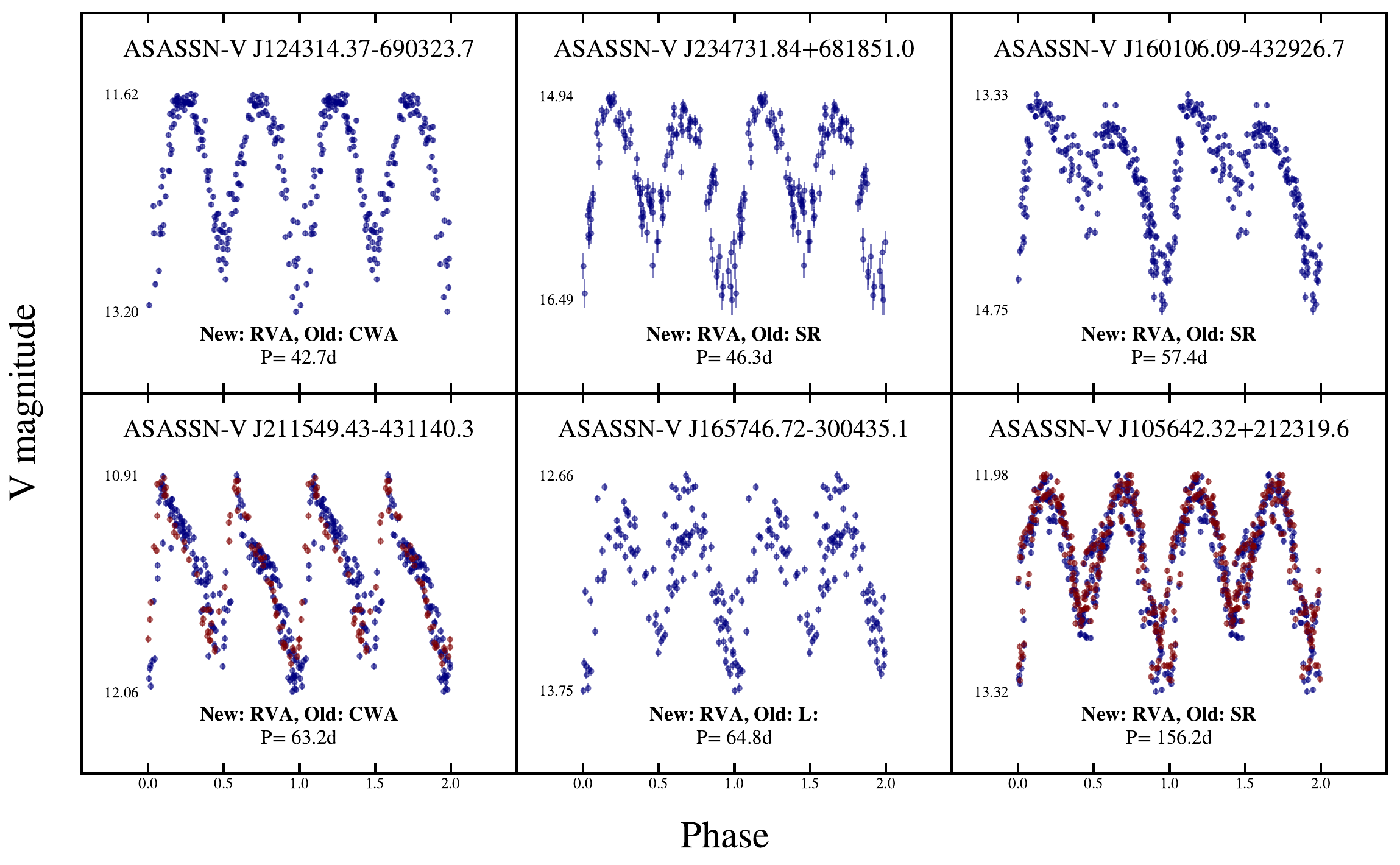}
    \caption{Phased light curves of new RV Tauri variables. The light curves are scaled by their minimum and maximum V-band magnitudes. Different colored points correspond to data from the different ASAS-SN cameras.}
    
    \label{fig:fig8}
\end{figure*}

We also suspected that RV Tauri variables are frequently classified as semi-regular variables. To examine this possibility, we ran all the semi-regular variables with classification probabilities $\rm Prob<0.75$, infrared colors $J-H<1$ mag, amplitudes $A>0.25$ mag, and periods $8 \leq P \leq 180$ d through the Cepheid pipeline. These cuts empirically identify atypical (low probability), yellow semi-regular variables that can then be further refined to isolate RV Tauri variables. If the variable falls into the defined classical/Type II PLR regions ($\S3.4.3$), we change their classification to reflect their membership as Cepheids and then check to see if they are RV Tauri variables. RV Tauri variables of type b (RVB) are not distinguished from RV Tauri variables of type a (RVA). Examples of newly identified RV Tauri variables are shown in Figure \ref{fig:fig8}. We reclassified 61 variables initially classified into the SR/IRR class as RV Tauri variables.

A shortcoming of our Cepheid classification routine is the identification of anomalous Cepheids (ACEP). Anomalous Cepheids form PLRs (fundamental mode and overtone) with luminosities larger than CWB variables and less than DCEP variables of the same period. ACEP variables have periods $0.35 \leq P \leq 2.5$ d \citep{2018arXiv180502079C}, and therefore contaminate the RR Lyrae, CWB and DCEP classes. Some fraction of the short period CWB variables are indeed ACEP variables, but we have not implemented a classification scheme to distinguish ACEP variables from CWB variables. From the complete sample of variables classified in this work, we calculate the misclassification rate of ACEP variables in the CWB sub-class as ${\sim} 3\%$. Cepheids are largely located towards the Galactic disk and are likely to suffer from considerable reddening. This makes it a challenge to distinguish the PLRs of ACEPs from those of CWB and DCEP variables. In this context, it is also virtually impossible to distinguish short period ($P<0.8$ d) overtone anomalous Cepheids toward the Galactic disk from RRC/RRAB variables without visual inspection of the light curves.

\subsubsection{Rotational Variables}
Rotational variables have light curves that show evidence of rotational modulation. We choose to retain the generic classification of ROT for variables in this class without defining sub-groups and instead focus on refining the membership within this class. The ROT variables are likely to be a diverse sample of rotational variable types, including $\alpha^2$ Canum Venaticorum variables (ACV), RS Canum Venaticorum-type (RS) binary systems, BY Draconis-type variables (BY), FK Comae Berenices-type variables (FKCOM), rotating ellipsoidal variables (ELL) and spotted T Tauri stars showing periodic variability (TTS/ROT).

\begin{figure}
	\includegraphics[width=0.5\textwidth]{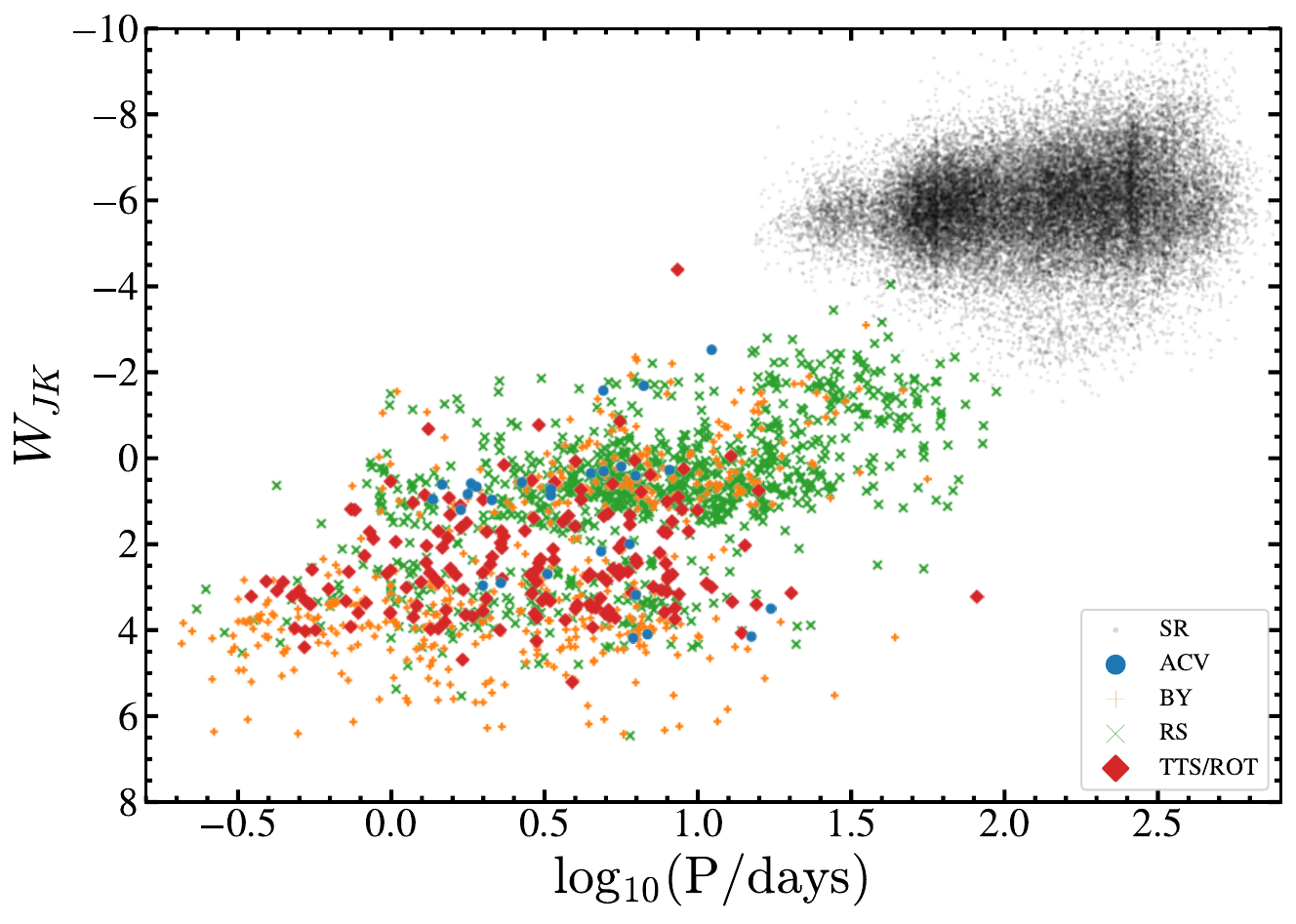}
    \caption{Wesenheit $W_{JK}$ vs $\log \rm P$ for the rotational variables grouped by their existing VSX type. The SR variables are also shown in black.}
    
    \label{fig:fig9}
\end{figure}

The location of these rotational variables in the Wesenheit $W_{JK}$ period-luminosity space is shown in Figure \ref{fig:fig9}, grouped into the sub-groups BY, ACV, RS and TTS/ROT based on the existing VSX classifications. Different clusters are visible --- main sequence/pre-main sequence dwarfs (e.g., BY, TTS/ROT) cluster at $W_{JK}{\sim}4$ mag and rotating giants (e.g., RS) cluster both at $W_{JK}{\sim}1$ mag and $W_{JK}{\sim}-1$ mag. Note that there seem to be significant numbers of misclassified BY variables that are giants rather than dwarfs. The distinction between the rotational variables and semi-regular variables (black points) at similar periods is evident.

In order to refine the sample of rotational variables, we set the classification of variables classified as ROT with periods $P>30$ d that fall into the region occupied by red giants ($\S3.4.6$) to semi-regular variables (SR). Rotational variables are those sources that are periodic, with $0.2 \leq \rm P \leq 150$ d. This broad range in period was chosen to match the definitions for rotational variables used by the VSX catalog. Outbursting Be stars and other Gamma Cassiopeia variables (GCAS) are commonly classified into the broad ROT class by our V1 RF classifier. Therefore, we looked at the non-periodic sources in this group to look for GCAS variables. We change the classification of a non-periodic ROT variable to GCAS if it falls along the region defined for GCAS variables in $\S 3.4.6$ and has $J-K_s<1.1$ mag, $G_{BP}$ - $G_{RP}<1.4$ mag, $J-H<0.5$ mag and $A>0.25$ mag. We also check for young stellar objects (YSOs) with rotational modulation using the criteria defined in $\S 3.4.6$. An uncertainty flag is added to the remainder of the sources that do not meet any of these criteria (i.e, ROT:).

We caution the reader that the ROT: class is a `catch-all' class for sources with amplitudes of variability below the ASAS-SN threshold, including eclipsing binaries in crowded fields with artificially shallow eclipses due to blending. The smallest detectable variability amplitude in ASAS-SN varies as a function of mean magnitude, from $A_{min}{\sim}0.02$ mag at $V{\sim}12$ mag, to $A_{min}{\sim}0.1$ mag at $V{\sim}16$ mag (see Figure 6 in Paper I). While we removed sources with variability amplitudes $A\lesssim0.05$ mag regardless of magnitude, sources with variability below the detection threshold at a given mean magnitude (usually rotational variables and eclipsing binaries) tend to be classified into the ROT: class in our pipeline, largely based on their Wesenheit magnitudes and colors. Sources that have poor ASAS-SN light curves with outliers are also likely to be included in this class. Care must be taken when interpreting the variability results for these sources.

\subsubsection{Eclipsing Binaries}
Eclipsing binaries allow observers to probe fundamental physical characteristics, including the masses and radii of stars and they span a diverse range of stellar systems (see \citealt{2010A&ARv..18...67T}, and references therein). Most main-sequence stars have a companion, hence eclipsing binaries are abundant compared to most kinds of variable stars. Recently, the OGLE survey cataloged a combined ${\sim}500,000$ eclipsing binaries in the Magellanic Clouds and the Galactic bulge \citep{2016AcA....66..421P,2016AcA....66..405S}. Eclipsing binaries are also used to estimate extragalactic distances \citep{2013Natur.495...76P,2003AJ....126..175B} .

We aim to categorize the light curves of the variables that are classified as eclipsing binaries (ECL) by the RF classifier into the VSX photometric (sub-)classes: EW, EB and EA. EW (W UMa) binaries have light curves with minima of similar depths and EB ($\beta$-Lyrae) binaries have minima of significantly different depths. Both EW (contact) and EB (contact/semi-detached) binaries transition smoothly from the eclipse to the out-of-eclipse state. EA (Algol) binaries are detached systems where the exact onset and end of the eclipses are easily defined. EA systems may or may not have a secondary minimum.

Eclipsing binaries have also been divided into the contact (EC), semi-detached (ESD) and detached binary (ED) configurations with the use of a Fourier model \citep{2002AcA....52..397P,2006MNRAS.368.1311P}. In Paper I, we noted the difficulty of separating the photometric classes through visual review alone. In order to improve classification and reduce human intervention, we implement an additional RF classifier dedicated to the classification of eclipsing binaries into the three photometric classes.

We updated our automated period doubling routine described in Paper I for this work. Each light curve is divided into 25 bins, phase folded by the best period with the primary minimum at phase 0, and normalized by the global minimum and maximum. We search for a secondary minimum in the phase interval $[0.2,0.8]$, updated from the interval of $[0.3,0.7]$ used in Paper I to improve our sensitivity to eccentric binaries. We identify the local minimum and check to see if the derivative $d(\rm mag)/d(\rm phase)$ within $\pm0.15$ in phase changes sign and is at least $0.25\sigma$ from the mean value of the light curve. If such a minimum is not found, we automatically double the period for that source. We visually reviewed a sample of eclipsing binaries that had their period doubled through this approach and find that it works very well except when the light curves are exceptionally noisy.

\begin{figure*}
	\includegraphics[width=\textwidth]{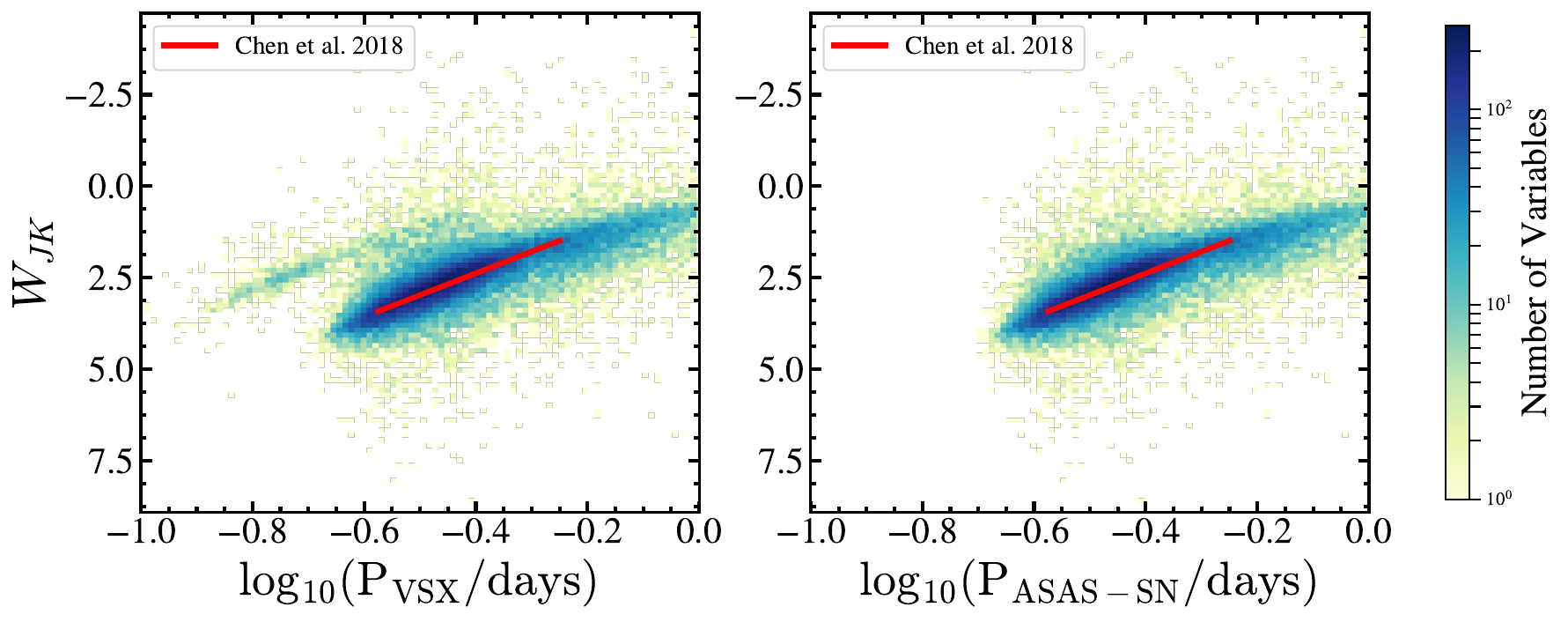}
    \caption{The Wesenheit $W_{JK}$ period-luminosity relationship (PLR) for EW binaries. The PLR for the eclipsing binaries is shown as a system density. The analytic expression for the PLR derived by \citet{2018ApJ...859..140C} is shown in red. \textit{Left}: The existing VSX periods for the EW binaries, \textit{Right}: The same sample of EW binaries using the updated periods. }
    \label{fig:fig10}
\end{figure*}

Figure \ref{fig:fig10} illustrates these results. The analytic expression for the PLR derived by \citet{2018ApJ...859..140C} is shown in red for reference. The left panel illustrates the existing VSX periods for the EW type binaries. A second locus of binaries is seen displaced in period by a factor of two. The results of our automated period doubling routine are shown in the right panel. The secondary locus of EW binaries with half their true period has vanished. 

In addition, the periods of over ${\sim} 15,000$ eclipsing binaries were visually reviewed by the authors in order to better familiarize themselves with the shapes of the light curves. We selected a sub-sample of ${\sim}5,000$ sources, removed noisy light curves, and manually classified them into the three photometric classes (EW, EB and EA) to obtain a training set of ${\sim}4,300$ eclipsing binaries. 

\begin{figure}
	\includegraphics[width=0.5\textwidth]{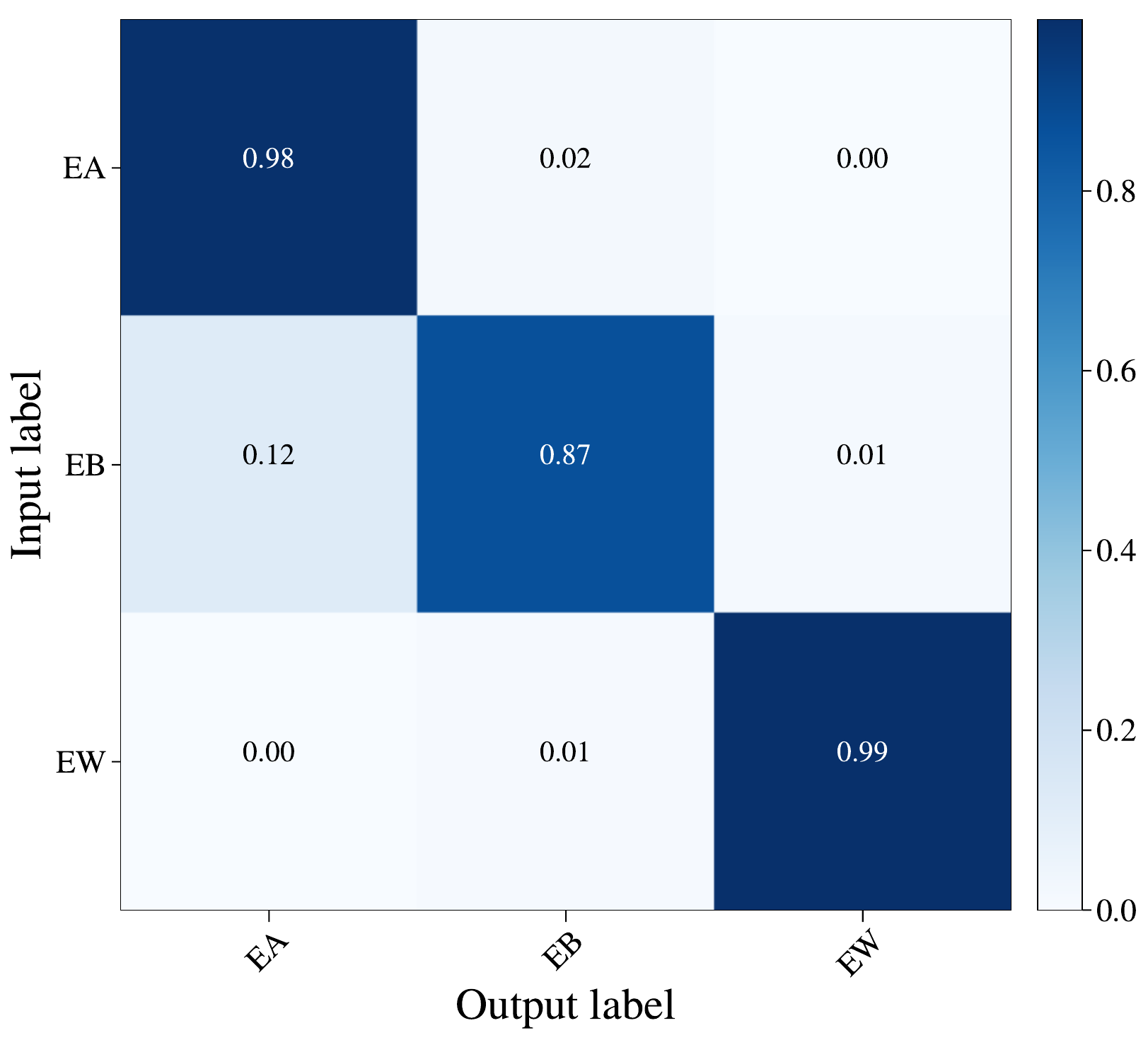}
    \caption{The normalized confusion matrix derived from the trained eclipsing binary random forest classifier. The y-axis corresponds to the `input' classification given to a variable, while the x-axis represents the `output' prediction obtained from the trained random forest model.}
    \label{fig:fig11}
\end{figure}

We trained a dedicated RF classifier for the eclipsing binaries using this training set. We use the following features in the ECL classifier, where the importance of each feature is listed in parentheses (see Table \ref{tab:features} for descriptions): $\rm Skew$ (8\%), $\rm Kurt$ (10\%), $A_{\rm HL}$ (3\%),  $\log P$ (15\%),$R_{41}$ (5\%), $R_{21}$ (10\%), $R_{32}$ (5\%), $R_{42}$ (31\%) and the ratio of minima $R_{\rm minima}$ (14\%). The features derived from the Fourier model play a heavy role in the classifier, with a combined importance of 51\%. The precision, recall and $F_1$ scores for each binary type is summarized in Table \ref{tab:ebscores}. The EA and EW classes are very well classified with $F_1 \gtrsim 95\%$. The EB class has a lower score of $F_1=91\%$ due to confusion with EW binaries, as illustrated in the confusion matrix shown in Figure \ref{fig:fig11}. The eclipsing binary classifier has an overall $F_1$ score of 95.1\%.

\begin{figure*}

	\includegraphics[width=\textwidth]{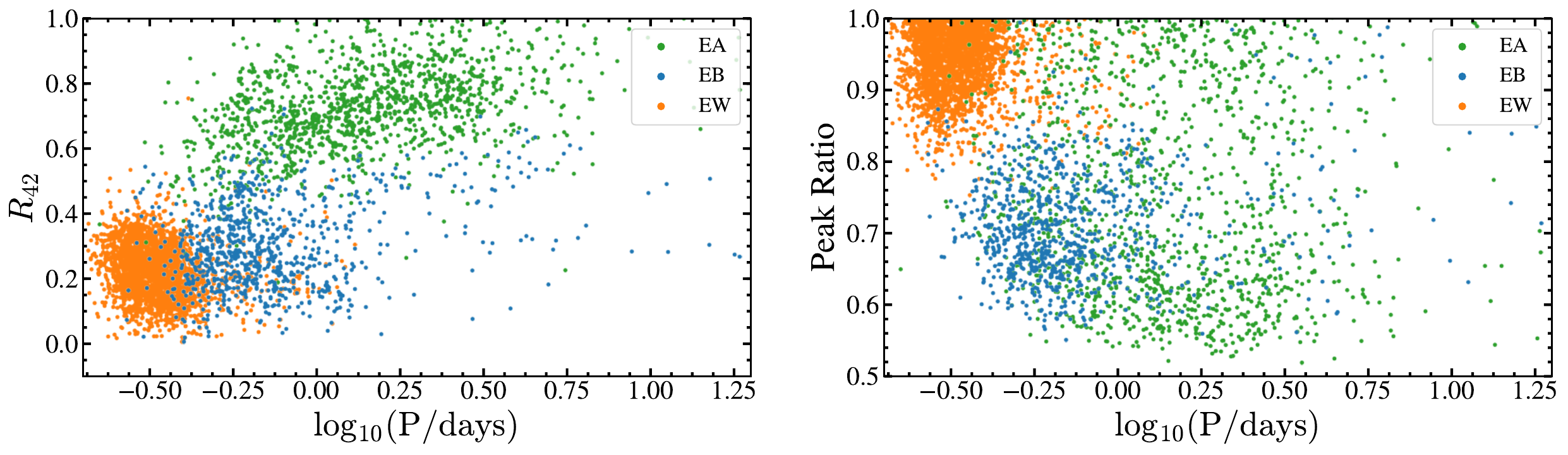}
    \caption{ Distribution of the eclipsing binary training set in $R_{42}$ vs $\log P$ (\textit{ left}), and the peak ratio $R_{\rm minima}$ vs $\log P$ (\textit{right}). The points are colored by the eclipsing binary class assigned to them.}
    \label{fig:fig12}
\end{figure*}

Figure \ref{fig:fig12} shows the distribution of the training set binaries in the three most important classification features --- $R_{42}$, $\log P$ and $R_{\rm minima}$. The left panel shows the distribution in $R_{42}$ and $\log P$, and it is immediately clear why this is the most important feature in this classifier. EW variables are strongly clustered at $(\log P,R_{42}){\sim}(-0.5,0.2)$, with the EB variables offset towards longer periods. EA variables are easily distinguished from both EW and EB types, as they predominantly occupy the region with $R_{42}\gtrsim0.5$. The right panel illustrates the distribution of the peak ratio $R_{\rm minima}$ against $\log P$. Once again, EW variables are strongly clustered towards $(\log P,R_{\rm minima}){\sim}(-0.5,0.95)$, as is expected given their eclipse profiles. EB variables are offset towards longer periods and typically have $R_{\rm minima}\lesssim 0.8$. EA variables are ``randomly distributed'' in this space since it is not unusual for detached eclipsing binaries to have minima of very similar depths.

\begin{figure*}

	\includegraphics[width=\textwidth]{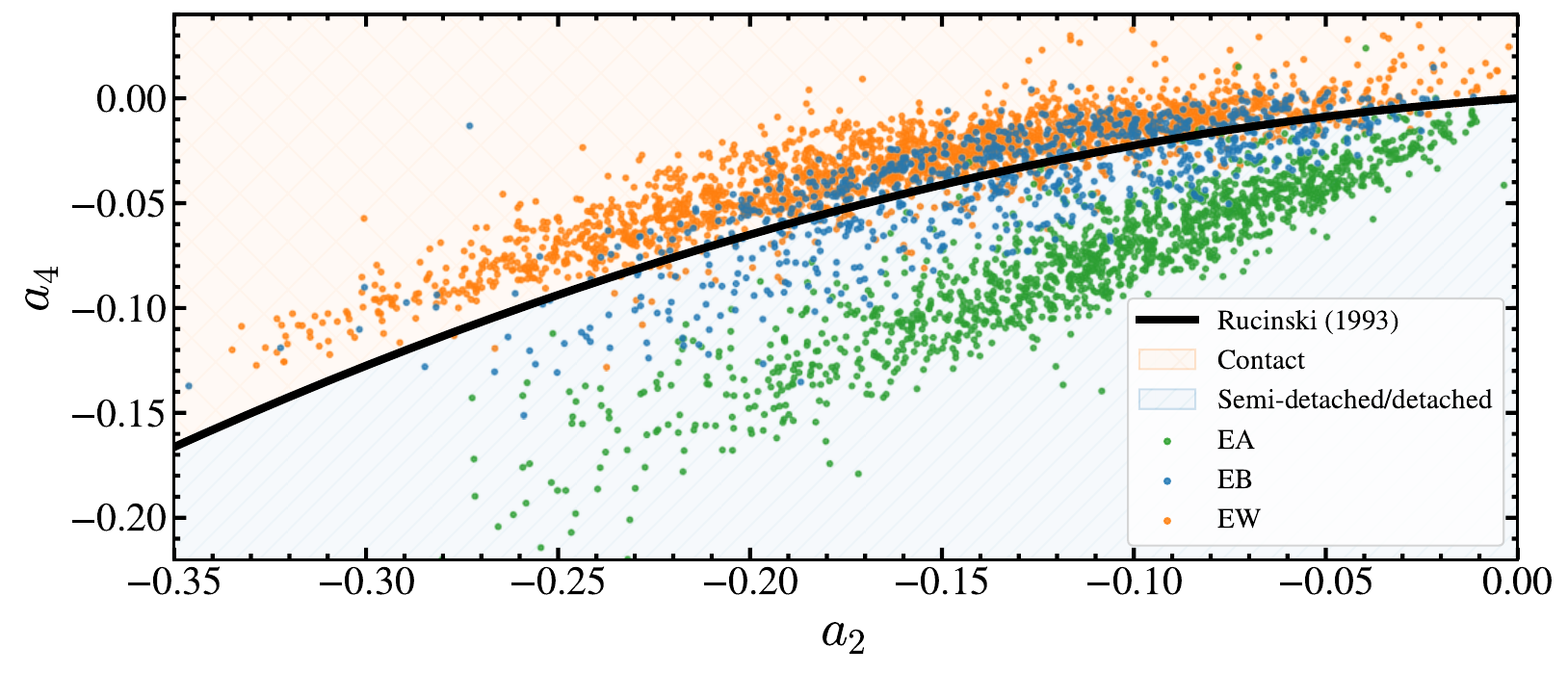}
    \caption{Distribution of the training sample of eclipsing binaries in the Fourier components $a_2$ and $a_4$. The relationship used by \citet{1993PASP..105.1433R} to seperate contact binaries from semi-detached/detached configurations is shown in black. The regions occupied by contact and semi-detached/detached binaries based on this relationship are shaded in orange and blue respectively.}
    \label{fig:fig13}
\end{figure*}

In Figure \ref{fig:fig13}, we show the distribution of the eclipsing binaries in our training sample in the Fourier components $a_2$ and $a_4$. These are commonly used to classify eclipsing binaries into the three physical configurations. \citet{1993PASP..105.1433R} defined a relationship to separate contact binaries from semi-detached/detached binaries as \begin{equation}
 a_4=a_2(0.125-a_2),
\label{eq:a24} \end{equation} which is shown in black. We have also shaded the locii of the contact and semi-detached/detached configurations in orange and blue respectively. This relationship is able to seperate contact binaries (EW) efficiently. EB types live in both the contact and semi-detached/detached regions, suggesting that the EB class is an intermediary class to both the EW and EA classes with respect to the physical configuration. EA binaries all lie in the semi-detached/detached space as is expected.

\begin{figure*}
	\includegraphics[width=\textwidth]{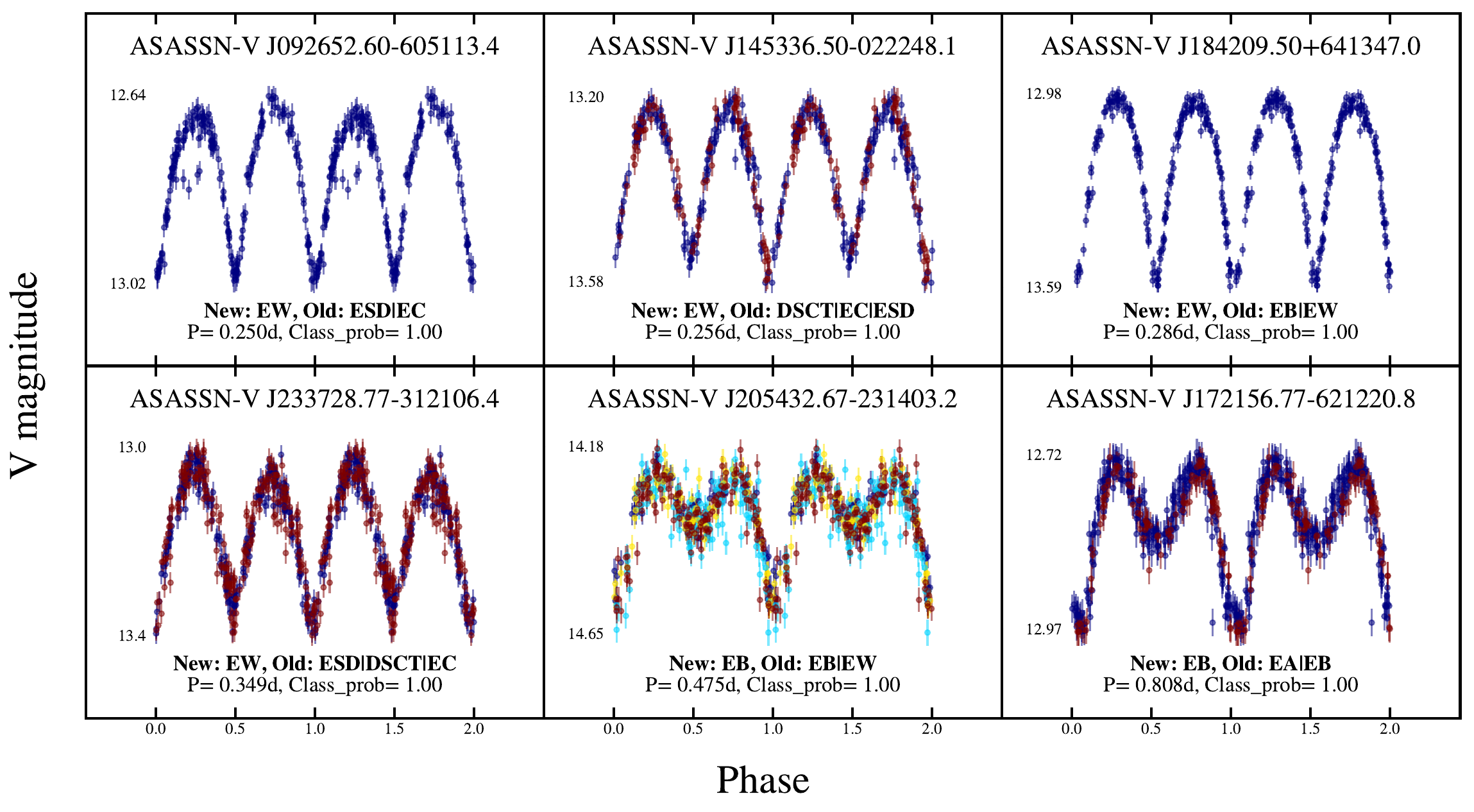}
    \caption{Phased light curves of eclipsing binaries with refined classifications. The format is the same as Figure \ref{fig:fig8}.}    
    \label{fig:fig14}
\end{figure*}

Both classification schemes (i.e., the photometric classes and the physical configurations) have been used by previous variability studies, and the mappings between the classifications are not homogeneous. With the use of this eclipsing binary classifier, we are able to homogeneously refine the classifications given to eclipsing binaries in the VSX catalog. Example light curves are shown in Figure \ref{fig:fig14}. It is clear that our pipeline is able to break common degeneracies in eclipsing binary classifications. For example, the ASAS survey had a large number of classifications with degenerate classifications. We have provided updated classifications in many such cases.

\begin{figure*}
	\includegraphics[width=\textwidth]{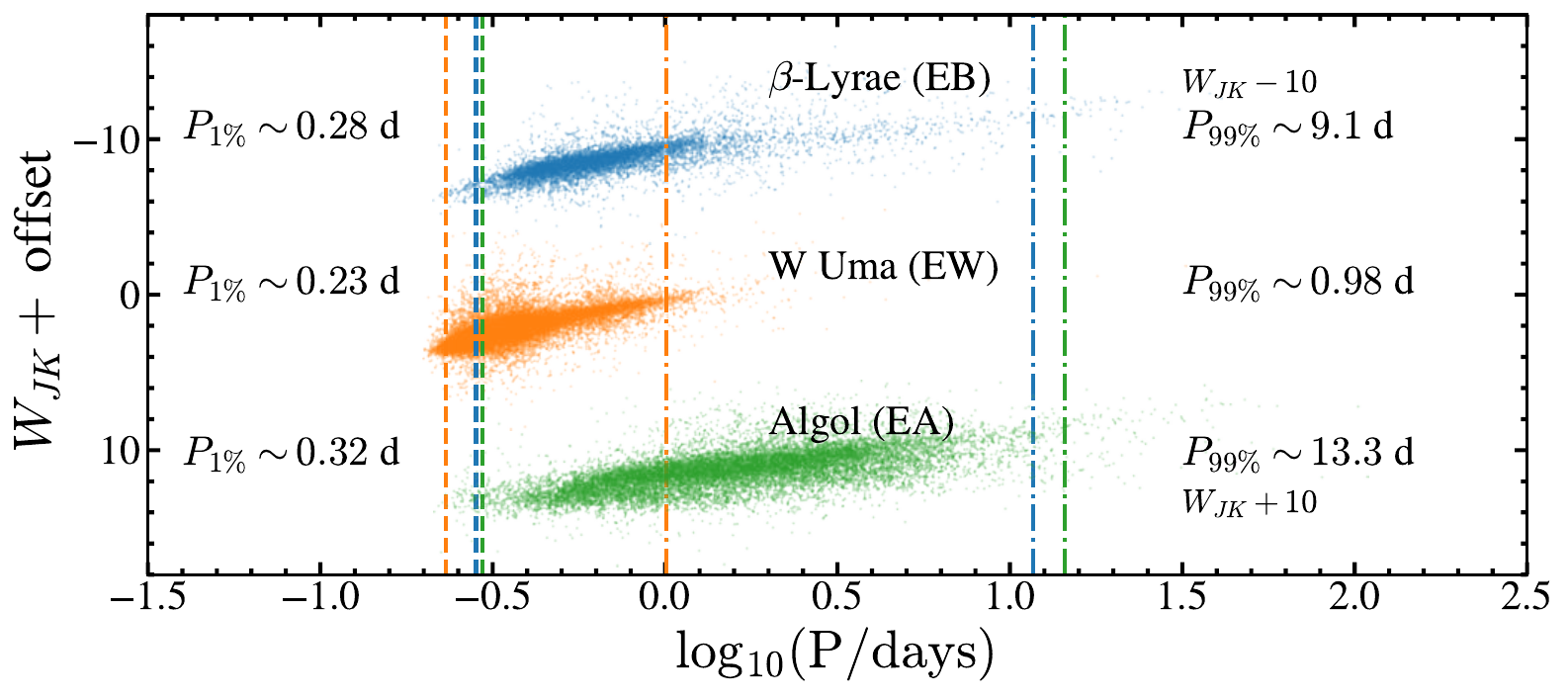}
    \caption{The Wesenheit $W_{JK}$ PLR for the different classes of eclipsing binaries. Lines denoting the $1^{\rm st}$ and $99^{\rm th}$ percentiles in period are drawn for each class.}
    
    \label{fig:fig15}
\end{figure*}

We show the Wesenheit $W_{JK}$ PLR for each class of eclipsing binary using our complete sample of ${\sim} 53,000$ binaries in Figure \ref{fig:fig15}. Lines denoting the $1^{\rm st}$ and $99^{\rm th}$ percentiles in period are drawn for each class. EW binaries span periods of $0.23\leq \rm P_{1-99} \leq 0.98$ d. The distribution of EB binaries start at slightly longer periods and extend to to periods longer than a day with $0.28\leq \rm P_{1-99} \leq 9.1$ d. We see a similar distribution in periods for the EA types, with these spanning the range $0.32\leq \rm P_{1-99} \leq 13.30$ d. The PLRs for EW and EB types are distinctly sharper and better defined than the PLR of EA binaries. The period distribution of eclipsing binaries in the training sample suggests that long period photometric binaries are exceedingly rare, and we visually verified that the vast majority of long period sources ($P>100$ d) that are classified as eclipsing binaries are actually misclassified semi-regular/irregular variables. Therefore, variables that are classified as eclipsing binaries with $P>100$ d and $\rm Prob <0.9$ are reclassified as semi-regular variables if they fall into the red giant region ($\S3.4.6$), otherwise they are assigned the generic variability class (`VAR'). 
\begin{table}
	\centering
	\caption{Precision, Recall and $F_1$ score Values by class for the eclipsing binary classifier.}
	\label{tab:ebscores}
	\begin{tabular}{lccc} % four columns, alignment for each
		\hline
		Class & Precision & Recall & $F_1$ score\\
		\hline
		EA & 93$\%$ & 98$\%$ & 95$\%$\\
		EB & 95$\%$ & 87$\%$ & 91$\%$\\
		EW & 100$\%$ & 99$\%$ & 99$\%$ \\        
		\hline
	\end{tabular}
\end{table}

\subsubsection{Semi-regular and Irregular Variables}

Most semi-regular variables are pulsating red giants that show varying levels of periodicity in their light curves. Multi-periodic behavior is commonly seen in the light curves of semi-regular variables and can be used to study the dynamics of stellar interiors \citep{1999A&A...346..542K}. OGLE discovered a new class of low amplitude semi-regular variable in the Magellanic clouds, the OGLE small amplitude variable red giants (OSARGs; \citealt{2004AcA....54..129S}), that follow a set of period-luminosity relations \citep{2007AcA....57..201S}. GCAS variables are eruptive irregular variables with early spectral types (O9-A0 III-Ve) and have mass outflows from their equatorial zones \citep{2006SASS...25...47W,2018MNRAS.479.2909B}. SRD variables are yellow semi-regular variables that are giants and supergiants with spectral types of F/G/K.

\begin{figure}
	\includegraphics[width=0.5\textwidth]{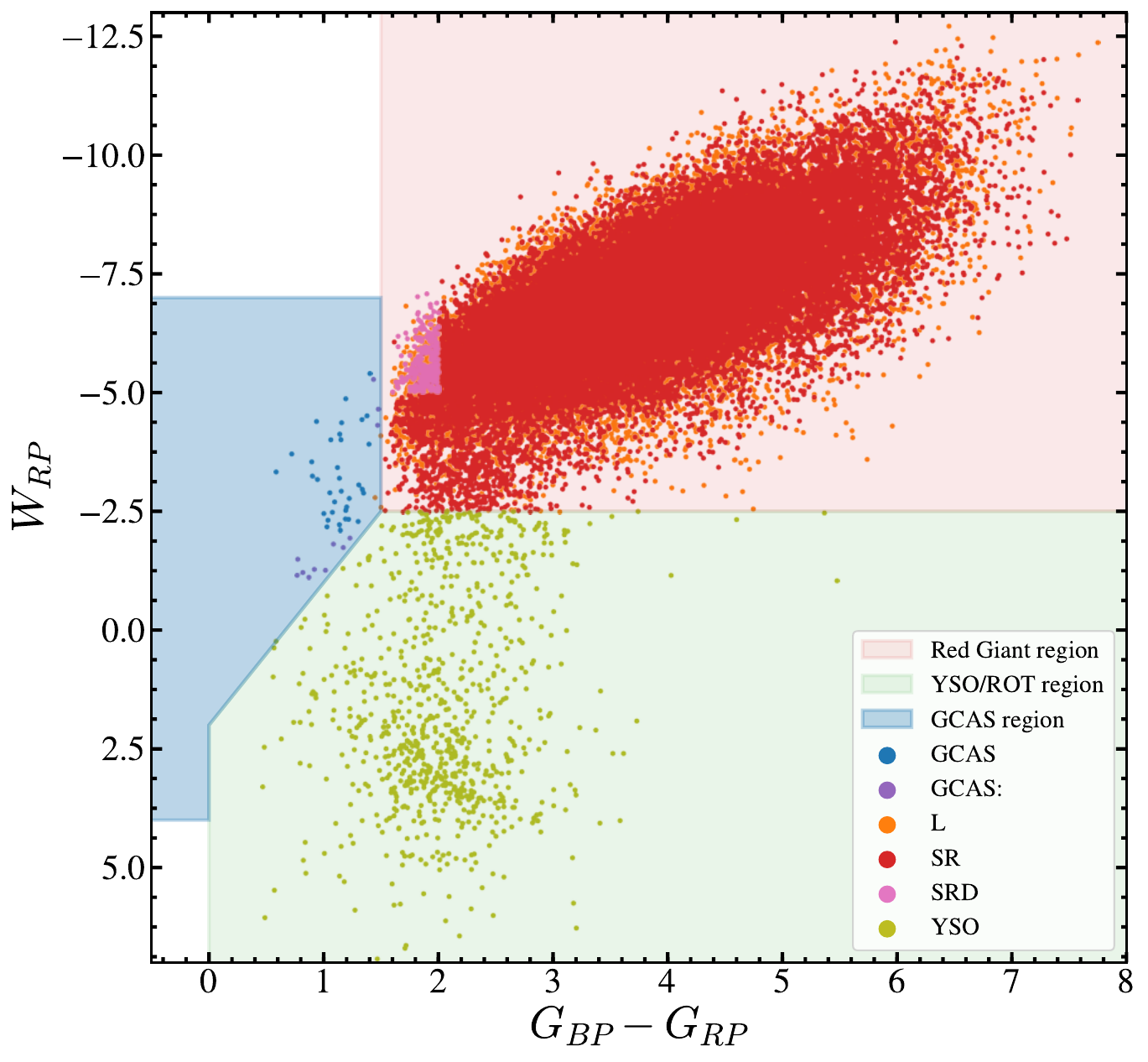}
    \caption{ The Wesenheit $W_{RP} $ vs. $G_{BP}-G_{RP}$ color-magnitude diagram used in the classification of semi-regular variables. The empirically defined regions for red giants, YSO/ROT variables, SRD variables and GCAS variables are shaded in red, green, pink and blue respectively. The points are colored according to the final classifications assigned to each variable.}
    \label{fig:fig16}
\end{figure}

In order to refine the classifications of semi-regular/irregular variables, we look at their positions in the Wesenheit $W_{RP}$ vs. $G_{BP}-G_{RP}$ color-magnitude diagram (Figure \ref{fig:fig16}). We empirically define regions for red giants, YSO/ROT variables, and GCAS variables and shade these in red, green, and blue respectively. Similar regions were found by \citet{2018arXiv180502035M} for YSOs and red giants. Variables that are classified as SR/IRR by the RF classifier are first sorted into one of these three sub-groups.

For the SR/IRR sources that lack Gaia data but have complete 2MASS photometry, we use the following criteria for classification. Variables are assigned the red giant class if they have $J-K_s>1.1$ mag and $J<10$ mag. Variables are given the GCAS classification if $J-K_s<1.1$ mag and $J-H<0.2$ mag. Variables are assigned the YSO VSX class if they have $J-K_s>1.1$ mag and $J>10$ mag OR $J-K_s<1.1$ mag and $J-H>0.2$ mag. If these sources have WISE photometry, we implement the additional criteria of $W1-W2>0.15$. Sources with $W1-W2<0.15$ are classified as VAR variables. Any other sources that do not specifically fall into these classes are also assigned the VAR classification.

To further refine the periods of the sources classified as red giants, we re-run the GLS periodogram in the restricted range $5\leq P \leq\,t_{\rm base}$ days, where $t_{base}$ is the time baseline of the ASAS-SN light curve. The resulting best GLS period is assigned to each variable. Periodic variables are classified as semi-regular (SR) variables while the sources that have no period are classified as a red irregular variables (L). Periodic variables that fall into the red giant region are classified as SRD variables if $W_{RP}<-5$ mag, $J-K_s<1.1$ mag, and $G_{BP}-G_{RP}<2$ mag. 

Variables need to meet the following criteria in order to be classified as GCAS variables: $J-K_s<1.1$ mag, $G_{BP}-G_{RP}<1.4$ mag, $J-H<0.5$ mag and $A>0.25$ mag. If they have $J-K_s>1.1$ mag and $G_{BP}-G_{RP}>1.4$ mag, they are classified using the criteria defined for red giants. The remainder of the sources in this region are assigned an uncertainty flag (i.e., GCAS:).

\begin{figure*}
	\includegraphics[width=\textwidth]{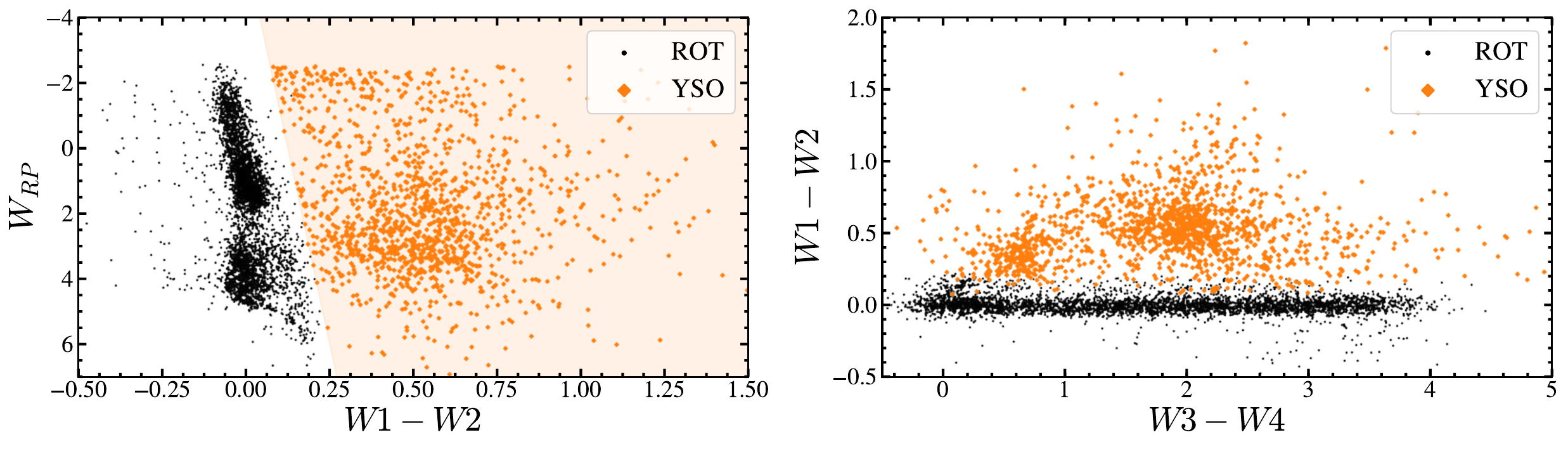}
    \caption{ \textit{Left}: The Wesenheit $W_{RP} $ vs. $W1-W2$ color-magnitude diagram used in the classification of YSO variables. The empirically defined locus for the YSO variables is shaded in orange, \textit{Right}: The WISE $W3-W4$ vs $W1-W2$ color-color diagram highlighting the YSO variables. The ROT class shown here is derived from the rotational variables classified separately from the SR/IRR sources in order to highlight the distinction between YSO and ROT variables. The YSO variables shown here were drawn from the SR/IRR class.}
    \label{fig:fig17}
\end{figure*}

We use AllWISE photometry to separate main sequence rotational variables from young stellar objects on the pre-main sequence. Variables that are grouped into the YSO/ROT sub-class are empirically sorted into the YSO class based on their position in the $W_{RP}$ vs. $W1-W2$ color-magnitude diagram (Figure \ref{fig:fig17}). In the left panel, rotational variables form two clusters comprised of dwarfs ($W_{RP}{\sim}4$ mag) and giants ($W_{RP}{\sim}0$ mag). In the right panel, YSOs form two clusters largely comprising of Class II YSOs ($W3-W4{\sim}0.8$ mag) and Class I/III YSOs ($W3-W4{\sim}2.2$ mag). We essentially look for evidence of an infrared excess for the YSO variables. This is similar to our approach in Paper I but is significantly more accurate due to the use of WISE and Gaia DR2 data. When comparing the AllWISE color-color diagram shown in the right panel in Figure \ref{fig:fig17} (this work) to Figure 7 of from \citet{2014ApJ...791..131K} , we see that this method is very useful in identifying YSOs. We successfully classify YSOs of different types, including sources that are consistent with the YSO classes I/II/II and YSOs with transition disks \citep{2014ApJ...791..131K}. If the YSO sources are periodic, we check to see if the period satisfies $1 \leq \rm P\leq 100$ d and that we have at least two cycles in the ASAS-SN light curve before assigning a period. YSOs that do not meet this periodicity criteria are not assigned a period. Variables in the YSO/ROT region that are not classified as YSOs are classified as rotational variables (ROT). The criteria used to refine these rotational variables are the same as in Section $\S 3.4.4$.

Variables with missing photometric information and variables that do not meet any of the classification criteria described above are assigned the VAR classification. This makes the VAR variable class a catch all for unusual sources that do not fit into standard variable star classification criteria, sources with bad nearest neighbor cross-matches to other catalogs and sources with one or more pieces of missing information.
    
\begin{figure}
	\includegraphics[width=0.5\textwidth]{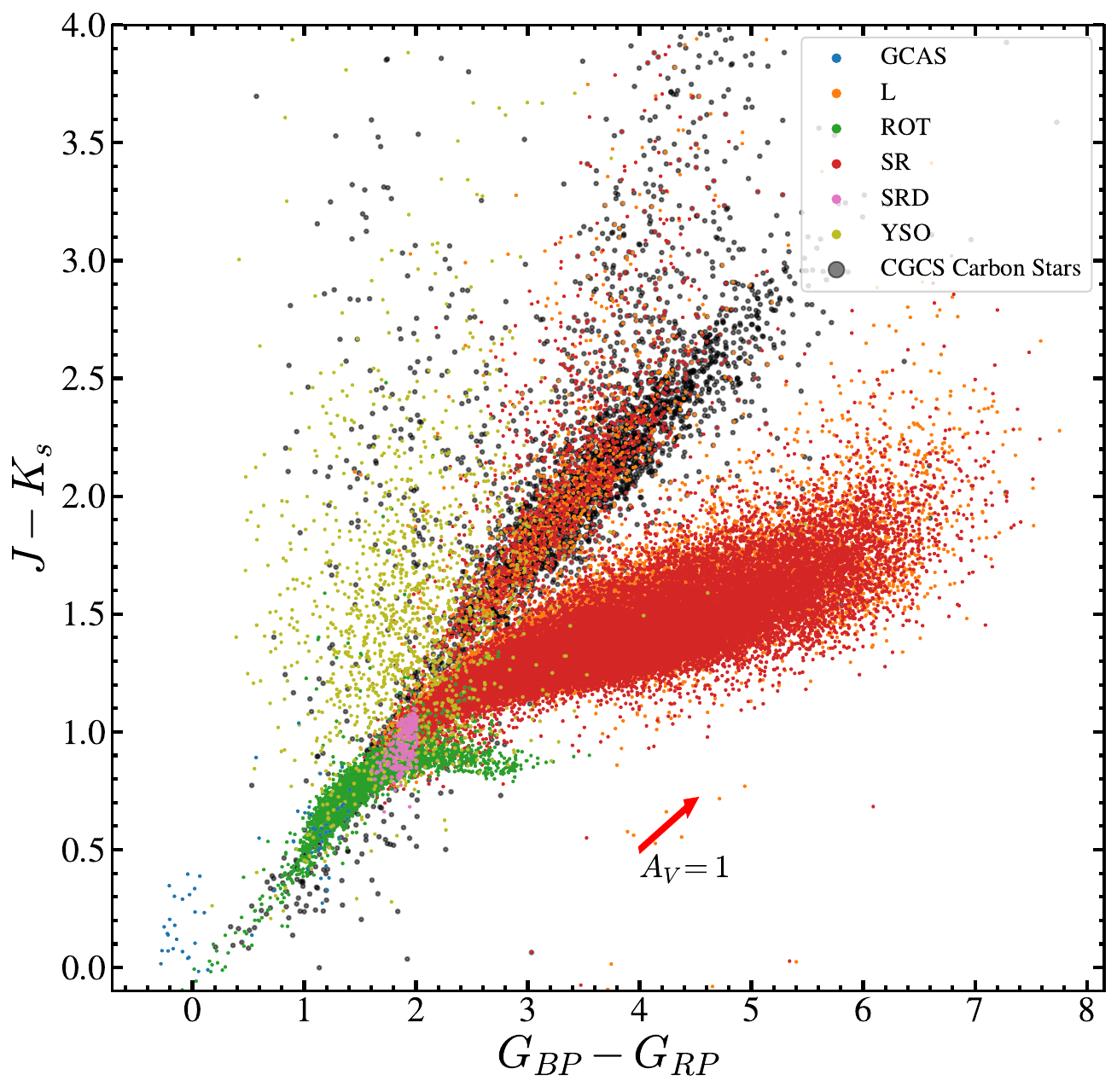}
    \caption{The Gaia DR2-2MASS $G_{BP}-G_{RP}$ vs $J-K_s$ color-color diagram for the rotational and semi-regular/irregular variables. Carbon stars from \citet{2001BaltA..10....1A} are plotted in black. The reddening vector corresponding to an extinction of $A_V=1$ mag is shown in red.}
    \label{fig:fig18}
\end{figure}

In Figure \ref{fig:fig18}, we look at the distribution of rotational and semi-regular/irregular variables in the Gaia DR2-2MASS $G_{BP}-G_{RP}$ vs $J-K_s$ color-color diagram. For reference, we also show the sources from the Catalog of Galactic Carbon Stars \citep{2001BaltA..10....1A} in black. The carbon stars form a sharp locus in this color-color space and have larger values of $J-K_s$ for a given $G_{BP}-G_{RP}$ once $G_{BP}-G_{RP}\gtrsim2$ mag than any other semi-regular/irregular variable source. Some fraction of the semi-regular and irregular sources fall along this Carbon star locus, suggesting that these are carbon enriched sources. The positions of the semi-regular and irregular sources in this space are clearly distinct from that of the carbon stars, which implies a difference in stellar chemistry for the majority of these sources when compared to the carbon rich sources. At bluer colors ($G_{BP}-G_{RP}\lesssim2$ mag), it becomes harder to distinguish the different variable classes. The position of the rotational variables is particular interesting --- it remains distinct from the locus of semi-regular/irregular sources and plateaus at a color of $J-K_s{\sim}0.8$. Most GCAS variables also lie away from the ROT locus, with a large number of these sources having $G_{BP}-G_{RP}<0.5$ mag and $J-K_s<0.5$ mag.

\subsubsection{Mira Variables}
Mira variables (M) are asymptotic giant branch (AGB) stars that show high amplitude (typically $A\gtrsim2.5$ mag in the V-band and decreasing with wavelength) variability with typical periods $P>100$ d. Mira variables have also been found to follow a period-luminosity relationship \citep{2008MNRAS.386..313W} and are useful distance indicators owing to their large intrinsic luminosities. However, Mira variables can undergo changes in their period over time \citep{1999PASP..111...94P}, so one must be cautious about using these variables when deriving precise distances. Recently, oxygen-rich Mira variables have also been used to study age gradients in Galactic populations, noted in \citet{2018arXiv180409186G}.

As in Paper I, some of the Mira variables can dip below the detection limits for ASAS-SN ($V\gtrsim17$ mag), resulting in underestimated variability amplitudes. This can also result in inadequately sampled light curves that result in incorrect periods. Mira variables in crowded fields, particular towards the bulge and Galactic disk, can be misclassified as semi-regular variables owing to the reduction in variability amplitudes due to blending \citep{2018arXiv180502035M}. 

To refine the classifications of Mira variables, we use the red giant region defined in Section $\S 3.4.6$. A Mira variable (M) is one which falls into the red giant region and has an amplitude $A>2$ mag in the V-band. We choose to relax this amplitude criteria from the standard of $A>2.5$ mag in the V-band to better deal with blended and underestimated amplitudes. Variables that are classified as MIRA through the RF classifier and fall into the red giant region but have amplitudes $A<2$ mag are classified as semi-regular (SR) variables. Other sources that are given the MIRA class but do not fall into the red giant region are assigned the generic VAR class. Mira variables without a period are given an uncertainty flag.

\subsection{Properties of the Final Training Sample}

All of the refinements from $\S3.4$ are applied to the enlarged training set from $\S3.3$ before building V2 of the RF variability classifier. The main refinements made to the broad variability classes output by the RF classifier are summarized in Table \ref{tab:refsum}. Sources that were assigned uncertain or generic variability classifications and those with V1 probabilities $\rm Prob<0.8$ were removed from the training sample, resulting in a final training set of $166,000$ variables. 

\begin{figure}
	\includegraphics[width=0.5\textwidth]{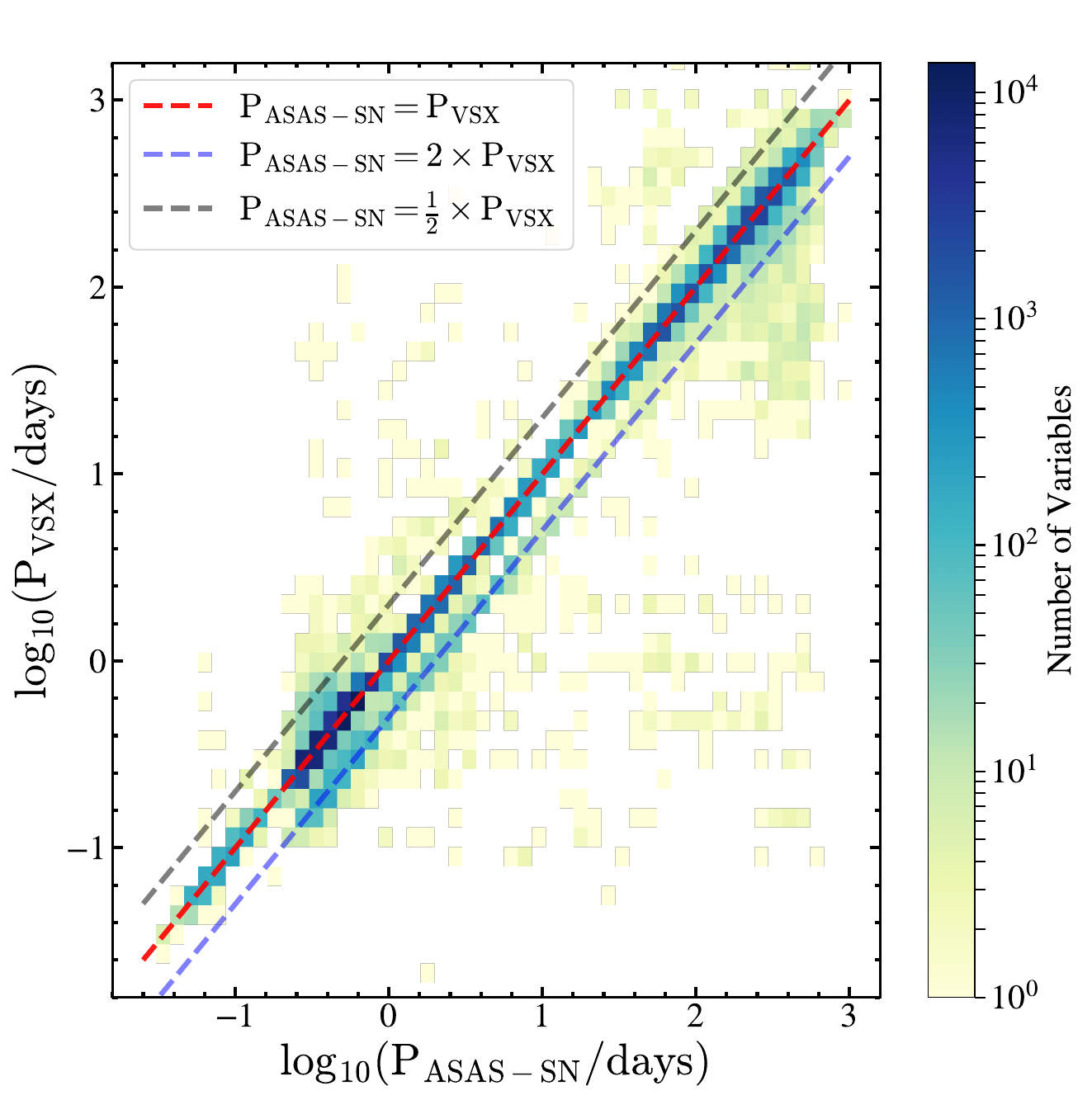}
    \caption{ A comparison of the ASAS-SN periods $P_{\rm ASAS-SN}$ to the existing, VSX periods $P_{\rm VSX}$ for the variables in the training set. Dotted line are shown to represent ASAS-SN periods that are the same (red), twice (blue) and half (black) the VSX period.}
    \label{fig:fig19}
\end{figure}

\begin{figure}
	\includegraphics[width=0.5\textwidth]{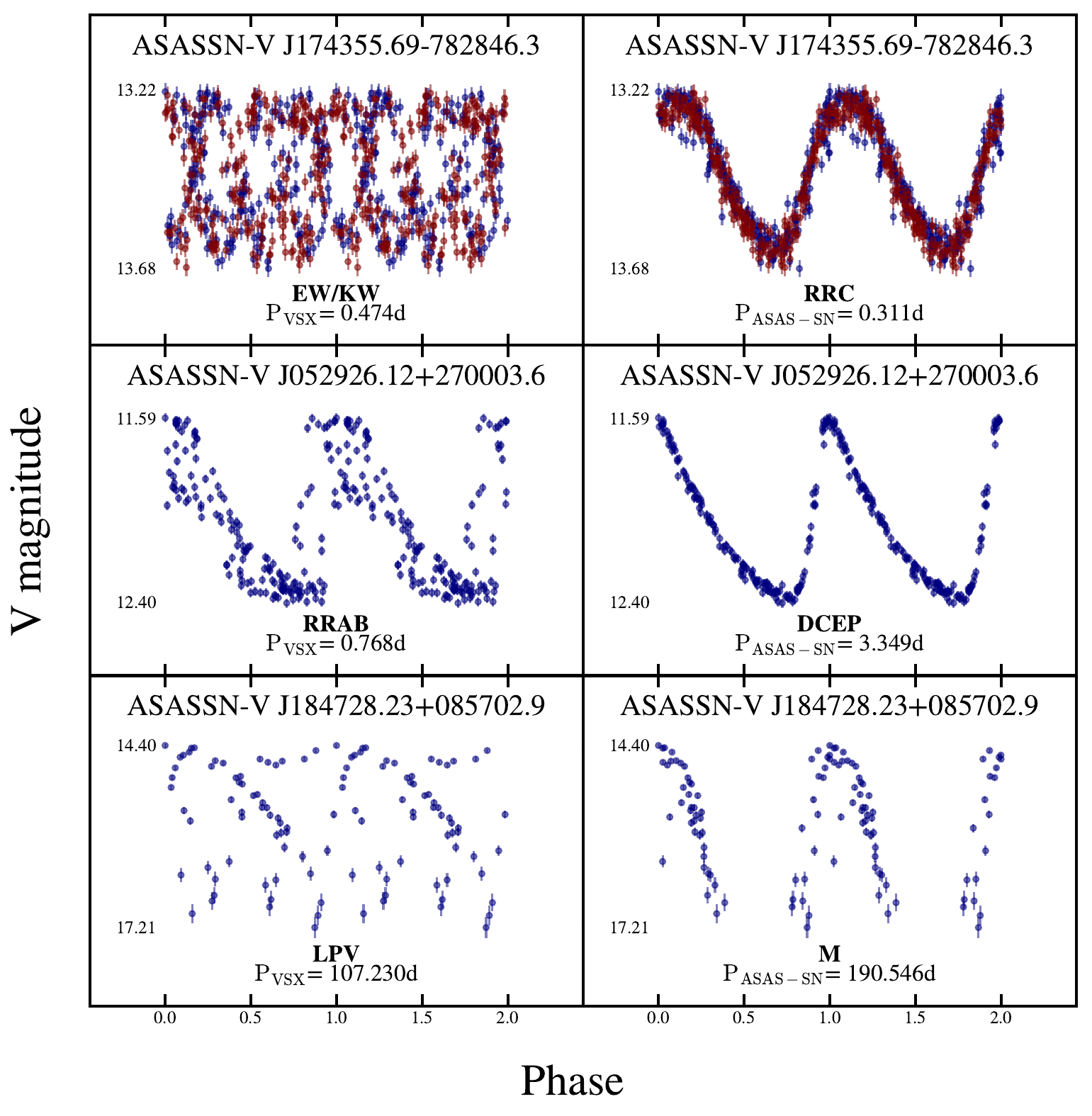}
    \caption{Examples of variables with different VSX (left) and ASAS-SN (right) periods. The format is the same as Figure \ref{fig:fig8}.}    
    \label{fig:fig20}
\end{figure}

In Figure \ref{fig:fig19}, we compare the periods derived through the ASAS-SN variability analysis pipeline ($P_{\rm ASAS-SN}$) to the periods in the VSX catalog ($P_{\rm VSX}$) for the training set. A large majority ($62\%$) of the sources have ASAS-SN periods within $\pm5\%$ of the VSX periods. Most of the differences are for longer ($P>100$ d) period variables that are typically semi-regular and Mira variables. The light curves of semi-regular variables are likely to be multi-periodic and have loosely defined periods. However, 2\% are assigned ASAS-SN periods that are twice the VSX period, largely due to our better treatment of eclipsing binaries. We also see some evidence that ASAS-SN provided better periods for sources that have aliases of a sidereal day as the VSX period. Examples of sources with different ASAS-SN and VSX periods are illustrated in Figure \ref{fig:fig20}.

The classification of a given variable is strongly dependent on the period assigned to it from our pipeline. In some cases, an incorrect period will result in a less reliable classification. However, given the robustness of the RF classifier and the numerous features that are used in the classification process other than the period, variables with incorrect periods will have lower classification probabilities than the variables with similar, but accurate periods. We have attempted to correct for multiplicity in the derived periods of the eclipsing binaries and RV Tauri variables as discussed in $\S3.4$.  While these processes are reliable in identifying the cases where the period needs to be doubled, we estimate the number of eclipsing binaries where the period is incorrectly doubled as $\lesssim4\%$ .

\begin{figure*}
	\includegraphics[width=\textwidth]{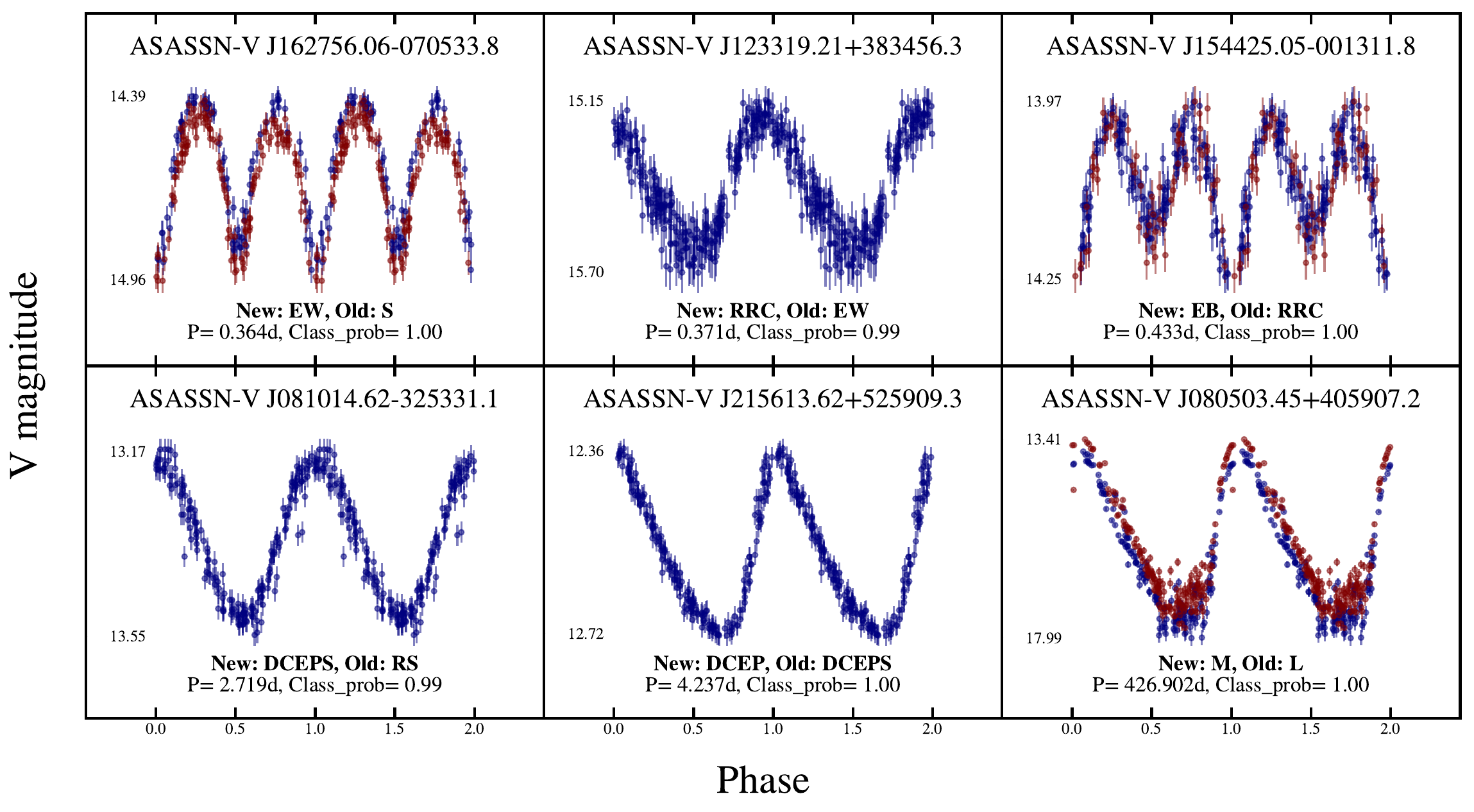}
    \caption{Examples of variables with new classifications. The format is the same as Figure \ref{fig:fig8}.}    
    \label{fig:fig21}
\end{figure*}

Due to the refinements in $\S3.4$, ${\sim} 5,000$ sources out of the ${\sim} 166,000$ sources were assigned to a different broad class. Of these, ${\sim}1,800$ variables were reclassified as eclipsing binaries, with the vast majority being reclassified as EW binaries. Reclassifications to and from the broad classes RRAB and RRC/RRD are also common, with the vast majority being RRAB variables reclassified as RRD variables. Examples of these reclassified sources are shown in Figure \ref{fig:fig21}. 

\begin{table*}
	\centering
	\caption{Summary of the variability refinement criteria for each variable class.}
	\label{tab:refsum}
\begin{tabular}{lrr}
		\hline
		Class & Sub-classes & Summarized refinement Criteria \\\\
				\hline
Delta Scuti ($\S3.4.1$)   &  & $W_{JK}$ PLR\\		
   & DSCT &  $A<0.15$ mag   \\
   & HADS &  $A>0.15$ mag   \\
		\hline
RR Lyrae ($\S3.4.2$)   &  & $-3<W_{JK}<5$ mag  \\		
   & RRAB &  $0.3 \leq \rm P \leq 1.2$ d   \\
   & RRC &  $0.2 \leq \rm P \leq 0.5$ d + $W_{RP}$-$G_{BP}-G_{RP}$ \\
   & RRD &  $0.72<{P_{1O}}/{P_{FO}}<0.78$ + $0.45<{P_{FO}}<0.60$ d   \\   
   \hline

Cepheids ($\S3.4.3$)   &  & $W_{JK}$ PLR  \\	
   & DCEP &  $P > 1$ d  \\ 
   & DCEPS &  $\rm P \leq 7$ d + $R_{21}$-$\log \rm P$ \\   
   & CWB &  $\rm P \leq 8$ d   \\
   & CWA &  $\rm P > 8$ d   \\   
   & RVA &  $16 \leq \rm P \leq 180$ d + $A>0.25$ mag + $J-H<1$ mag + Period Doubling \\
   \hline
Rotational Variables ($\S3.4.4$)   &  & \\		
   & ROT &  $A<0.15$ mag + $0.2 \leq \rm P \leq 150$ d   \\
		\hline   
Eclipsing Binaries ($\S3.4.5$)   &  & ECL RF Classifier + Period Doubling \\		
   & EA &  $P < 100$ d   \\
   & EW &  $W_{RP}$-$G_{BP}-G_{RP}$ + $W_{JK}$ PLR \\
   & EB &  $P < 100$ d  \\   
   \hline  
Semiregular and Irregular Variables ($\S3.4.6$)   &  & $W_{RP}$-$G_{BP}-G_{RP}$ \\		
   & SR &  $P > 5$ d   \\
   & SRD &  $P > 5$ d + $W_{RP}<-5$ mag + $J-K_s<1.1$ mag \\   
   & L &  No significant period \\
   & YSO &  $W_{RP}$-$W1-W2$ + $W_{JK}$ PLR \\   
   & GCAS & $J-K_s<1.1$ mag +  $J-H<0.5$ mag + $A>0.25$ mag \\
   \hline  
Mira Variables ($\S3.4.7$)   &  & $W_{RP}$-$G_{BP}-G_{RP}$ \\		
   & M &  $P > 80$ d  + $A>2$ mag \\
   \hline     
\end{tabular}
\end{table*}

\begin{figure*}
	\includegraphics[width=\textwidth]{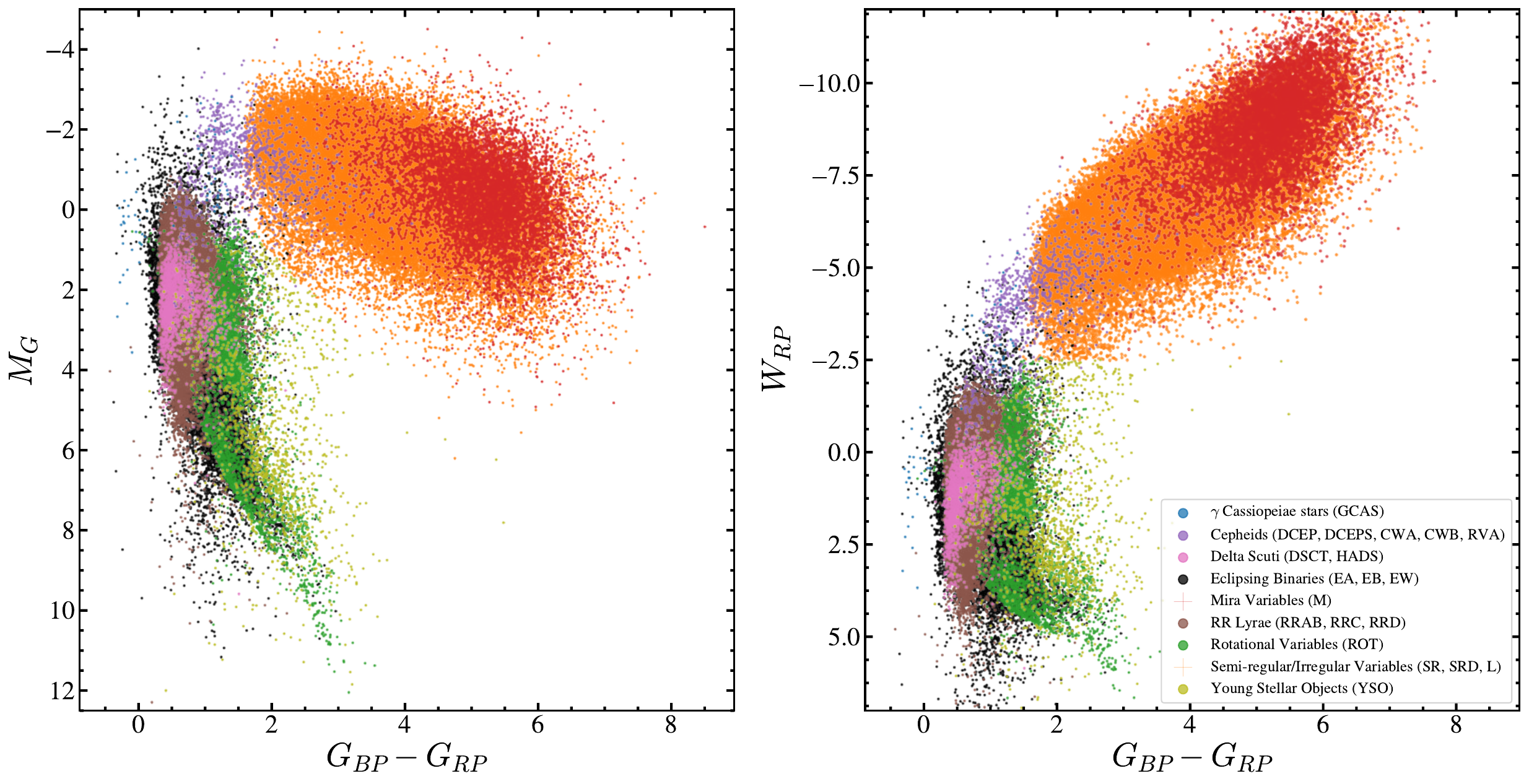}
    \caption{The Gaia DR2 $M_G$ (left) and $W_{RP}$ (right) vs. $G_{BP}-G_{RP}$ color-magnitude diagram for the variables in the training sample. The points are sorted into the groups described in the legend, and are colored accordingly. While the effects of extinction are removed in the right panel by using $W_{RP}$, the $G_{BP}-G_{RP}$ color is still affected by extinction.}
    \label{fig:fig22}
\end{figure*}

\begin{figure*}
	\includegraphics[width=\textwidth]{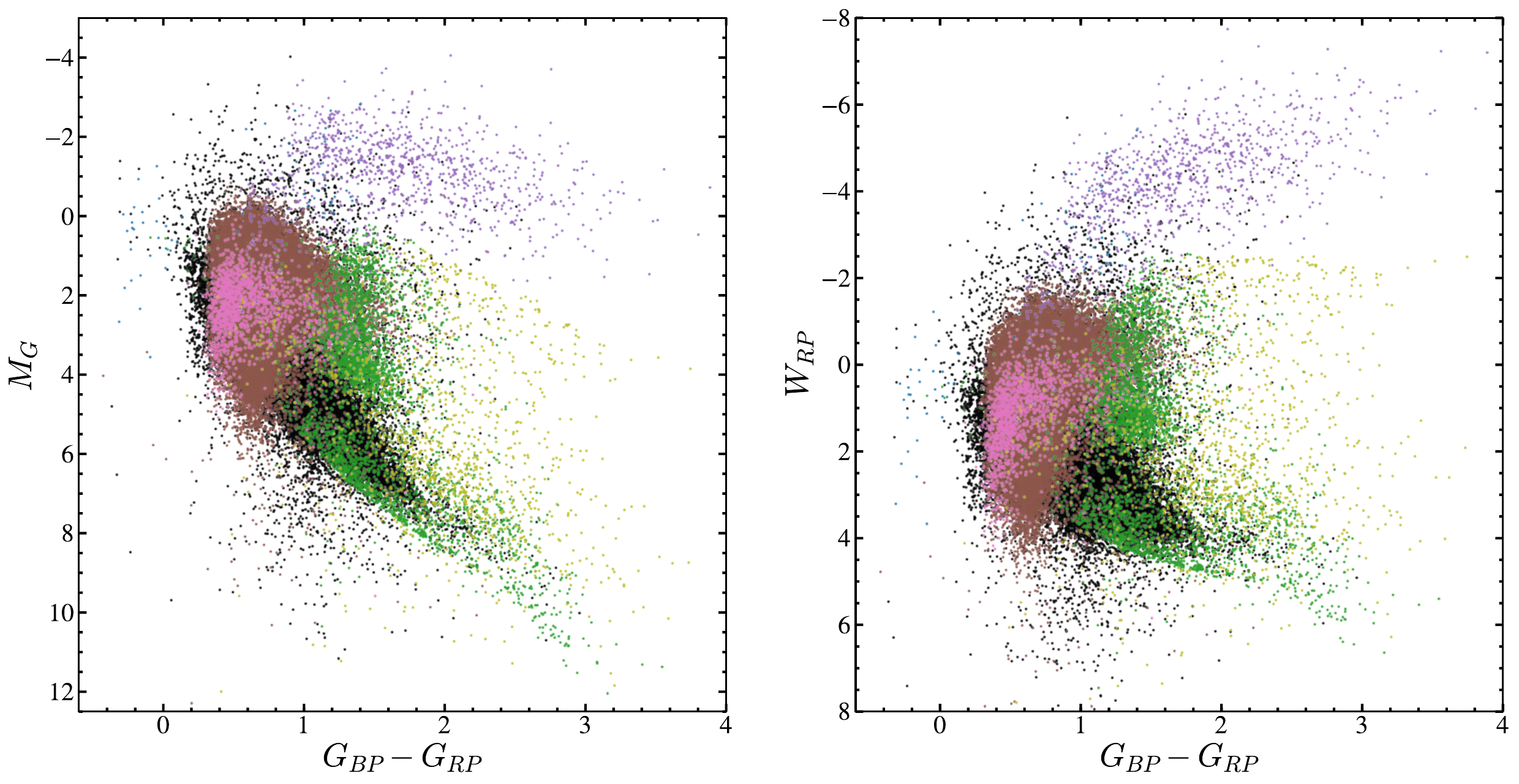}
    \caption{The Gaia DR2 $M_G$ (left) and $W_{RP}$ (right) vs. $G_{BP}-G_{RP}$ color-magnitude diagram for the variables in the training sample, excluding red giants and Mira variables. The points are colored as in Figure \ref{fig:fig22}.}
    \label{fig:fig23}
\end{figure*}

\begin{figure*}
	\includegraphics[width=\textwidth]{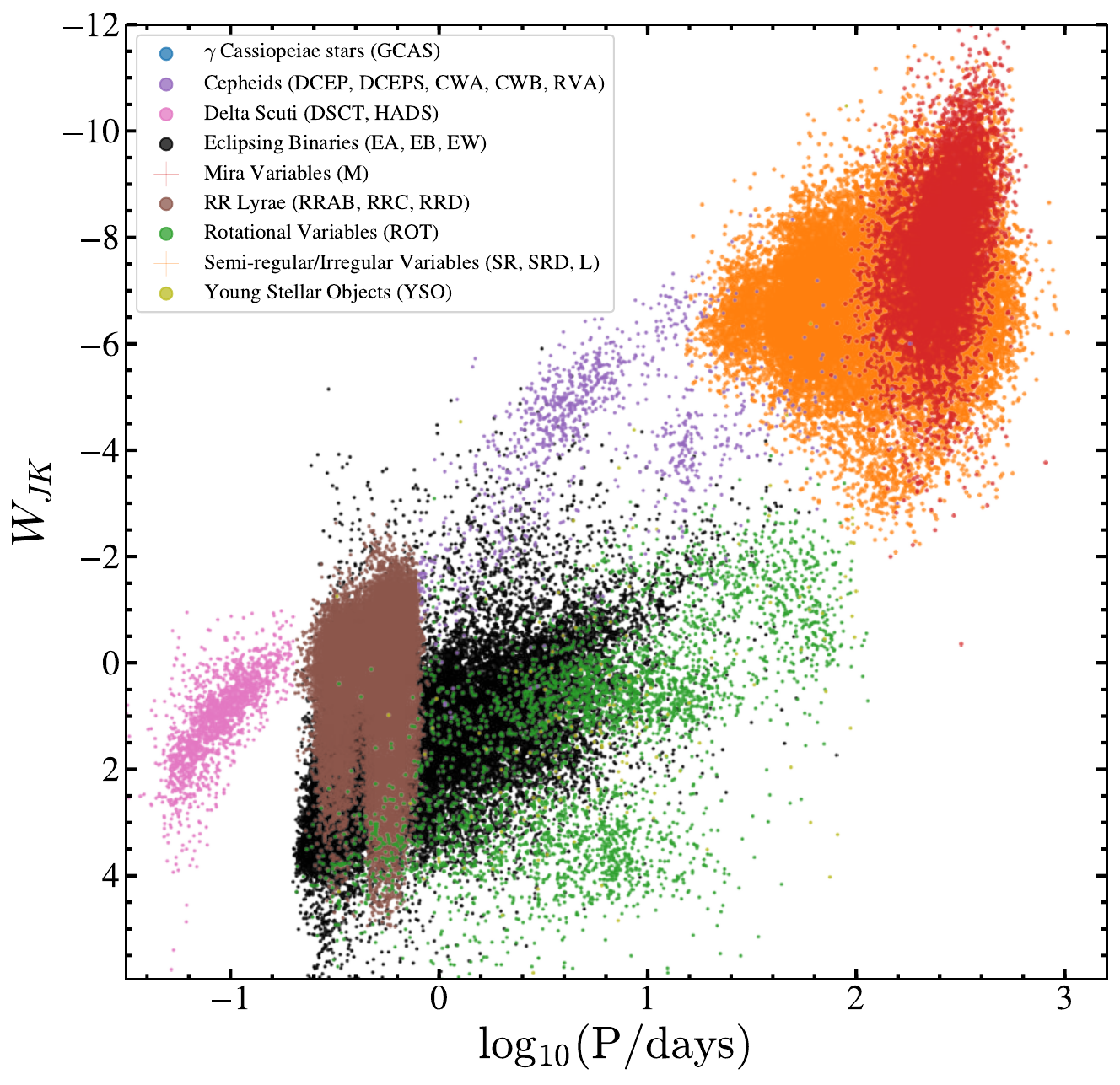}
    \caption{The Wesenheit $W_{JK}$ PLR for the different classes of periodic variables in our training sample. The points are colored as in Figure \ref{fig:fig22}.}
    \label{fig:fig24}
\end{figure*}

The Gaia DR2 $M_G$ and $W_{RP}$ vs. $G_{BP}-G_{RP}$ color-magnitude diagrams for all the sources in the training sample are shown in Figure \ref{fig:fig22}. We have sorted the variables into groups to highlight the different classes of variable sources after refinement. To highlight the variable types other than the red giants and Mira variables, we show the same color-magnitude diagrams in Figure \ref{fig:fig23}. Using the same color scheme, the combined Wesenheit $W_{JK}$ PLR diagram for the periodic variables in our training sample is shown in Figure \ref{fig:fig24}. 

\begin{figure}
	\includegraphics[width=0.5\textwidth]{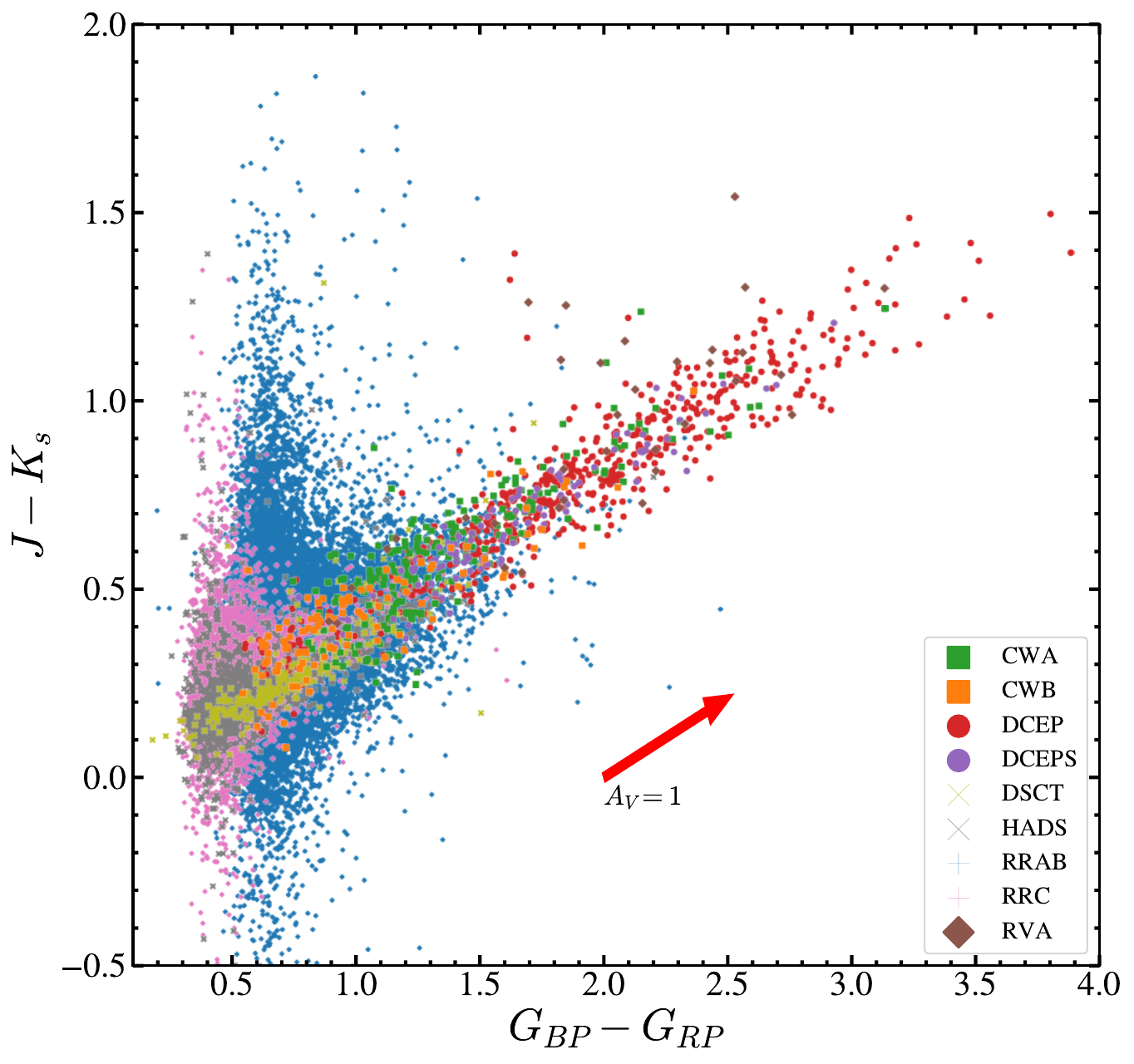}
    \caption{The Gaia DR2-2MASS $G_{BP}-G_{RP}$ vs $J-K_s$ color-color diagram for the RR Lyrae, Delta Scuti and Cepheid variables. The reddening vector corresponding to an extinction of $A_V=1$ mag is shown in red.}
    \label{fig:fig25}
\end{figure}

In Figure \ref{fig:fig25}, we show the distribution of RR Lyrae, $\delta$ Scuti and Cepheid variables in the Gaia DR2-2MASS $G_{BP}-G_{RP}$ vs $J-K_s$ color-color diagram. The locus of Cepheids (both classical and Type II) is sharp and near $G_{BP}-G_{RP}{\sim}0.5$. There is some overlap with the RR Lyrae locii. RRC, DSCT and HADS variables are typically bluer in their $G_{BP}-G_{RP}$ colors than RRAB variables. The locus of HADS variables is slightly different from that of the DSCT variables.


\begin{figure*}
	\includegraphics[width=\textwidth]{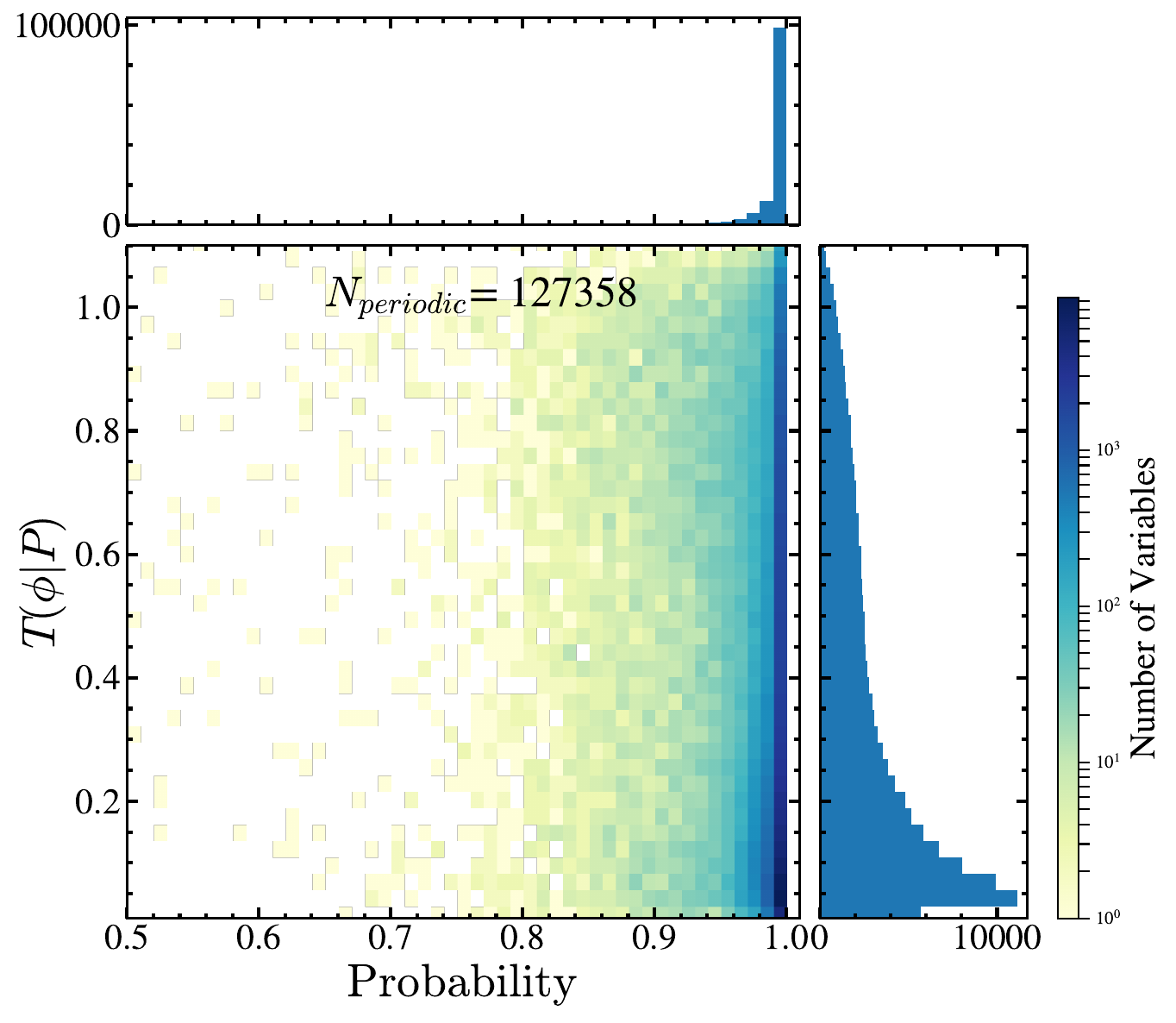}
    \caption{The distribution of the periodic variables in the training sample in $T(\phi | P)$ and classification probability. }
    \label{fig:fig26}
\end{figure*}

\begin{figure*}
	\includegraphics[width=\textwidth]{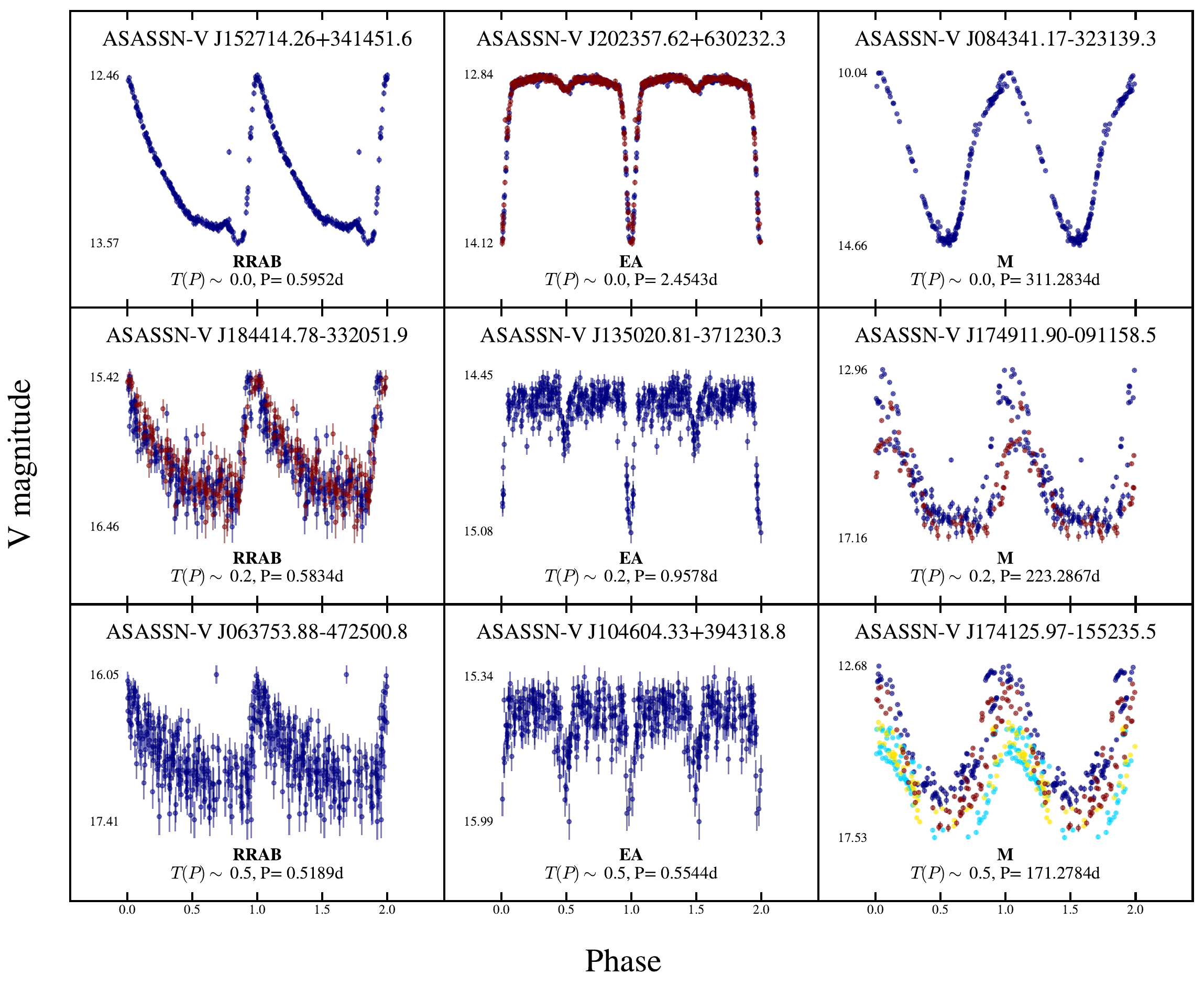}
    \caption{Examples of RRAB (left), EA (middle) and Mira (right) variables with varying $T(\phi | P)$. The format is the same as Figure \ref{fig:fig8}.}    
    \label{fig:fig27}
\end{figure*}

Figure \ref{fig:fig26} shows the distribution of the training sample in the space of $T(\phi | P)$ and classification probability $\rm Prob$ for the ${\sim}127,000$ periodic variables in the training set. Values for $T(\phi | P)$ or $T(t)$ and $\rm Prob$ are provided on the ASAS-SN Variable Stars Database for users to fine-tune their queries and retrieve high confidence classifications. $T(\phi | P)$ and $T(t)$ are very sensitive to outliers and is a probe of light curve quality, especially for strictly periodic variable types like RR Lyrae and Cepheids. $T(\phi | P)$ will be larger for sources with noisy light curves, those with outliers and less regular periodic variables. We illustrate the sensitivity of $T(\phi | P)$ to outliers and noise in Figure \ref{fig:fig27}. While we choose to use all ${\sim}127,000$ variables in our classifier, a very clean sub-sample of ${\sim}90,000$ periodic sources having high classification probabilities and well phased light curves with minimal noise/outliers can be obtained by selecting the variables with $\rm Prob>0.9$ and $T(\phi | P)<0.5$. 

Table \ref{tab:tp} shows how $T(\phi | P)$ and $T(t)$ depend on the type of variability. For periodic variables, $T(\phi | P)$ is calculated on the light curve sorted by phase. While for non-periodic sources, $T(t)$ is calculated on the light curve sorted by time. It is clear that the variable types expected to have larger dispersions in the phased light curve (ex: ROT, RRD, SR) have larger median values for $T(\phi | P)$ when compared to variable types that have more regular light curves (ex: DCEP, DCEPS, RRAB). 
\begin{table*}
	\centering
	\caption{Behavior of $T(\phi | P)$ and $T(t)$ with variability type. For periodic variables, $T(\phi | P)$ is calculated on the light curve sorted by phase. For non-periodic sources, $T(t)$ is calculated on the light curve sorted by time. The median, and standard deviation for $T(P)$ in each class is shown.}
	\label{tab:tp}
\begin{tabular}{llcr}
		\hline
		VSX Type & Description & Median $T(\phi | P)$ or $T(t)$ & $\sigma$\\
		\hline
CWA   & W Virginis type variables with $P>8$ d & 0.03 & 0.14 \\
CWB   & W Virginis type variables with $P<8$ d & 0.06 & 0.14 \\
DCEP  & Fundamental mode Classical Cepheids& 0.03 & 0.18 \\
DCEPS & First overtone Cepheids & 0.04 & 0.22\\
DSCT  & $\delta$ Scuti variables & 0.76 & 0.31\\
EA    & Detached Algol-type binaries & 0.18 & 0.28 \\
EB    & $\beta$ Lyrae-type binaries & 0.13 & 0.24\\
EW    & W Ursae Majoris type binaries & 0.22 & 0.26\\
HADS  & High amplitude $\delta$ Scuti variables & 0.16 & 0.24\\
M     & Mira variables & 0.10  & 0.20\\
ROT   & Rotational variables & 0.69 & 0.26\\
RRAB  & RR Lyrae variables (Type ab) & 0.21 & 0.22\\
RRC   & First Overtone RR Lyrae variables & 0.25 & 0.23\\
RRD   & Double Mode RR Lyrae variables & 0.72 & 0.30\\
RVA   & RV Tauri variables (Subtype A) & 0.15 & 0.31\\
SR    & Semi-regular variables & 0.55 & 0.29\\
SRD   & Yellow semi-regular variables &0.46 & 0.25\\

\hline
L     & Irregular variables & 0.18 & 0.16\\
GCAS  & $\gamma$ Cassiopeiae variables & 0.18 & 0.18\\
YSO   & Young stellar objects & 0.50 & 0.28\\

\end{tabular}
\end{table*}

\begin{figure}
	\includegraphics[width=0.5\textwidth]{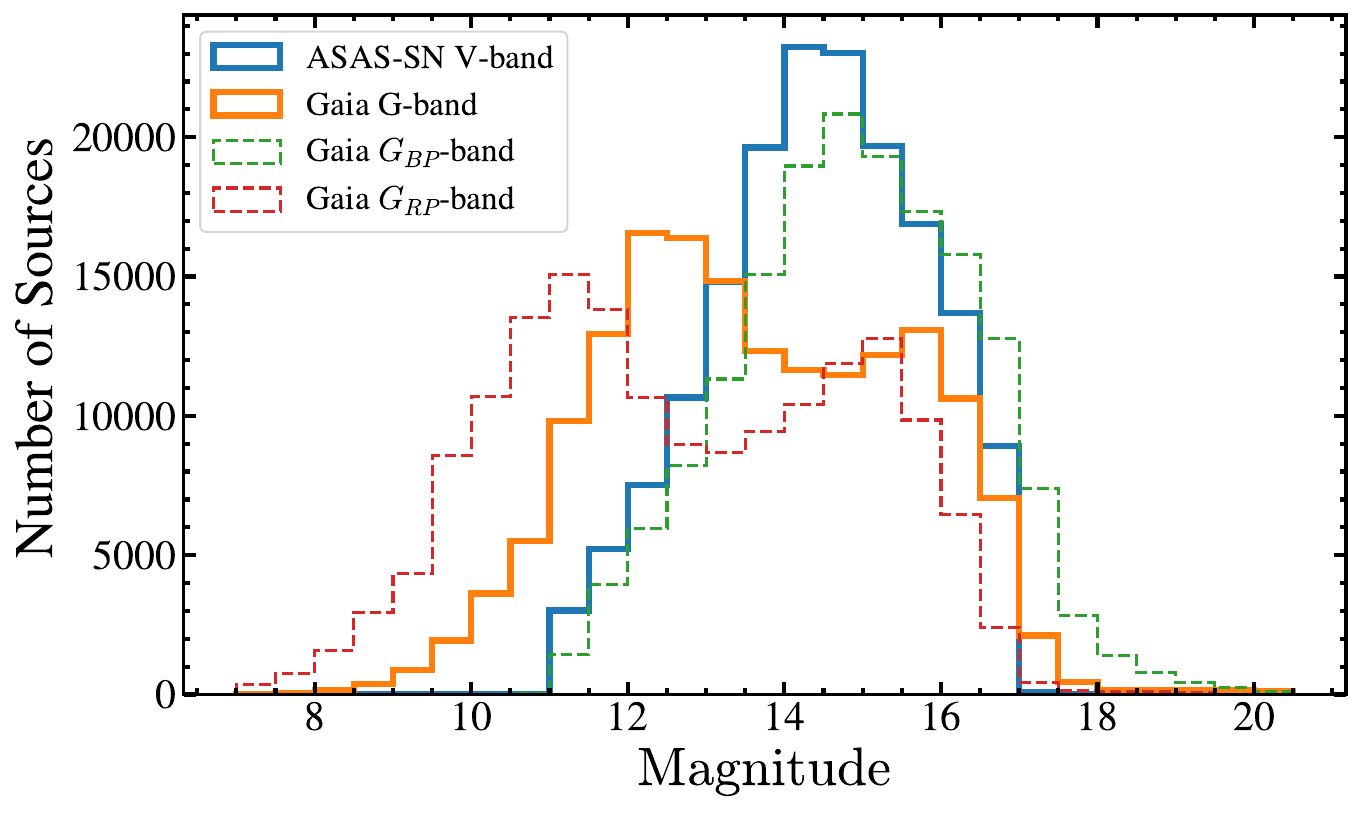}
    \caption{The distribution of the training sample in magnitude. }
    \label{fig:fig28}
\end{figure}

The distribution of the variables in our training sample with the ASAS-SN V-band, Gaia DR2 G-band, $G_{BP}$-band and $G_{RP}$-band magnitudes are shown in Figure \ref{fig:fig28}. The distribution of sources in the $G_{BP}$-band closely follows the distribution in the ASAS-SN V-band. A small fraction of these sources are consistent with issues from cross-matching. Nearest neighbor matching can result in errors, however through our refinement process, most of the sources with cross-matching errors are given a VAR or uncertain classification (denoted by ':').

\subsection{V2 of the ASAS-SN Variability Classifier}
\label{rfc}
Next, we create an updated version of our RF classifier. The random forest model parameters used to define the V2 RF classifier were the same as that used for V1. The variables in our training sample were assigned the same broad classes based on their refined classifications: CEPH, RRAB, RRC/RRD, SR/IRR, M, DSCT and ECL. To calculate Fourier features, we use a Fourier model of order 6. Distance estimates from \citet{2018AJ....156...58B} were used to calculate absolute Wesenheit magnitudes. The complete list of classification features and their importances for the V2 RF classifier is summarized in Table \ref{tab:features}. Overall, we only use 17 features here compared to the 28 used in the first version. Features that can be extracted without the use of external catalogs account for ${\sim}70\%$ of the importance value.  

As in Paper I, the training set was split for training ($80\%$) and testing ($20\%$) in order to evaluate its performance. The performance of the V2 classifier is summarized in Table \ref{tab:perf}. There is an improvement over V1 in $F_1$ scores across the board. All of the classes have $F_1$ scores greater than 99\%. The overall $F_1$ score for the V2 classifier is 99.4\%, a significant improvement from the the overall $F_1$ score of 93.3\% for V1. 

\begin{figure*}
	\includegraphics[width=\textwidth]{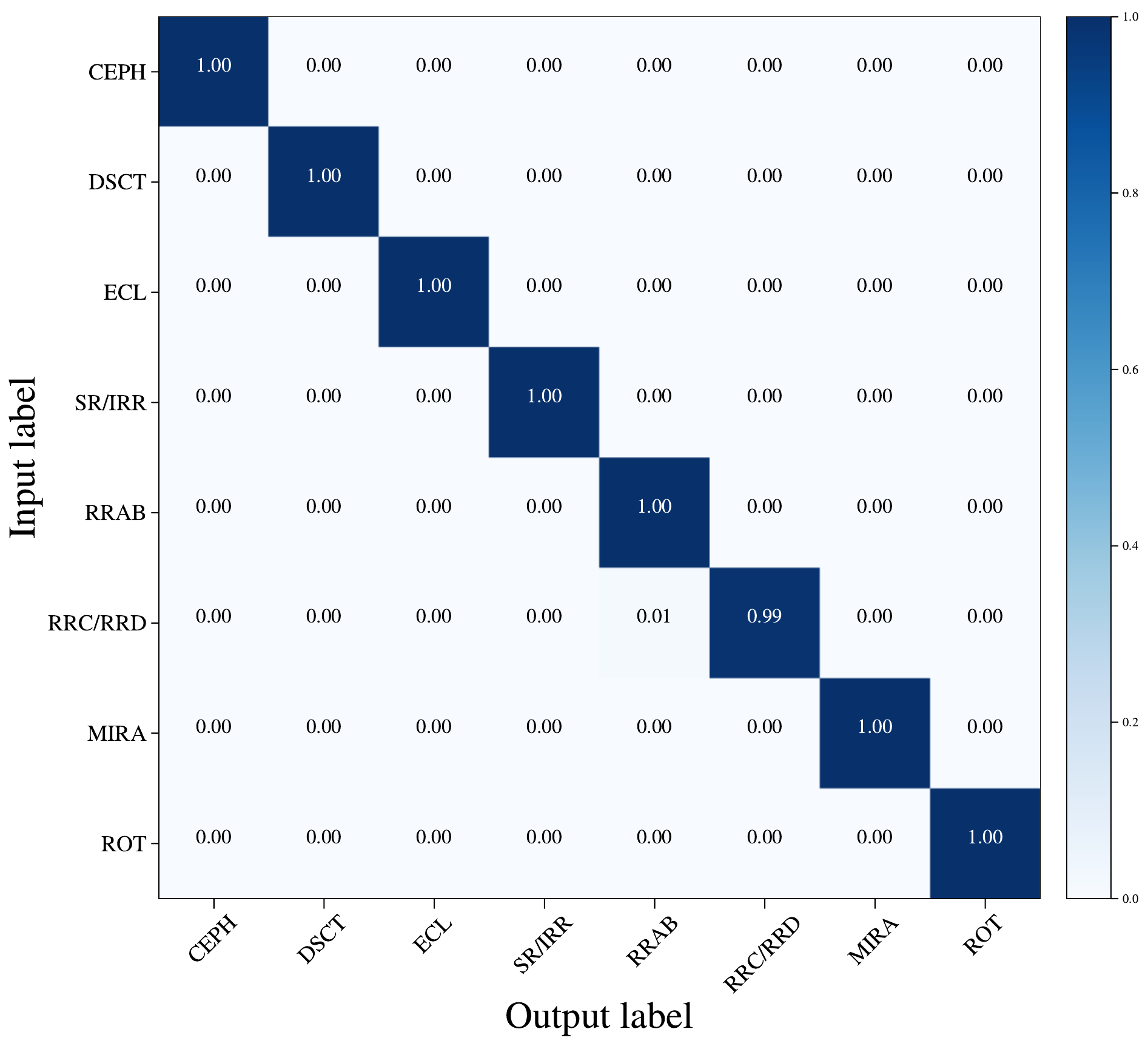}
    \caption{The normalized confusion matrix derived from the final version of the trained random forest classifier. The y-axis corresponds to the `input' classification, while the x-axis is the `output' prediction obtained from the trained random forest model.}
    \label{fig:fig29}
\end{figure*}

We illustrate the ability of the RF model to classify new objects with the confusion matrix shown in Figure \ref{fig:fig29}. The greatest confusion (1\%) is between the RRC/RRD and RRAB classes. After making this test, we rebuilt the classifier using all the variables in the training set for use on the full data set.

\begin{figure}
	\includegraphics[width=0.5\textwidth]{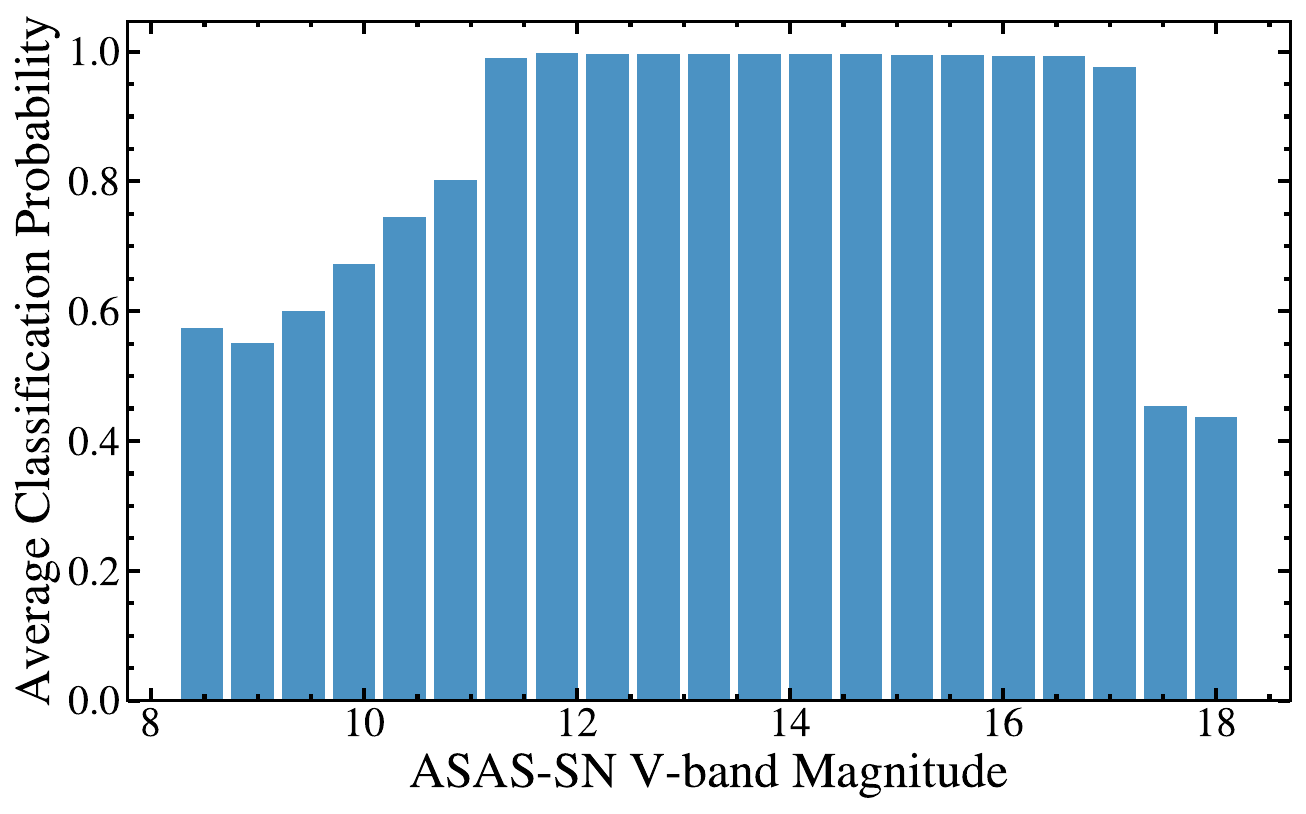}
    \caption{Average classification probability vs. ASAS-SN V-band magnitude for the ${\sim}412,000$ variables in the final catalog.}    
    \label{fig:fig30}
\end{figure}

Figure \ref{fig:fig30} illustrates the average classification probability against the mean ASAS-SN V-band magnitude of the ${\sim}412,000$ variables in the final catalog. As expected, the classification of sources in the saturated ($V<11$ mag) and noisy ($V>17$ mag) regimes is poor. This is largely due to the lack of sources with similar V-band magnitudes in the training set, and the declining quality of the ASAS-SN light curves in these regimes.

\section{Classifying other sources}
We integrated the V2 RF classifier described in Section $\S 3.6$ with the refinement criteria in Section $\S 3.4$ to create a coherent pipeline for variability classification. In this pipeline, periods are assigned to the sources if $T(\phi | P)<0.55$ for periods $\rm P <40$ d, and $T(\phi | P)<1.0$ for periods $\rm P >40$ d. We expect the SR variables that are common at longer periods to have larger values of $T(\phi | P)$. With the V2 RF classifier built, we can now classify the sources excluded when creating the training set. The variables excluded from the training set were those with miscellaneous classifications in VSX, those identified by the OGLE survey, the set of variables with low V1 probabilities (see $\S 3.3$) and those with mean magnitudes outside of the optimal ASAS-SN range ($11<V<17$ mag).

\subsection{Variables with Miscellaneous Classifications}
We applied our variability classifier to the ${\sim}94,000$ miscellaneous/generic variables with mean magnitudes in the range $11<V<17$ mag. This includes ${\sim}12,000$ variables identified by the KELT survey that do not currently have specific classifications \citep{2018AJ....155...39O}.

\begin{figure*}
	\includegraphics[width=\textwidth]{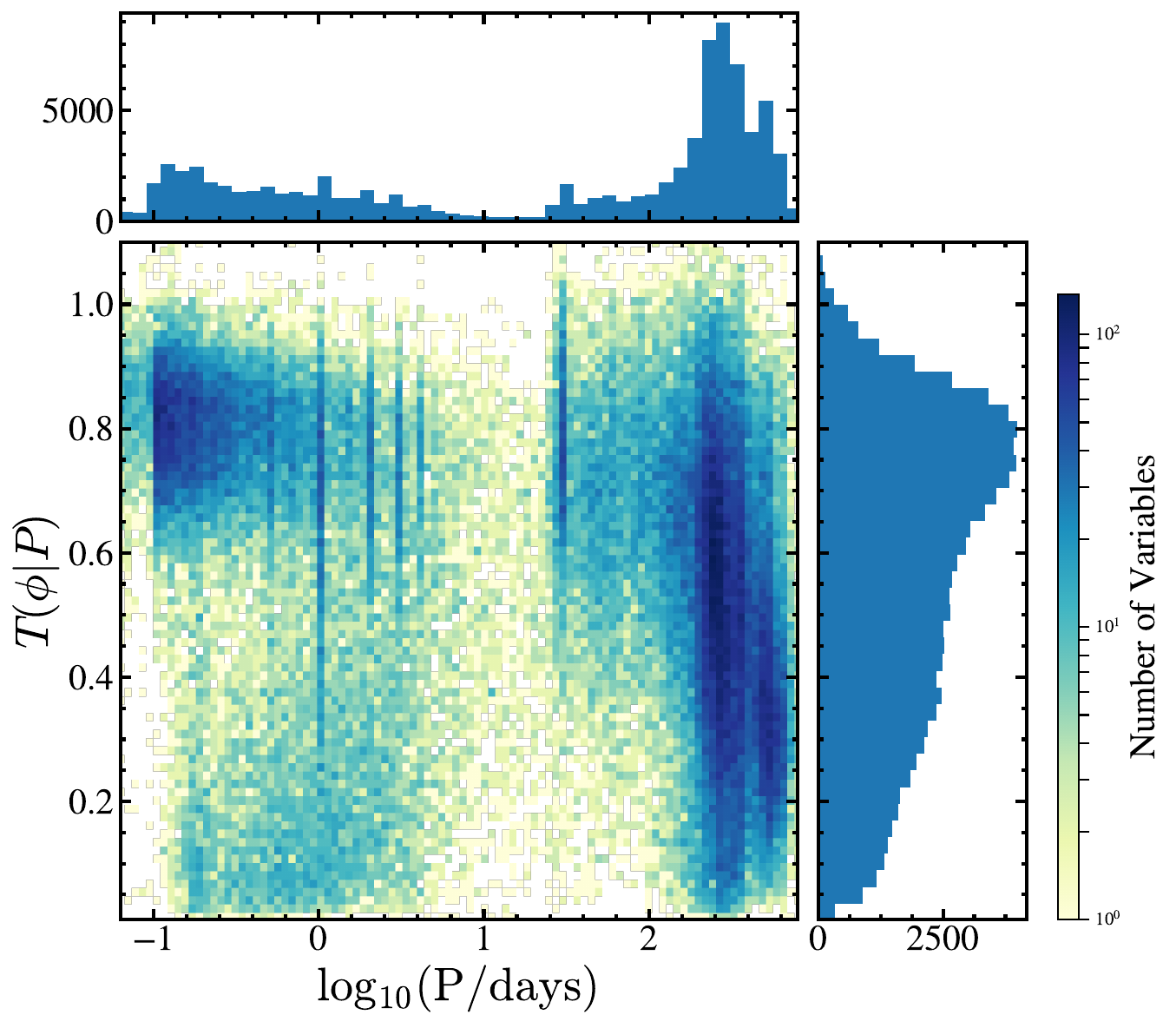}
    \caption{The distribution of the miscellaneous VSX variables with $T(\phi | P)$ and ASAS-SN $\log \rm P$.}
    \label{fig:fig31}
\end{figure*}

The distribution of the ASAS-SN period $\log \rm P$ and $T(\phi | P)$ for the miscellaneous variables is shown in Figure \ref{fig:fig31}. Features due to diurnal aliases are clearly visible. In some cases, when a light curve is phased with an alias, some spurious structure is present and minimizes $T(\phi | P)$ for that aliased period. In most such cases, these variables do not have strong periodicity. 

\begin{figure*}
	\includegraphics[width=\textwidth]{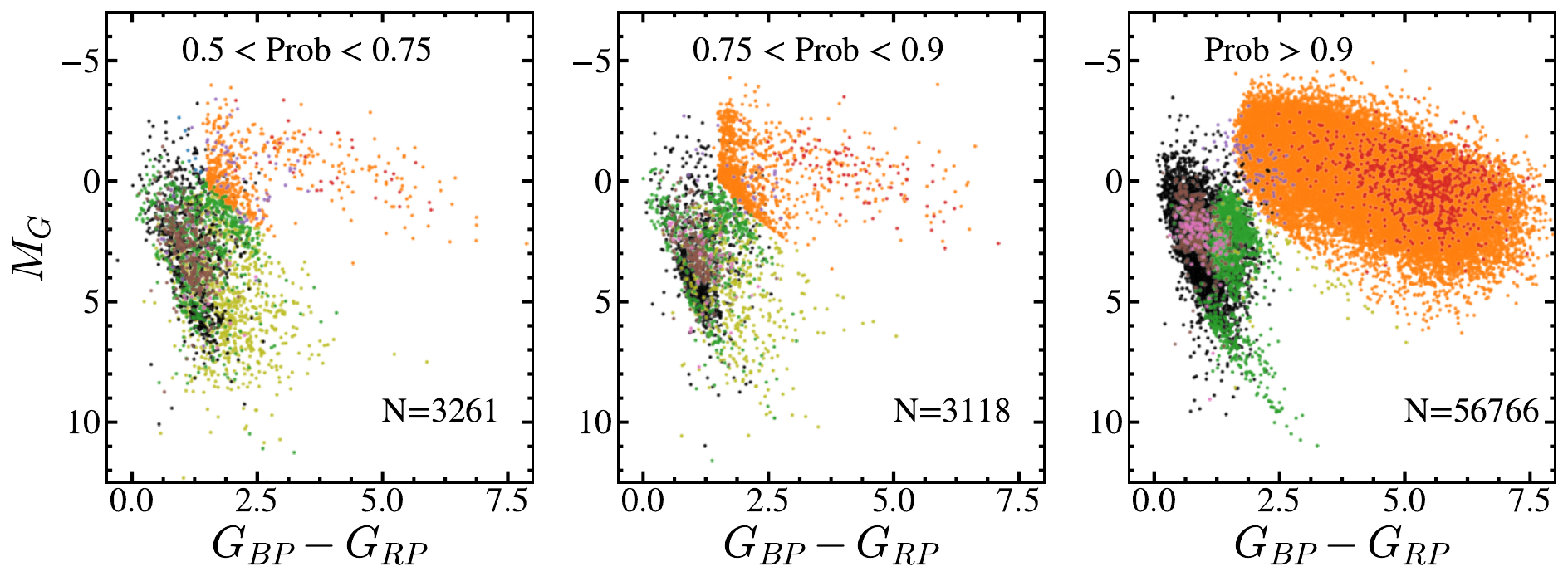}
    \caption{The Gaia DR2 $M_G$ vs. $G_{BP}-G_{RP}$ color-magnitude diagram for the variables with miscellaneous classifications in VSX. \textit{Left}: Variables with classification probability $0.5<\rm Prob <0.75$, \textit{Middle}: Variables with classification probability $0.75<\rm Prob <0.9$, \textit{Right}: Variables with classification probability $\rm Prob >0.9$. The points are colored as in Figure \ref{fig:fig22}.}
    \label{fig:fig32}
\end{figure*}

\begin{figure*}
	\includegraphics[width=\textwidth]{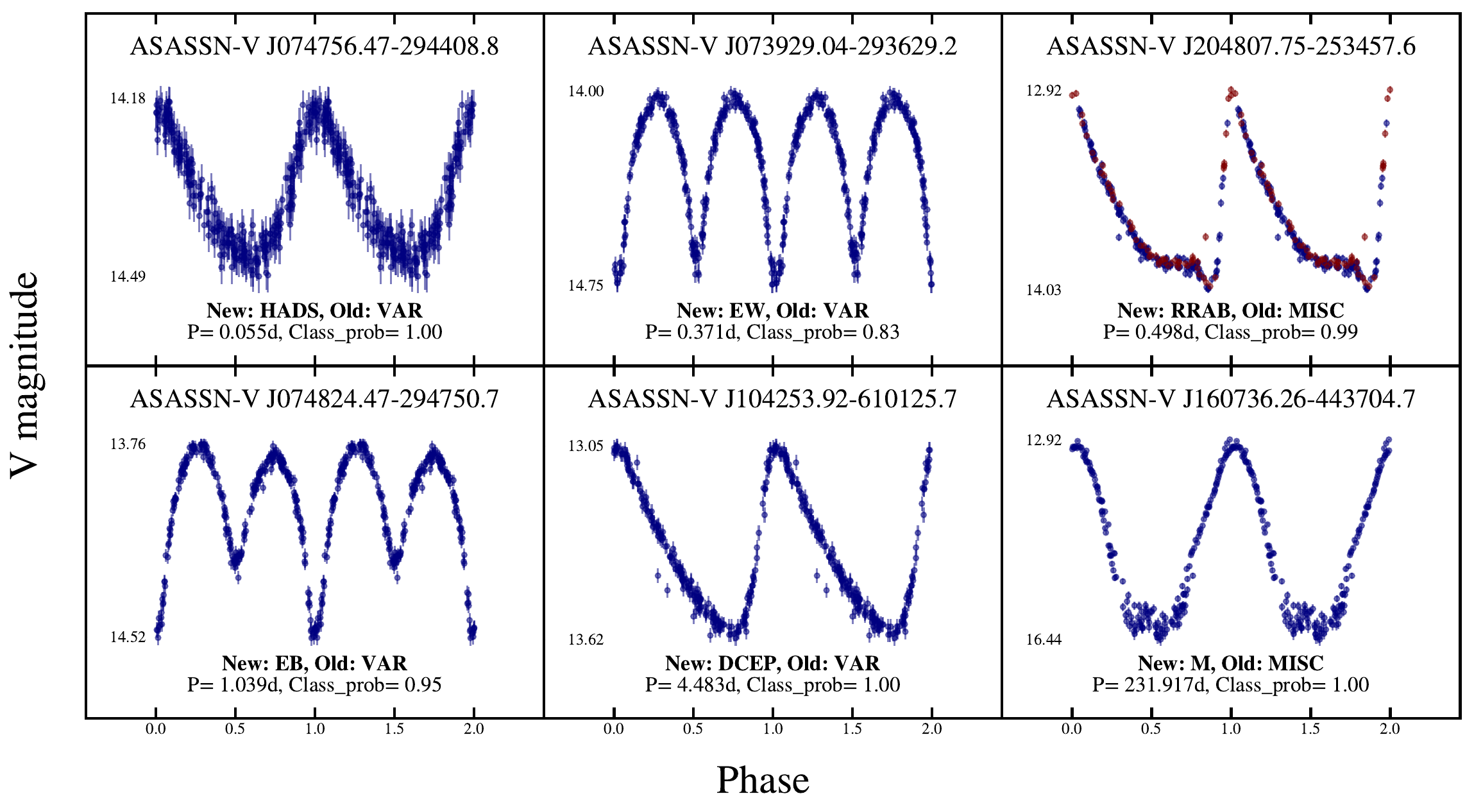}
    \caption{Examples of miscellaneous VSX variables with high probability classifications. The format is the same as Figure \ref{fig:fig8}.}    
    \label{fig:fig33}
\end{figure*}

In Figure \ref{fig:fig32}, we show these variables in the Gaia DR2 $M_G$ vs. $G_{BP}-G_{RP}$ color-magnitude diagram divided into arbitrary bins of classification probability for comparison with the distribution of the training sample shown in Figure \ref{fig:fig22}. We consider variables with classification probabilities $\rm Prob >0.9$ , $0.75<\rm Prob <0.9$, and $0.5<\rm Prob <0.75$, to have `excellent', `very good' and `good' classifications respectively. The VSX MISC category is clearly dominated by luminous, red variables. Examples of these newly classified miscellaneous variables are shown in Figure \ref{fig:fig33}. Most of these sources have good classifications: ${\sim}60\%$ (${\sim}64\%$) of the miscellaneous variables have classification probabilities of >0.9 (>0.75). However, ${\sim}28,000$ of the variables with miscellaneous VSX classifications do not have definite classifications in ASAS-SN, and are classified either to the VAR class or to one of the uncertain classifications (i.e., GCAS:, RRAB:, ROT:, and DSCT:). 

\subsection{OGLE Variables}
\begin{figure*}
	\includegraphics[width=\textwidth]{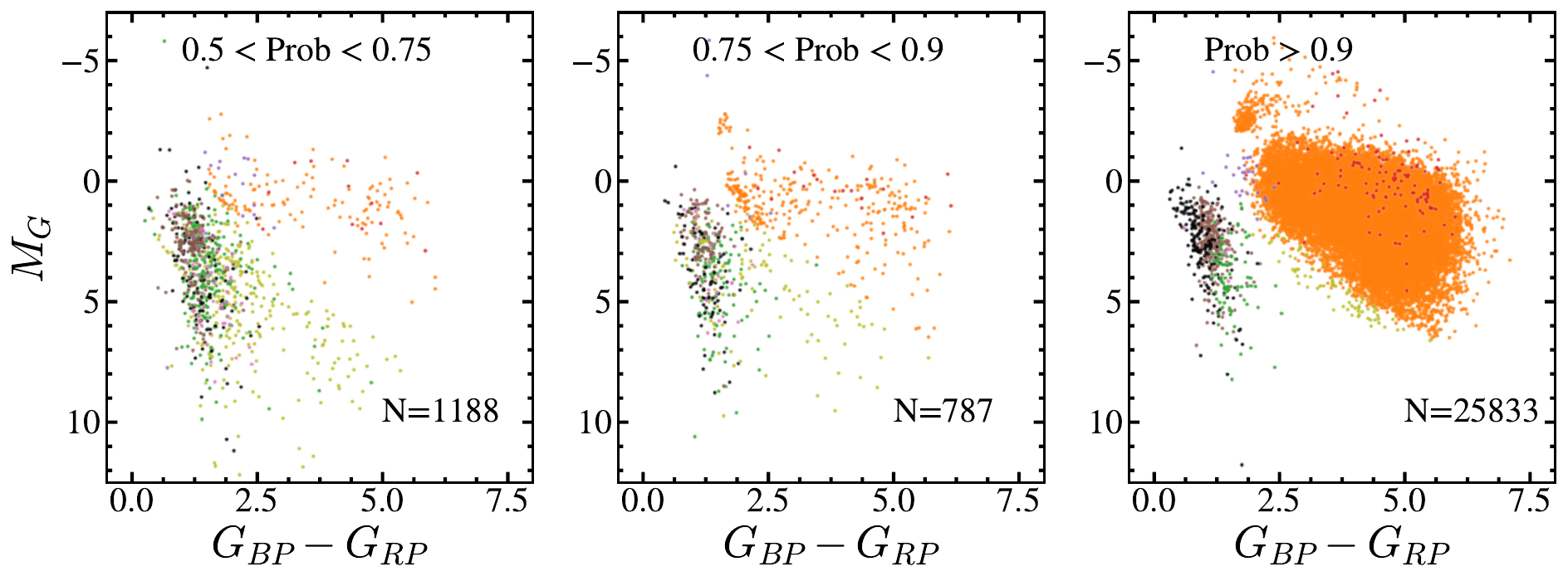}
    \caption{The Gaia DR2 $M_G$ vs. $G_{BP}-G_{RP}$ color-magnitude diagram for the sample of OGLE variables that were analyzed with the classification pipeline. \textit{Left}: Variables with classification probability $0.5<\rm Prob <0.75$, \textit{Middle}: Variables with classification probability $0.75<\rm Prob <0.9$, \textit{Right}: Variables with classification probability $\rm Prob >0.9$. The points are colored as in Figure \ref{fig:fig22}.}
    \label{fig:fig34}
\end{figure*}

We also apply the variability classification pipeline to the ${\sim}52,000$ OGLE variables in the VSX catalog. For the OGLE sources with an `LMC' (`SMC') identifier, we use a distance of $d=49.97$ (62.1) kpc \citep{2013Natur.495...76P,2014ApJ...780...59G}. We use the distance estimates from \citet{2018AJ....156...58B} for the remainder. These variables are illustrated in the Gaia DR2 $M_G$ vs. $G_{BP}-G_{RP}$ color-magnitude diagram (Figure \ref{fig:fig34}).

The OGLE sample is dominated by sources at low Galactic latitudes ($|b|<10$ deg) and sources in the Magellanic clouds that are heavily affected by crowding and blending. Multiple stellar sources are likely to be found within the ASAS-SN FWHM (${\sim} 16\farcs0$) towards these regions. The cross-matching error rate to external photometric catalogs also increases towards crowded fields. Thus, crowding/blending can significantly affect the classification of variables. 

The reclassification of these sources is more likely to be correct for those variables with high amplitude variability signals in the ASAS-SN light curves than for the low amplitude variables that are significantly affected by crowding/blending. This problem is evident in the small numbers of OGLE variables that have high classification probabilities--- only ${\sim}50\%$ (${\sim}51\%$) of these variables have classification probabilities of >0.9 (>0.75). Most of the variables with classification probabilities $\rm Prob>0.9$ are red giants with large variability amplitudes, as expected. Of the OGLE variables, 41\% have matching variability classes and 93\% of the OGLE variables with matching variability classes fall into the SR/IRR class. Of the OGLE variables with discrepant variable classes, 73\% have uncertain classifications in ASAS-SN. Of the OGLE variables with discrepant classes, 67\% are originally classified into the RRAB, RRC/RRD and ECL classes, further indicative of the blending and crowding issues towards these fields.

\subsection{Sources with low V1 classification probabilities}
\begin{figure*}
	\includegraphics[width=\textwidth]{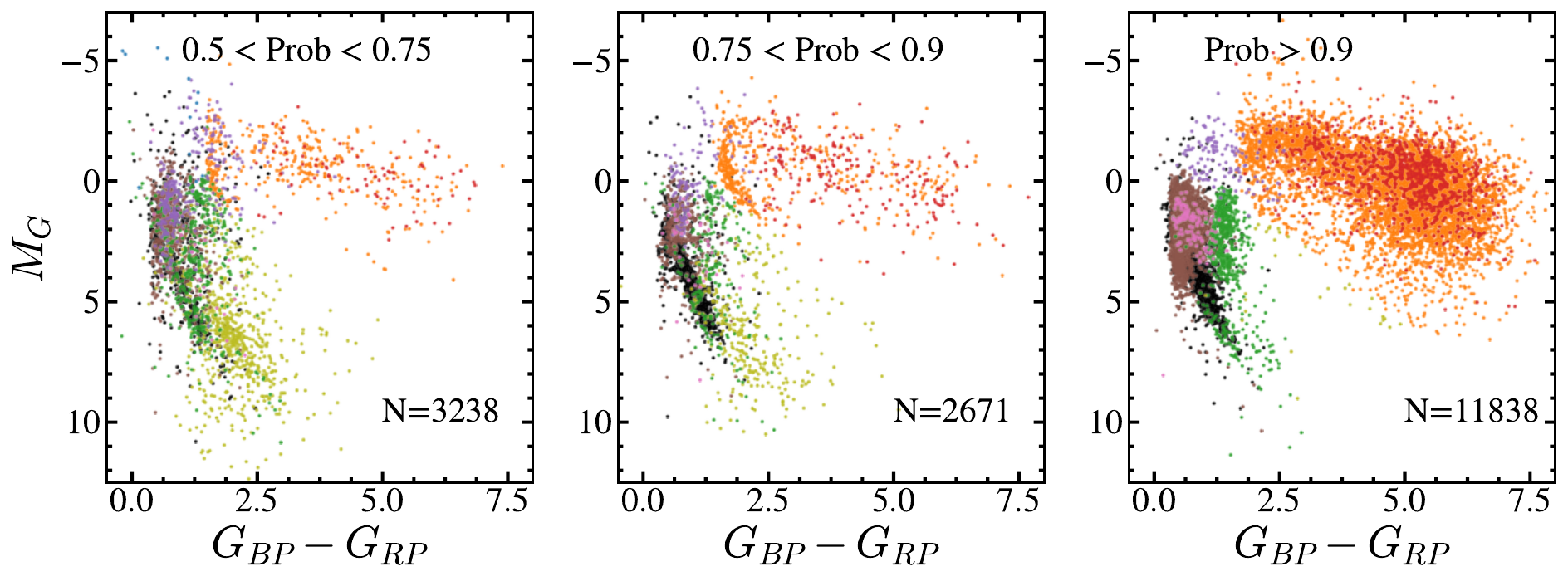}
    \caption{The Gaia DR2 $M_G$ vs. $G_{BP}-G_{RP}$ color-magnitude diagram for the sample of variables with low V1 classification probabilities. \textit{Left}: Variables with classification probability $0.5<\rm Prob <0.75$, \textit{Middle}: Variables with classification probability $0.75<\rm Prob <0.9$, \textit{Right}: Variables with classification probability $\rm Prob >0.9$. The points are colored as in Figure \ref{fig:fig22}.}
    \label{fig:fig35}
\end{figure*}

Finally, we analyzed the ${\sim} 47,000$ variables that had discrepant classifications between VSX and the V1 classifier or were excluded from the initial training set due to low V1 classification probabilities ($\S 3.5$). These variables are illustrated in the Gaia DR2 $M_G$ vs. $G_{BP}-G_{RP}$ color-magnitude diagram (Figure \ref{fig:fig35}). A large fraction of these variables are assigned uncertain classifications (37\%) by the V2 classifier and only 25\% (31\%) of these variables have classification probabilities of >0.9 (>0.75). A large fraction (40\%) of these sources are located towards low Galactic latitudes ($|b|<10$ deg) and so they will be more affected by crowding and extinction.

\section{Estimating Variability Amplitudes Through Random Forest Regression}

In Paper I, we calculated the variability amplitude $A$ as the difference between the $5^{\rm th}$ and $95^{\rm th}$ percentiles in magnitude. While this is well defined, it is not ideal as it ignores 10\% of the data. Simply consider the case of eclipsing binaries--- this approach will ignore the deepest parts of narrow eclipses and thus underestimate the variability amplitude. 

In our new analysis, we instead use Random Forest Regression to estimate the amplitudes. A Random Forest regressor uses decision trees (much like a random forest classifier) to fit a model to the data \citep{breiman,2018arXiv180502587K}. This technique is useful because it does not assume a particular model when fitting the data. Unlike Fourier models, one can use RF regression for multi-periodic or irregular sources. RF regression models are also likely to work better for the light curves of detached eclipsing binaries.

We use different RF regression models to fit the light curves of periodic and irregular sources. For the light curves of periodic variables, we fit a RF regression model to the phased light curve. In this case, we prune the trees at a depth of \verb"max_depth=10" and set \verb"min_samples_leaf=2", to minimize the effect of outliers. For the light curves of irregular variables, we fit a RF regression model to the temporal light curves. The trees were pruned at a depth of \verb"max_depth=16". The light curves of irregular variables are sparsely populated in time when compared to the phased light curves of periodic variables in phase space, so we choose to keep \verb"min_samples_leaf=1". Similarly, we choose to increase the depth of the trees in the irregular RF regression model to better fit complex light curves. In both RF regression models, the number of decision trees in the forest was set to \verb"n_estimators=800".

\begin{figure*}
	\includegraphics[width=\textwidth]{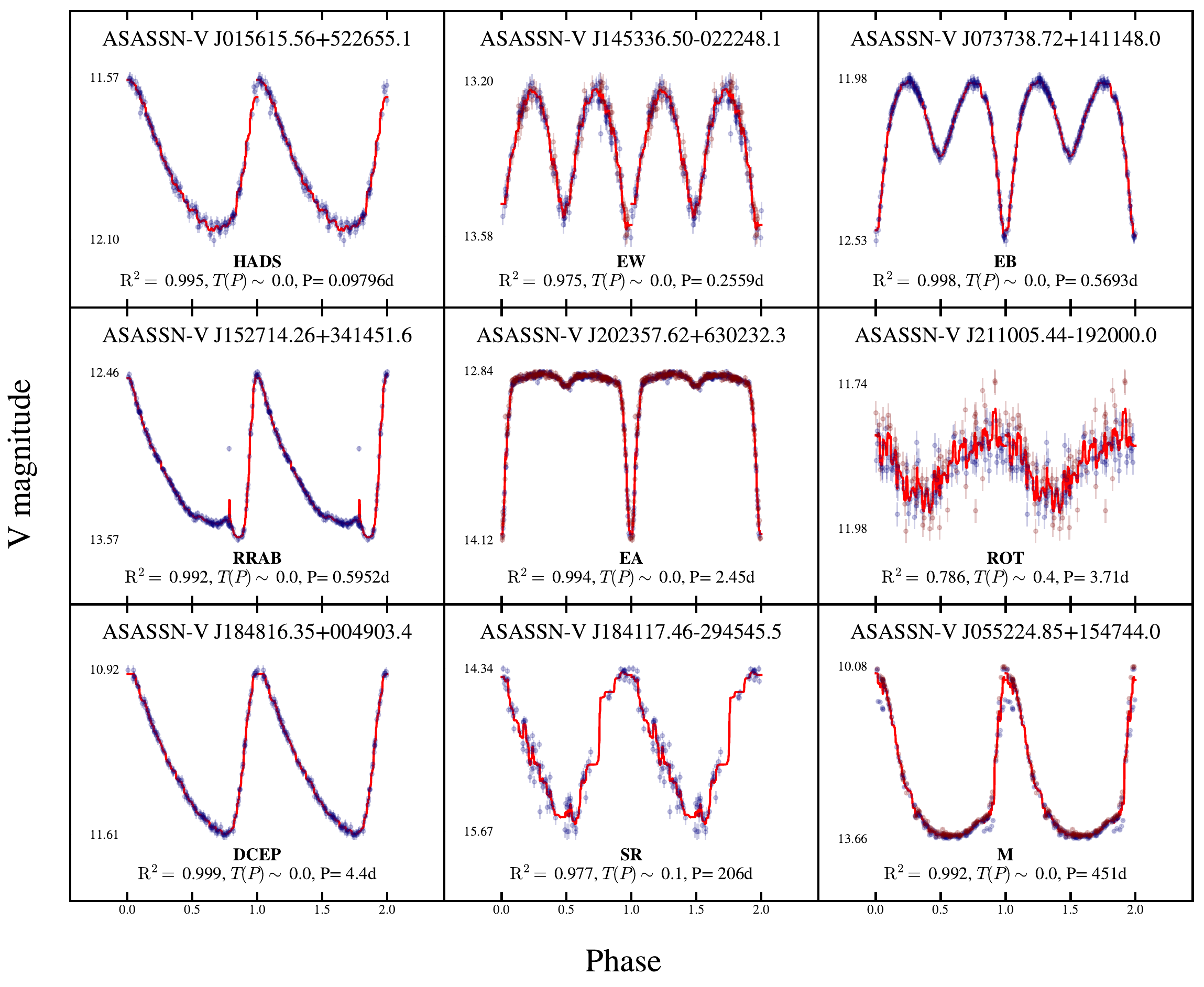}
    \caption{Random Forest Regression fits for periodic variables. The RF Regression model is plotted in red.}    
    \label{fig:fig36}
\end{figure*}

\begin{figure*}
	\includegraphics[width=\textwidth]{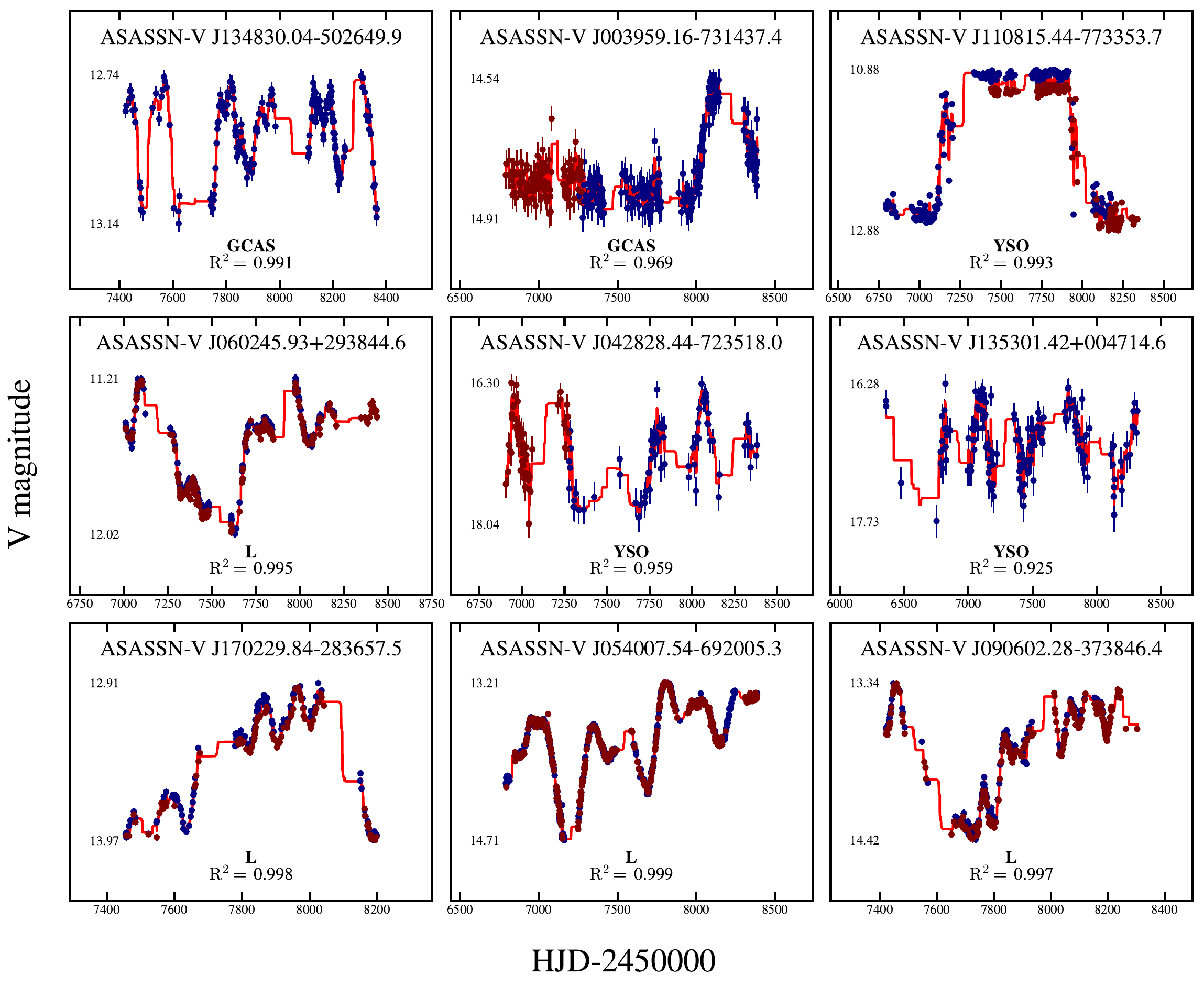}
    \caption{Random Forest Regression fits for irregular variables. The RF Regression model is plotted in red. The format is the same as Figure \ref{fig:fig36}.}    
    \label{fig:fig37}
\end{figure*}

The variability amplitude ($A_{\rm RFR}$) is calculated from a RF regression model $R(x)$ as \begin{equation}
    A_{\rm RFR}=R(x)_{max}-R(x)_{min},
	\label{eq:rfr}
\end{equation} where $x$ is either the phase or date. Each RF regression fit is assigned a score \begin{equation}
    R^2=1-\frac{\sum_{i=1}^{\rm N} \big[(m_{\rm true}-m_{\rm pred})\big]}{\sum_{i=1}^{\rm N} \big[(m_{\rm true}-\overline m_{\rm true})^2 \big]},
	\label{eq:rfs}
\end{equation}where $m_{\rm true}$ is the ASAS-SN V-band magnitude and $m_{\rm pred}$ is the predicted magnitude for the same point from the RF regression model. The RFR score is in the range $0\leq R^2 \leq1$, with a perfect model having a RFR score of $R^2=1$. Examples of these RF regression models for the light curves of periodic variables are shown in Figure \ref{fig:fig36}. For periodic variables, the variability amplitude derived from the RF regression model ($A_{\rm RFR}$) is dependent on the period. To investigate this further, we recalculated the variability amplitudes for the sources shown in Figure \ref{fig:fig36} with test periods ($\rm P_{\rm test}$) in the range [0.5$\rm P_{\rm true}$,2$\rm P_{\rm true}$] and an interval of 0.05$\rm P_{\rm true}$. When phased with 0.5$\rm P_{\rm true}$, we obtained fractional differences in $A_{\rm RFR}$ ranging from 0.6\% to 23\%. When phased with 2$\rm P_{\rm true}$, we obtained fractional differences in $A_{\rm RFR}$ ranging from 0.04\% to 40\%. The largest fractional difference in $A_{\rm RFR}$ across the entire range in $\rm P_{\rm test}$ for all the variables was 50\%. RF regression model fits for irregular sources are shown in Figure \ref{fig:fig37}. The models are not perfect, but they are correctly estimating the variability amplitudes even for the irregular light curves shown in Figure \ref{fig:fig37}.

\begin{figure*}
	\includegraphics[width=\textwidth]{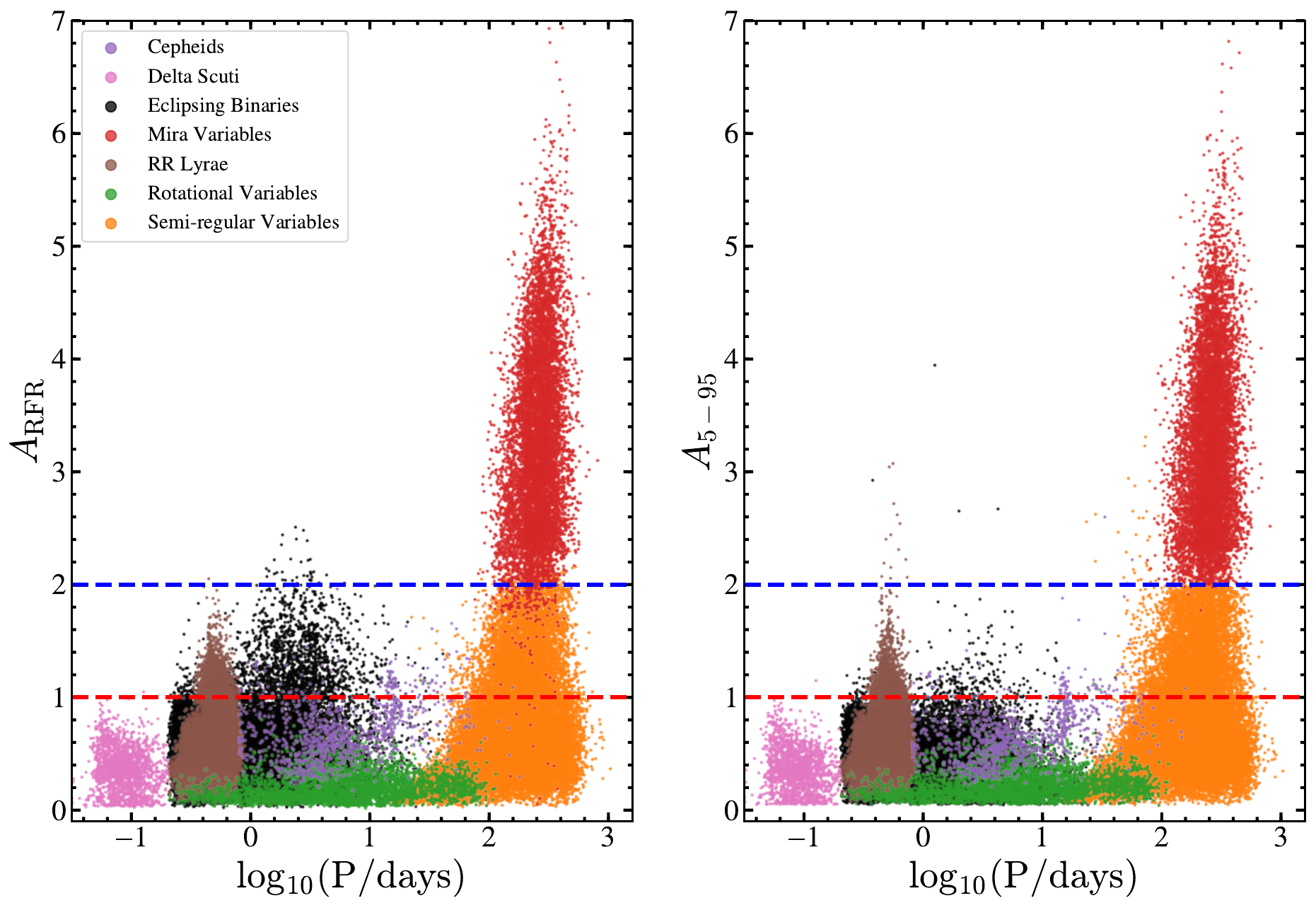}
    \caption{Period-amplitude plot for the variables in our training sample under the two methods of calculating amplitudes, the RFR method used in this paper ($A_{\rm RFR}$, left) and the 5-95\% spread used in Paper I ($A_{5-95}$, right). Reference amplitudes of 1 and 2 mag are shown in red and blue respectively. The points are colored according to the prescription in Figure \ref{fig:fig22}. The differences are most noticeable for eclipsing binaries.}    
    \label{fig:fig38}
\end{figure*}

We use the variability amplitudes derived from RF regression fits to illustrate the training sample in period-amplitude space (Figure \ref{fig:fig38}) grouped by the variable groups described in Figure \ref{fig:fig22}. The left panel illustrates the amplitudes calculated using RF regression, whereas the right panel shows the amplitudes for the same sources calculated based on the prescription in Paper I. It is clear that the biggest difference between these two methods lie in the calculation of variability amplitudes for eclipsing binaries. $A_{\rm RFR}$ clearly provides better estimates for the depths of eclipses than the competing method. We do not see major differences between the two methods for other variable types.

\begin{table}
	\centering
	\caption{Distribution of $A_{\rm RFR}$, in magnitudes, with variability type for the V2 classifier training sample. The median, standard deviation, $5^{\rm th}$ percentile and $95^{\rm th}$ percentile for $A_{\rm RFR}$ for each class is given.}
	\label{tab:arfr}
\begin{tabular}{lrrrr}
		\hline
		VSX Type & Median  & $\sigma$ & $5^{\rm th}$ percentile & $95^{\rm th}$ percentile\\
		\hline
CWA   & 0.87 & 0.25 & 0.47 & 1.23 \\
CWB   & 0.81 & 0.25 & 0.41 & 1.28 \\
DCEP  & 0.65 & 0.21 & 0.4  & 1.05 \\
DCEPS & 0.35 & 0.08 & 0.26 & 0.51 \\
DSCT  & 0.08 & 0.03 & 0.04 & 0.14 \\
EA    & 0.5  & 0.35 & 0.18 & 1.31 \\
EB    & 0.39 & 0.20  & 0.17 & 0.8  \\
EW    & 0.41 & 0.17 & 0.18 & 0.75 \\
HADS  & 0.39 & 0.16 & 0.19 & 0.7  \\
M     & 3.18 & 0.82 & 2.13 & 4.7  \\
ROT   & 0.17 & 0.12 & 0.06 & 0.44 \\
RRAB  & 0.81 & 0.25 & 0.37 & 1.19 \\
RRC   & 0.46 & 0.12 & 0.25 & 0.67 \\
RRD   & 0.72 & 0.21 & 0.36 & 1.08 \\
RVA   & 0.93 & 0.3  & 0.49 & 1.38 \\
SR    & 0.55 & 0.45 & 0.16 & 1.65 \\
SRD   & 0.37 & 0.35 & 0.15 & 1.21 \\
\hline
GCAS  & 0.36 & 0.29 & 0.29 & 0.99 \\
L     & 0.57 & 0.37 & 0.25 & 1.37 \\
YSO   & 0.85 & 0.55 & 0.14 & 1.91
\end{tabular}
\end{table}

Table \ref{tab:arfr} shows the dependence of the amplitude $A_{\rm RFR}$ with the VSX type for the variables in the V2 training sample. In general, rotational variables have smaller variability amplitudes, with $A_{\rm RFR}<0.5$ mag, than other variables at similar periods. The majority of the $\delta$-Scuti sample consists of HADS variables rather than the lower amplitude DSCT variables. Eclipsing binaries with deep eclipses usually consist of two stars with very different temperatures, and such systems are relatively rare. Only ${\sim}5\%$ of the detached EA systems have $A_{\rm RFR}>1.3$ mag. The distinction between semi-regular variables and Mira variables is simply set by the amplitude cut ($A_{\rm RFR}>2$ mag) used in this work. The amplitudes of Mira variables can be as high as $A_{\rm RFR}{\sim}5$ mag. We reclassify high-amplitude ($A_{\rm RFR}>2$ mag) semi-regular variables as Mira variables provided that they also have a period $P>80$ d.

We identify light curves that are likely to have problematic outliers if they satisfy either $\log \rm P<-0.7$ and $A_{\rm RFR}>1.1$ mag OR $-0.7 \leq \log \rm P \leq 1.9$ and $A_{\rm RFR}>2$ mag. These light curves are removed from our catalog if they also have an RFR score $R^2<0.8$ and $T(\phi| P)$>0.5. Only ${\sim}250$ light curves were removed through this procedure.

\begin{figure*}
	\includegraphics[width=\textwidth]{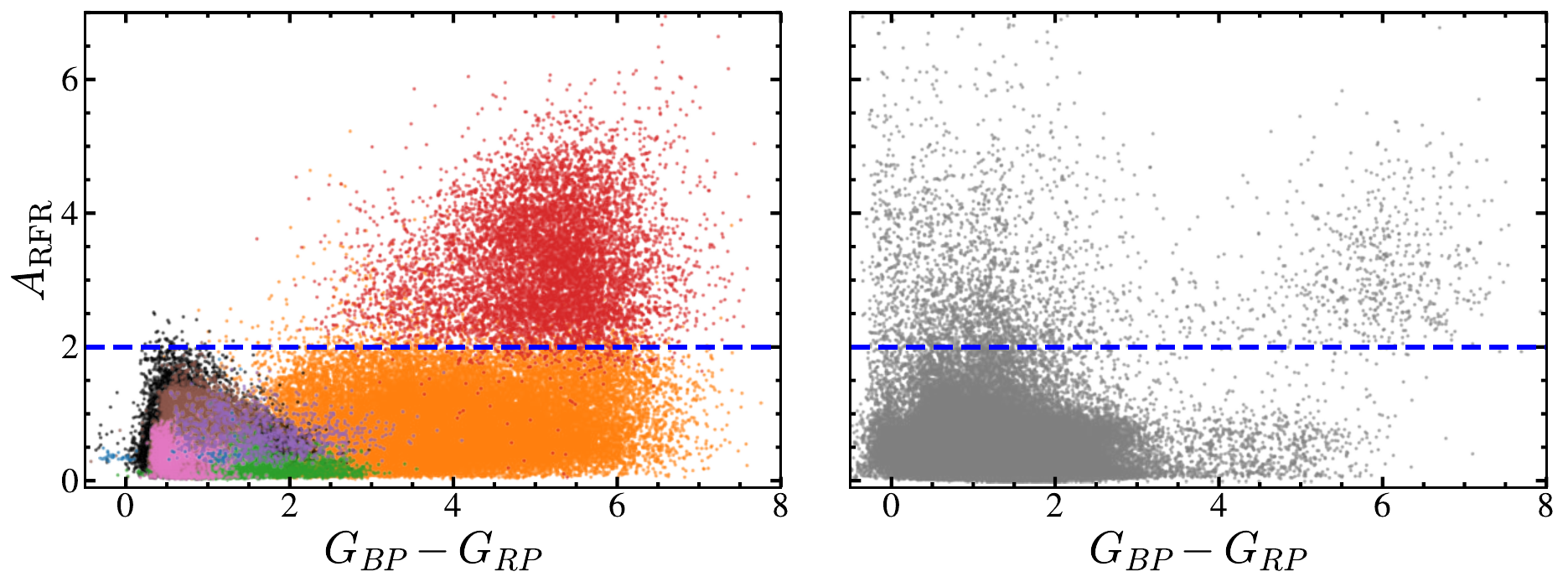}
    \caption{Color-amplitude plot for the variables with certain (left) and uncertain (right) classifications in our catalog. A reference amplitude of 2 mag is shown in blue. The points in the left panel are colored according to the prescription in Figure \ref{fig:fig22} and the variables with uncertain classifications are colored in gray.}    
    \label{fig:fig39}
\end{figure*}

We perform an additional check to identify problematic light curves by looking at the distribution of sources with uncertain classifications in the ($G_{BP}-G_{RP}$)-$A_{\rm RFR}$ color-amplitude space (Figure \ref{fig:fig39}). The left panel shows the sources in the training sample and the right panel illustrates the ${\sim}114,000$ sources with uncertain classifications in the complete catalog (including saturated and noisy sources). There is an abundance of sources that have uncertain classifications with $A_{\rm RFR}>2$ mag and $G_{BP}-G_{RP}<2.5$ mag. We visually verified that the light curves of these blue, high amplitude variables possess significant outliers. When looking at the distribution of the sources in the training sample in this region, we do not see this behavior. These variables were removed from our final catalog if their classification probabilities are $\rm Prob<0.9$. We removed ${\sim} 3,000$ such light curves through this procedure. The majority of the sources with $A_{\rm RFR}>2$ mag and $G_{BP}-G_{RP}>2.5$ mag are Mira variables without an ASAS-SN period.  

\section{Discussion}
We have homogeneously classified ${\sim} 412,000$ variables using our V2 variability classification pipeline. The classifier combines the ASAS-SN light curves with the data from Gaia DR2, 2MASS and AllWISE surveys. This catalog provides a set of variables suitable for training automated variability classifiers for both existing and future variability surveys. We have refined the periods and classifications of ${\sim} 356,000$ variables with V-band magnitudes in the range $11<V<17$ mag, some of which were cataloged over ${\sim} 50$ years ago. The homogeneity of our classifications provides a useful point of comparison to the results from other projects. We also make the V-band light curves of the variables in this catalog available online at the ASAS-SN Variable Stars Database (\url{https://asas-sn.osu.edu/variables}). We started with ${\sim} 440,000$ variables with $11<V<17$ mag in the VSX catalog, and the final sample of ${\sim} 356,000$ is contained in the database. This includes both the ASAS-SN light curves and the ancillary data used in the classifications. The ${\sim} 84,000$ VSX variables that were dropped includes the sources with $<30$ detections in the V-band, sources with excessively noisy or bad light curves, sources with variability types not included in this work (e.g., supernovae and exoplanet hosts), and the sources with variability amplitudes $<50$ mmag. We also classified the ${\sim} 56,000$ sources with V-band magnitudes outside the range $11<V<17$ (see Appendix A). In $\S 6.1$, we discuss the catalog of known variables with definite ASAS-SN classifications. We discuss the set of known variables with uncertain classifications in $\S 6.2$, the rare variables added to our catalog in $\S 6.3$ and examine the spatial distribution of the variables in $\S 6.4$. For the subset of sources with both ASAS-SN and Gaia DR2 classifications, we compare the results of our classifier to the Gaia DR2 results in $\S 6.5$ and we discuss the implications of interstellar extinction on the classification pipeline in $\S 6.6$.

\subsection{Variables with definite classifications}
Here we discuss the catalog of ${\sim} 278,000$ variables with definite ASAS-SN classifications and V-band magnitudes in the range $11<V<17$ mag. The catalog of known variables with V-band magnitudes outside this range is discussed in Appendix A. Definite classifications are the VSX types without the uncertainty flag `:'. Uncertain classifications include the variables classified as VAR and the VSX types with an uncertainty flag.

Table \ref{tab:ndistc} lists the variability types from our classifier and the distribution of the ${\sim} 278,000$ variables with definite classifications by variability type. We also provide the number of high probability classifications ($\rm Prob>0.9 $), good classifications ($\rm Prob>0.5 $), and low scatter light curves ($T(\phi| P)$ or $T(t) <0.5$). There are ${\sim} 255,000$ sources with $\rm Prob>0.9 $, which forms a sample that is more than adequate to create a training set for future variability studies. In addition, there are ${\sim} 170,000$ variables with very good quality light curves where $T(\phi| P)$ or $T(t) <0.5$.

\begin{table*}
	\centering
	\caption{Number distribution of the ${\sim} 278,000$ variables with definite classifications by quality cuts.}
	\label{tab:ndistc}
\begin{tabular}{lrrrr}
		\hline
		VSX Type & $N_{\rm tot}$ & $N(\rm Prob>0.9) $ & $N(\rm Prob>0.5) $ & $N[T(\phi| P)$ or $T(t) <0.5] $\\
		\hline
CWA   & 422   & 261   & 357   & 382   \\
CWB   & 511   & 76    & 309   & 422   \\
DCEP  & 753   & 588   & 707   & 714   \\
DCEPS & 160   & 129   & 150   & 152   \\
DSCT  & 372   & 331   & 372   & 145   \\
EA    & 17949 & 16126 & 17628 & 14325 \\
EB    & 9544  & 8856  & 9434  & 8194  \\
EW    & 37208 & 34685 & 36975 & 28381 \\
GCAS  & 208   & 0     & 23    & 103   \\
HADS  & 1472  & 1373  & 1469  & 1227  \\
L     & 55754 & 54784 & 55503 & 42602 \\
M     & 10239 & 9587  & 10222 & 9050  \\
ROT   & 7715  & 4945  & 7030  & 3301  \\
RRAB  & 29114 & 26556 & 28494 & 24526 \\
RRC   & 7741  & 6761  & 7637  & 6213  \\
RRD   & 448   & 353   & 437   & 149   \\
RVA   & 123   & 15    & 100   & 100   \\
SR    & 92230 & 88774 & 91118 & 27757 \\
SRD   & 228   & 200   & 228   & 125   \\
YSO   & 5306  & 322   & 4013  & 1479 
\end{tabular}
\end{table*}

\begin{table*}
	\centering
	\caption{Summary of the variables with definite classifications}
	\label{tab:defvarsum}
\begin{tabular}{lrr}
		\hline
		Group & Subset & $N_{\rm tot}$ \\\\
				\hline
Definite ASAS-SN Classifications   &  & ${\sim} 278,000$  \\		
   & $\rm Prob >0.9$ &  ${\sim} 255,000$   \\
   & $T(\phi| P)$ or $T(t) <0.5$ &  ${\sim} 170,000$   \\
		\hline
No VSX period   &  & ${\sim} 52,000$  \\		
   & $\rm Prob >0.9$ &  ${\sim} 41,000$   \\
   & $T(\phi | P)<0.5$ &  ${\sim} 20,000$   \\
   & $T(\phi | P)<0.2$ &  ${\sim} 8,000$   \\
   & VSX VAR &  ${\sim} 36,000$   \\   
   &  ASAS-SN SR &  ${\sim} 33,000$   \\ 
   &  ASAS-SN ECL &  ${\sim} 7,000$   \\   
   \hline
Variables with definite VSX classifications   &  & ${\sim} 200,000$  \\		
   &  Different variability class in ASAS-SN &  ${\sim} 65,000$   \\
   &   + $\rm Prob >0.9$ &  ${\sim} 18,000$   \\    
   & Definite ASAS-SN classifications &  ${\sim} 153,000$   \\   
   & + different variability class &  ${\sim} 23,000$   \\   
   & + $\rm Prob >0.9$ &  ${\sim} 17,000$   \\     
   \hline

ASAS-SN Variables from Paper I   &  & ${\sim} 66,000$  \\	
   & Definite classifications from Paper I &  ${\sim} 61,000$   \\
   &  + different variability class &  ${\sim} 5,000$   \\   
   & + $\rm Prob >0.9$ &  ${\sim} 1,400$   \\    
   \hline   
 
VSX eclipsing binaries assigned new photometric classes   &  & ${\sim} 6,000$  \\		
   \hline      
\end{tabular}
\end{table*}

A number of sources in the VSX catalog do not have previously derived periods, and we derive periods for ${\sim} 52,000$ such sources. Of these, 78\% have V2 classification probabilities $\rm Prob >0.9$ and 38\% (15\%) have $T(\phi | P)<0.5$ ($T(\phi | P)<0.2$). In the VSX catalog, most (68\%) of these sources were classified as VAR variables, however, 64\% of the ${\sim} 52,000$ variables without a VSX period are classified as SR variables through our V2 classifier. A significant fraction (14\%) were classified as eclipsing binaries.

Of the ${\sim} 291,000$ VSX variables in our catalog with $11<V<17$ mag (excluding the variables from Paper I), ${\sim} 200,000$ have definite VSX classifications. Out of the variables with definite VSX classifications, ${\sim}65,000$ sources have different broad variability classes than the ones assigned to them in the VSX catalog. ${\sim} 18,000$ of these discrepant sources have high classification probabilities of $\rm Prob>0.9$.

Out of this set of ${\sim} 200,000$ variables with definite VSX classifications, if we only consider just the ${\sim} 153,000$ variables with definite ASAS-SN classifications, ${\sim} 23,000$ sources have discrepant broad classes and ${\sim}17,000$ of these have classification probabilities of $\rm Prob>0.9$. We find that ${\sim} 6,500$ variables that were previously classified as Mira variables are reclassified as semi-regular/irregular variables in ASAS-SN. It is likely that the majority of these sources are actually Mira variables but have diminished amplitudes in ASAS-SN due to blending. Using Gaia DR2 data, \citet{2018arXiv180502035M} noted this effect for ${\sim} 1,200$ sources in the sample of semi-regular variables identified by ASAS-SN in Paper I.

In addition, we have homogenized the sample of eclipsing binaries in this catalog by classifying their light curves into the photometric eclipsing binary classes defined in the VSX catalog (i.e, EA, EB and EW). We assigned new VSX classes to ${\sim} 6,000$ eclipsing binaries that were classified based on their physical configuration (i.e, EC, ESD and ED).

Of the ${\sim} 66,000$ ASAS-SN variables identified in Paper I, ${\sim}61,000$ were classified into variable types other than VAR using our previous classification scheme. From these ${\sim}61,000$ sources, ${\sim}5,000$ have different broad classes in this work, out of which ${\sim} 1,400$ have classification probabilities $\rm Prob>0.9$. We also reclassify the ${\sim}5,000$ VAR variables in Paper I to other variable types in this work. We have also broken the degeneracy in the EA|EB classification and refined the classification of Cepheids and YSOs from Paper I. 

We have summarized the information discussed in this section in Table \ref{tab:defvarsum}.

\subsection{Variables with uncertain classifications}

Of the ${\sim} 80,000$ sources assigned uncertain classifications, ${\sim} 2,000$ were removed because of problems in the ASAS-SN light curves (see $\S 5$) to leave ${\sim} 78,000$ variables with uncertain classifications. Table \ref{tab:ndistun} presents their distribution by classification along with the number that do not have a Gaia DR2 parallax, a 2MASS match, a WISE match, a period in ASAS-SN, or have faint Gaia DR2 $G_{BP}$-band magnitudes ($G_{BP}>19$ mag). The number of sources with $G_{BP}>19$ mag is an indicator of the frequency of errors in cross-matching. Figure \ref{fig:fig28} suggests that the $G_{BP}$-band closely follows the ASAS-SN V-band magnitudes, hence very faint $G_{BP}$-band magnitudes likely arise from matches to the wrong source. The ROT: and VAR classes have the largest number of sources with $G_{BP}>19$ mag, suggesting that these are the classes with the greatest amount of cross-matching errors.

\begin{table*}
	\centering
	\caption{Number distributions of the ${\sim} 78,000$ variables with uncertain classifications.}
	\label{tab:ndistun}
\begin{tabular}{lrrrrrr}
		\hline
		VSX Type & $N_{\rm tot}$ & No Parallax &  No 2MASS &  No AllWISE & No Period &  $N(G_{BP}>19) $\\
		\hline
DSCT: & 358   & 169   & 8    & 13   & 0     & 3    \\
GCAS: & 742   & 0     & 0    & 8    & 385   & 2    \\
M:    & 1071  & 0     & 0    & 18   & 601   & 63   \\
ROT:  & 45641 & 1     & 1    & 2789 & 40265 & 3107 \\
RRAB: & 45    & 5     & 3    & 10   & 0     & 3    \\
VAR   & 30636 & 11709 & 1187 & 2583 & 23376 & 1624
\end{tabular}
\end{table*}

\begin{figure}
	\includegraphics[width=0.5\textwidth]{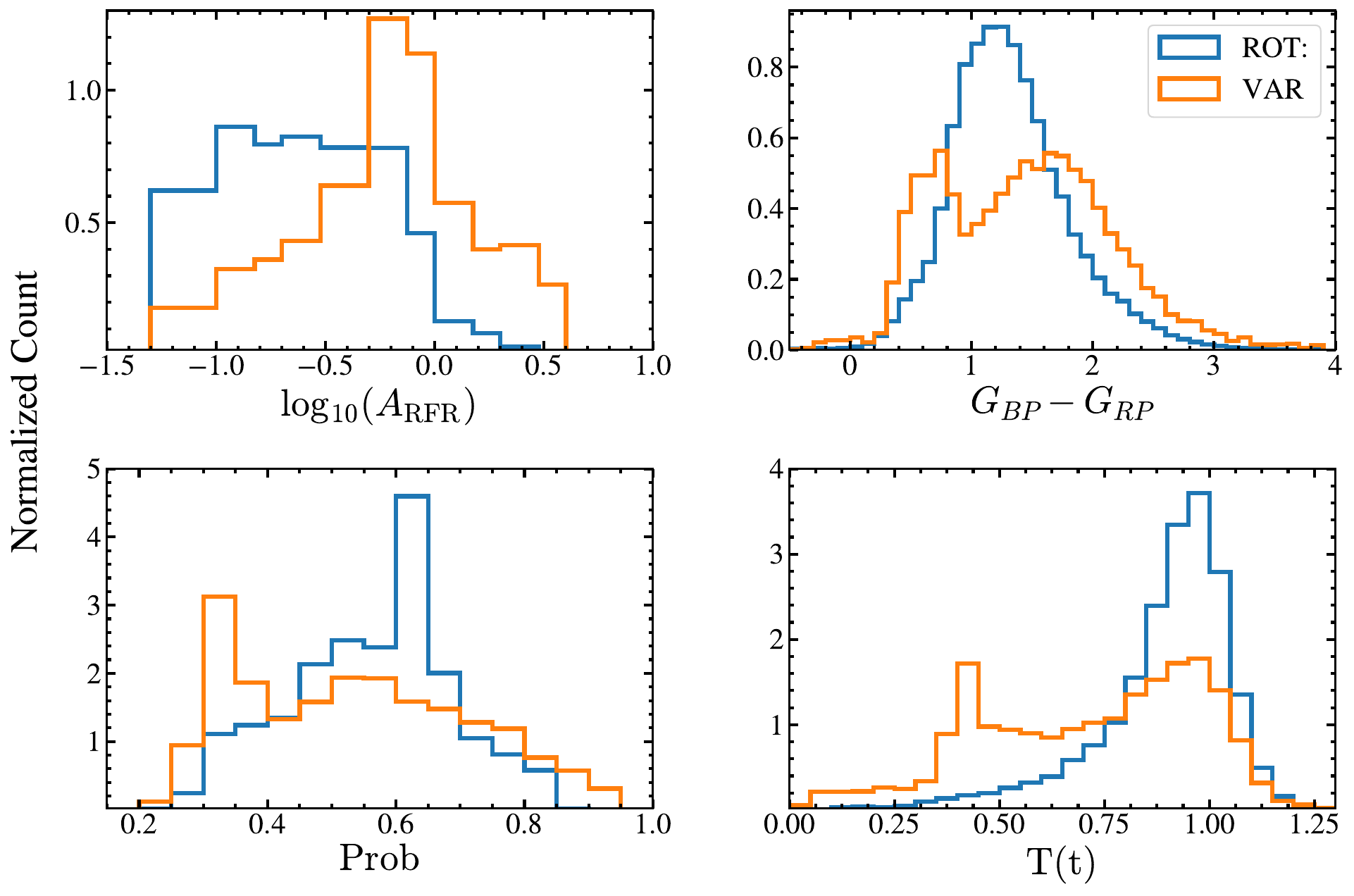}
    \caption{Distribution of the sources classified as ROT: and VAR in $A_{\rm RFR}$, $G_{BP}-G_{RP}$, $\rm Prob$ and $\rm T(t)$.}    
    \label{fig:fig40}
\end{figure}

We further investigate the sources classified as ROT: and VAR by looking at their distributions in the $A_{\rm RFR}$, $G_{BP}-G_{RP}$, $\rm Prob$ and $\rm T(t)$ parameter spaces (Figure \ref{fig:fig40}). The distribution of these sources in the amplitude $A_{\rm RFR}$ suggests that the ROT: class predominantly consists of low amplitude sources when compared to the VAR class that spans a wider distribution in amplitude. The VAR class shows a bimodal distribution in the color $G_{BP}-G_{RP}$ and $\rm T(t)$. This suggests that this group largely consists of two populations, including a set of red sources with long term variations in their light curves and a set of bluer sources with noisy light curves. The distribution of $\rm T(t)$ for the ROT: class suggests that a significant fraction of these sources do not have structure in their light curves that is evident of long term variations. Both the ROT: and VAR classes span a wide range in the classification probability, with 477 VAR sources having $\rm Prob >0.9$. There are no ROT: sources with $\rm Prob >0.9$. $40\%$ of the sources classified as VAR do not have a Gaia DR2 parallax. Without the Gaia DR2 distance, we are unable to carry out the empirical checks described in $\S3.4$, and ultimately these sources are classified as VAR. In contrast, almost all the ROT: sources have Gaia DR2 parallax measurements. These sources are likely to have variability signals below the ASAS-SN detection threshold, and thus the amplitude of the scatter in their light curves resemble the low amplitude variability associated with ROT variables. Based on these differences between the two classes, we have decided to retain these groupings without merging them into a single class.

\subsection{Rare Variables}

\begin{figure*}
	\includegraphics[width=\textwidth]{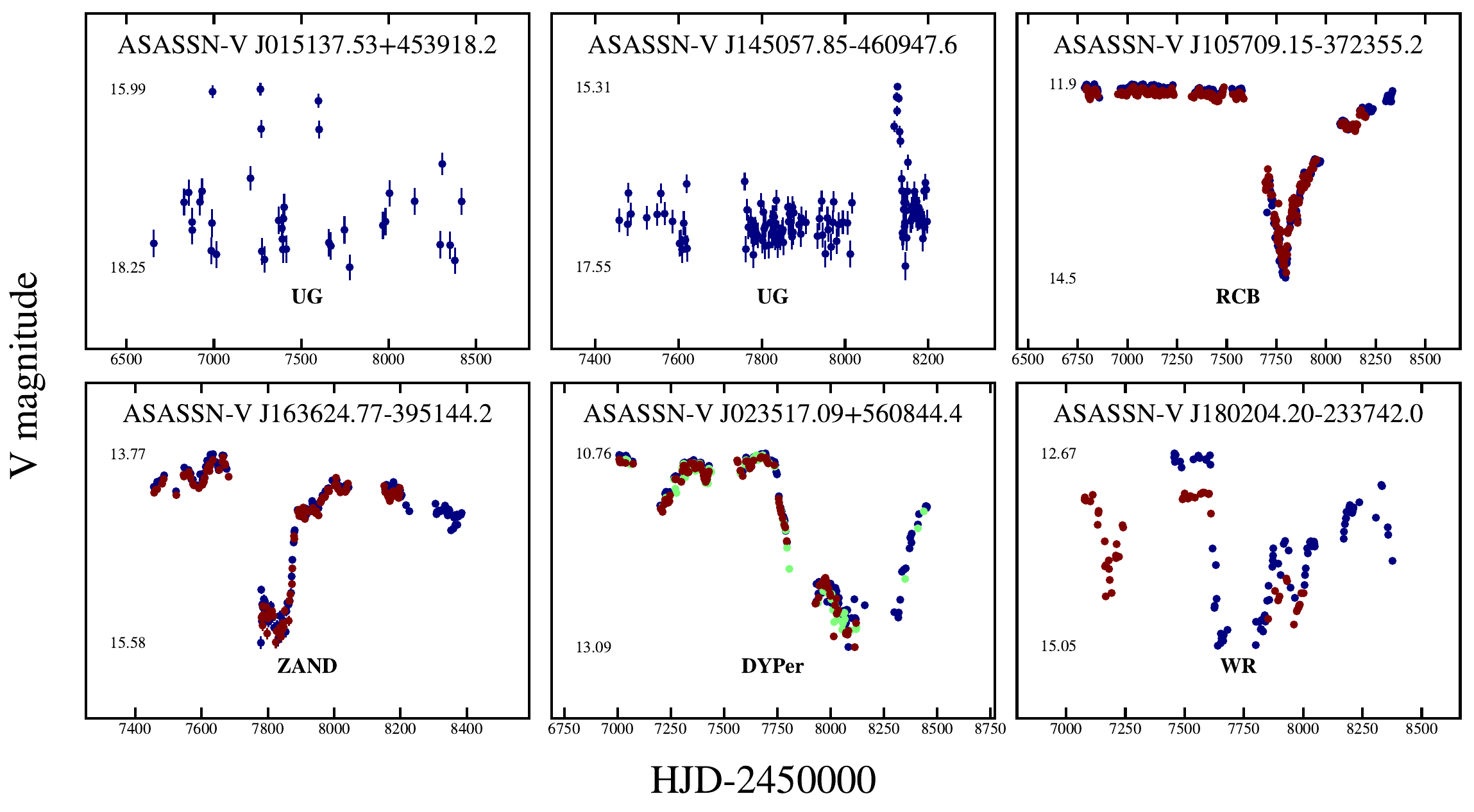}
    \caption{Light curves of sources included in the ASAS-SN variable stars database that were not classified in the ASAS-SN pipeline. The format is the same as Figure \ref{fig:fig8}. The classifications used here are from the VSX catalog.}    
    \label{fig:fig41}
\end{figure*}

We also include the light curves of ${\sim} 4,000$ rare or transient variable types. This includes the light curves of cataclysmic variables, R Coronae Borealis variables, flare stars, variable white dwarfs and symbiotic variables. These have not been classified by our pipeline, however we have run periodograms, searched for best periods, and computed variability amplitudes through RF regression. We generally just retain the VSX classification, but opt to simplify degenerate classifications (ex: ZAND|EA) into a single classification (i.e, ZAND) for simplicity. Examples of these variables are shown in Figure \ref{fig:fig41}.

During this work, we also identified a long period detached eclipsing binary system that was previously misclassified as an irregular variable \citep{2018RNAAS...2c.125J}. ASASSN-V J192543.72+402619.0 is one of the longest period detached eclipsing binary systems known, with an orbital period of $P{{\sim}} 2679$ d. The systematic variability analysis of the ${\sim} 50$ million $V<17$ mag APASS sources in ASAS-SN will likely identify similar long period eclipsing binary systems. We also identified two other R Coronae Borealis (RCB) candidates during the analysis of the SR/IRR class (ASASSN-V J 043259.32+415854.0, ASASSN-V J 175700.51-213934.5). These sources and 17 other RCB candidates discovered in ASAS-SN are discussed in \citet{2018arXiv180904075S}.

\subsection{Spatial Distribution of the Variable Stars}
\begin{figure*}
	\includegraphics[width=\textwidth]{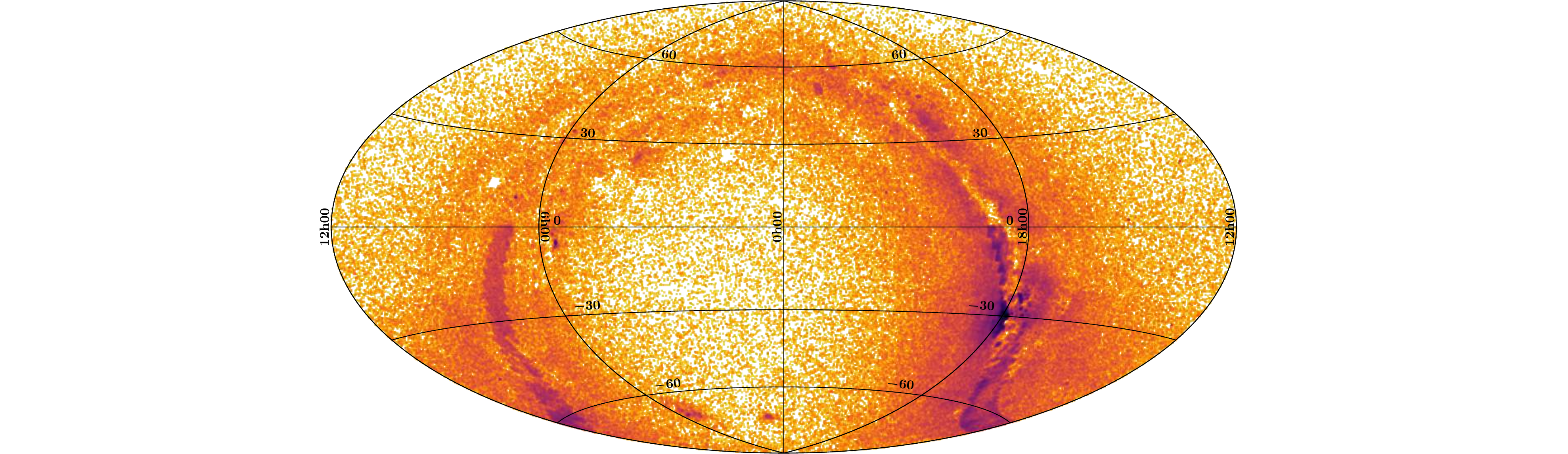}
    \caption{Sky density plot for the ${\sim} 298,000$ variables with definite ASAS-SN classifications, in equatorial coordinates. }    
    \label{fig:fig42}
\end{figure*}

\begin{figure*}
	\includegraphics[width=\textwidth]{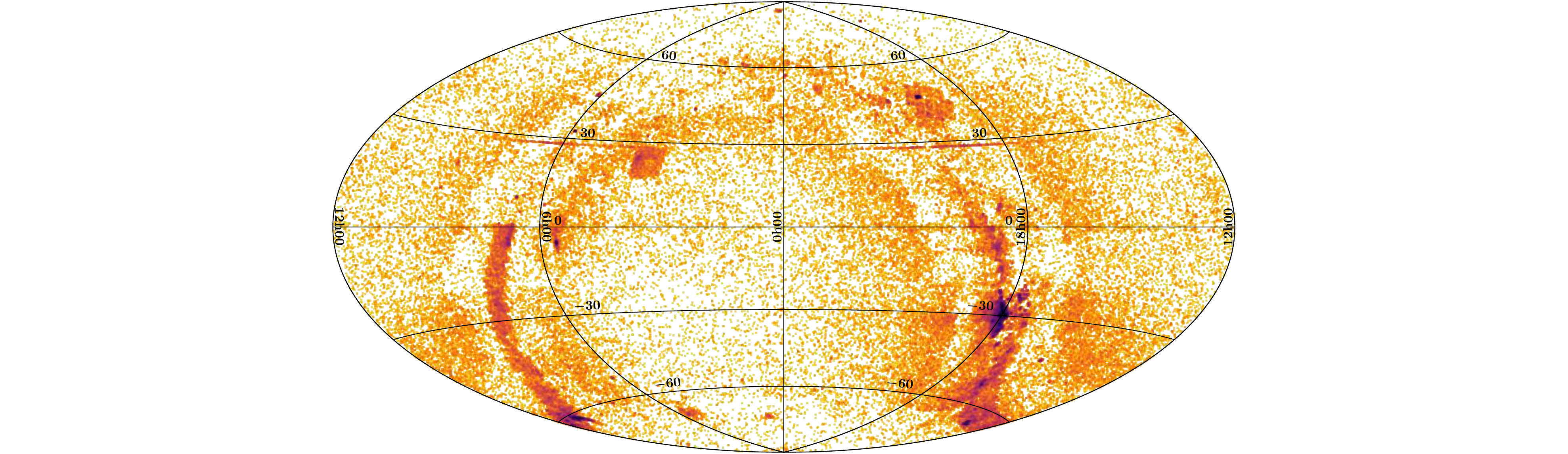}
    \caption{Sky density plot for the ${\sim} 114,000$ variables with uncertain ASAS-SN classifications, in equatorial coordinates. }    
    \label{fig:fig43}
\end{figure*}

The spatial distribution of the ${\sim} 298,000$ variables with definite ASAS-SN classifications is shown as a sky density plot in Figure \ref{fig:fig42}. Similarly, the sky density plot for the ${\sim} 114,000$ variables with uncertain ASAS-SN classifications is shown in Figure \ref{fig:fig43}.

There is an abundance of sources towards the Galactic plane, and ASAS-SN's ability to provide truly all-sky coverage is evident. We see that most of the uncertain classifications are clustered towards the Galactic plane and the Magellanic clouds where the effects of crowding hamper more accurate classifications. We also find that the spatial distribution of the variables with uncertain classifications highlights the footprints of multiple sky surveys (Figure \ref{fig:fig43}). In particular, the FOV of the \textit{Kepler} space telescope is clearly visible. 

\subsection{Comparison to the Gaia DR2 Catalog of Variable Stars}

In order to compare the classifications for the variables identified by Gaia DR2 and the ASAS-SN classifications, we cross-matched the ${\sim} 412,000$ variables in this work to the list of ${\sim} 391,000$ Gaia DR2 variables listed in the Specific Object Studies (SOS) variables list using the Gaia DR2 \verb"source_id" \citep{2018arXiv180409373H}. The mean $G_{BP}$-band magnitude for these sources ranges from $G_{BP}{\sim}16.4$ mag to $G_{BP}{\sim}18.2$ mag depending on the class of variable, suggesting that a large fraction of these sources are outside the ASAS-SN V-band detection threshold \citep{2018arXiv180409373H}. We identified ${\sim} 23,000$ variables that had both ASAS-SN and Gaia DR2 classifications. We group their classifications into broad classes for comparison and list the percentage of sources with matching ASAS-SN and Gaia DR2 classifications in Table \ref{tab:gaia}. The SR/IRR and MIRA variables were grouped into a single class since Gaia DR2 provided general classifications that did not distinguish between them.

Most of the Gaia DR2 classifications agree very well with the ASAS-SN classifications, however the CEPH (71 $\%$) and RRC/RRD (84$\%$) classes show large discrepancies. The discrepancy for the CEPH class is largely due to sources with periods close to $P{\sim}1$ d. Our classification pipeline does not enforce a lower bound on the periods of Cepheids at $P{\sim}1$ d, but allows for sources falling on the Cepheid PLRs that are initially classified as RRAB with $P\gtrsim0.9$ d to be reclassified as Cepheids. These sources remain as RRAB in the Gaia DR2 catalog, and account for 64\% of the discrepant classifications. However, we note that it is challenging to disentangle the PLRs of all-sky RR Lyrae and short period Cepheids near $P{\sim}1$ d owing to the distance uncertainties in the current Gaia DR2 catalog \citep{2018arXiv180502079C}. The remainder of the discrepancies are due to the classification of long period Cepheids as semi-regular/Mira variables in the Gaia DR2 catalog. The Gaia DR2 light curves are sparsely sampled when compared to the ASAS-SN light curves, thus it is likely that our classification of these long period variables as semi-regular/Mira variables is accurate. The discrepancy in the RRC/RRD class is largely due to sources classified as RRAB in the ASAS-SN pipeline.

\begin{table}
	\centering
	\caption{ASAS-SN variability classification matches for the ${\sim} 23,000$ variables also listed in the Gaia DR2 catalog}
	\label{tab:gaia}
\begin{tabular}{lrr}
		\hline
		Variability Class & $N_{\rm tot}$ & Matched\\
		\hline
CEPH & 219   & 71\%   \\
SR/IRR+MIRA & 14273 &  99\%   \\
DSCT    & 173  & 88\%   \\
RRAB & 6866 & 100\%  \\
RRC/RRD   & 906  & 84\%   \\
ROT & 43 & 100\%  \\
\end{tabular}
\end{table}

\subsection{Effects of Interstellar Extinction on the ASAS-SN Variability Classifications}
Our variability classification pipeline uses observed colors and Wesenheit magnitudes. While we account for the effects of reddening in the absolute magnitudes with the use of Wesenheit magnitudes, we do not correct for extinction in the colors. 

To evaluate the effects of extinction on our classifications, we selected a sample of ${\sim} 167,000$ variables with estimates of the extinction in Gaia DR2 \citep{2018arXiv180409365G,2018A&A...616A...8A}. The extinctions in the $J$, $H$, $K_s$, $W1$ and $W2$ bands are calculated using the Gaia DR2 extinction estimate $A_{\rm G}$ as \begin{equation}
    A_\lambda=A_G\bigg(\frac{A_V}{A_G}\bigg)\bigg(\frac{A_\lambda}{A_V}\bigg),
	\label{eq:ext}
\end{equation}
where $A_G/A_V=0.85926$, $A_J/A_V=0.29434$, $A_H/A_V=0.18128$, $A_{K_s}/A_V=0.11838$, $A_{W1}/A_V=0.07134$ and $A_{W2}/A_V=0.05511$. We note that the relationship between $A_G$ and $A_V$ is non-trivial \citep{2018A&A...616A...8A}, but we have made this assumption to allow a simple evaluation of the problem.

Using the same variability classification pipeline, we reclassified these variables after correcting their $G_{BP}-G_{RP}$, $J-H$, $J-K_s$, and $W1-W2$ colors for extinction. We only consider the ${\sim}108,000$ variables with $\rm Prob >0.9$ in order to select a test sample with well defined variable sources. The fraction of classifications which do not change is listed in Table \ref{tab:extin}. The RR Lyrae (RRAB, RRC, RRD), Cepheids (DCEPS, DCEP, CWA, CWB), eclipsing binaries (EA, EB, EW), $\delta$ Scuti (HADS), semi-regular variables (SR), Mira variables (M), rotational variables (ROT) and red, irregular variables (L) all have excellent recovery rates ($\gtrsim96\%$).

On the other hand, sources with recovery rates $<96\%$ include yellow semi-regular (SRD) and YSO variables. GCAS and RVA variables often have low RF classification probabilities and so were not included in the test sample. YSOs are typically located towards Galactic star forming regions, which by their nature have high extinction. Hence, the observed magnitudes for YSO variables are more likely to suffer from extinction when compared to other variability types. Furthermore, the Gaia DR2 extinction estimates for the YSO variables effectively also include the circumstellar extinction which drives the near/mid-infrared colors used to identify them. Hence, it is not surprising that the identification of YSOs is strongly affected by this test.
\begin{table}
	\centering
	\caption{Recovery rate of the ${\sim} 108,000$ variables with high classification probability after correcting for extinction in their colors using the Gaia DR2 extinction estimates.}
	\label{tab:extin}
\begin{tabular}{lr}
		\hline
		VSX Type & Recovery Rate \\
		\hline
CWA   & 98\%   \\
CWB   & 100\% \\
DCEP  & 98\% \\
DCEPS & 100\%\\
DSCT  & 91\%\\
EA    & 100\%\\
EB    & 99\%\\
EW    & 100\%\\
HADS  & 99\%\\
L     & 99\% \\
M     & 100\% \\
ROT   & 97\%\\
RRAB  & 100\% \\
RRC   & 99\%\\
RRD   & 96\%\\
SR    & 96\%\\
SRD   & 86\% \\
YSO   & 75\% 

\end{tabular}
\end{table}

\section{Conclusions}
\label{conclude}

We uniformly analyzed the ASAS-SN light curves of ${\sim} 412,000$ known variables from the VSX catalog. We analyzed all the sources for periodicity and were able to determine periods for ${\sim} 52,000$ VSX variables lacking them. We homogeneously estimated the variability amplitudes using random forest regression. Using the RF classifier from Paper I and a series of classification corrections based in large part on using Gaia DR2 parallax information, we built a clean training sample of ${\sim} 166,000$ variables. We used this to build a refined RF classifier. The new classifier has an overall $F_1$ score of 99.4\%, a considerable improvement over Paper I. We then use this V2 classifier on the complete sample of variables and provide new classifications for ${\sim}94,000$ variables with miscellaneous classifications. We also classified the variables with mean magnitudes outside the magnitude range $11<V<17$ mag, but these should be used with caution.

We provide definite classifications for ${\sim} 278,000$ VSX variables, of which ${\sim} 17,000$ have different broad variability classifications than VSX to high confidence. Most of these changes are from the refinements in classifying RR Lyrae, $\delta$-Scuti, eclipsing binaries and the semi-regular and irregular variables. We also include the light curves of ${\sim} 4,000$ rare and transient variables on the ASAS-SN Variable Stars Database (\url{https://asas-sn.osu.edu/variables}). We have updated the ASAS-SN Variable Stars Database with new photometric, Gaia DR2 distance, classification quality and light curve quality information. This sample provides an excellent, homogeneously classified training set of variables for both current and future all-sky surveys. A full variability analysis of the ${\sim}50$ million APASS sources brighter than $V{\sim}17$ mag is currently underway.

\section*{Acknowledgements}
We thank the anonymous referee for useful comments. We thank the Las Cumbres Observatory and its staff for its
continuing support of the ASAS-SN project. We also thank the Ohio State University College of Arts and Sciences Technology Services for helping us set up and maintain the ASAS-SN variable stars database.

ASAS-SN is supported by the Gordon and Betty Moore
Foundation through grant GBMF5490 to the Ohio State
University and NSF grant AST-1515927. Development of
ASAS-SN has been supported by NSF grant AST-0908816,
the Mt. Cuba Astronomical Foundation, the Center for Cos-
mology and AstroParticle Physics at the Ohio State Univer-
sity, the Chinese Academy of Sciences South America Center
for Astronomy (CAS- SACA), the Villum Foundation, and
George Skestos. 

TAT is supported in part by Scialog Scholar grant 24216 from the Research Corporation. Support for JLP is provided in part by FONDECYT through the grant 1151445 and by the Ministry of Economy, Development, and Tourism's Millennium Science Initiative through grant IC120009, awarded to The Millennium Institute of Astrophysics, MAS. SD acknowledges Project 11573003 supported by NSFC. Support for MP and OP has been provided by the PRIMUS/SCI/17 award from Charles University. 

This work has made use of data from the European Space Agency (ESA)
mission {\it Gaia} (\url{https://www.cosmos.esa.int/gaia}), processed by
the {\it Gaia} Data Processing and Analysis Consortium (DPAC,
\url{https://www.cosmos.esa.int/web/gaia/dpac/consortium}). Funding
for the DPAC has been provided by national institutions, in particular
the institutions participating in the {\it Gaia} Multilateral Agreement.

This publication makes use of data products from the Wide-field Infrared Survey Explorer, which is a joint project of the University of California, Los Angeles, and the Jet Propulsion Laboratory/California Institute of Technology, funded by the National Aeronautics and Space Administration.

This research was made possible through the use of the AAVSO Photometric All-Sky Survey (APASS), funded by the Robert Martin Ayers Sciences Fund. This publication makes use of data products from the Two Micron All Sky Survey, which is a joint project of the University of Massachusetts and the Infrared Processing and Analysis Center/California Institute of Technology, funded by the National Aeronautics and Space Administration and the National Science Foundation. 

This research has made use of the VizieR catalogue access tool, CDS, Strasbourg, France. The original description of the VizieR service was published in A\&AS 143, 23. 

This research made use of Astropy, a community-developed core Python package for Astronomy (Astropy Collaboration, 2013).

%%%%%%%%%%%%%%%%%%%%%%%%%%%%%%%%%%%%%%%%%%%%%%%%%%



%\bibliographystyle{mnras}

\bibliographystyle{plainnat}

%%%%%%%%%%%%%%%%%%%%%%%%%%%%%%%%%%%%%%%%%%%%%%%%%%

%%%%%%%%%%%%%%%%% APPENDICES %%%%%%%%%%%%%%%%%%%%%

\appendix

\section{Saturated/Faint Sources}

Variables that maybe saturated ($V<11$ mag) or faint ($V>17$ mag) in ASAS-SN are not considered part of the primary analysis. However, we do include these ${\sim} 56,000$ sources in the ASAS-SN Variable Stars Database with classifications. We caution the reader to interpret these classifications carefully --- saturation artifacts and noise hinder accurate variable classifications. 

\begin{figure*}
	\includegraphics[width=\textwidth]{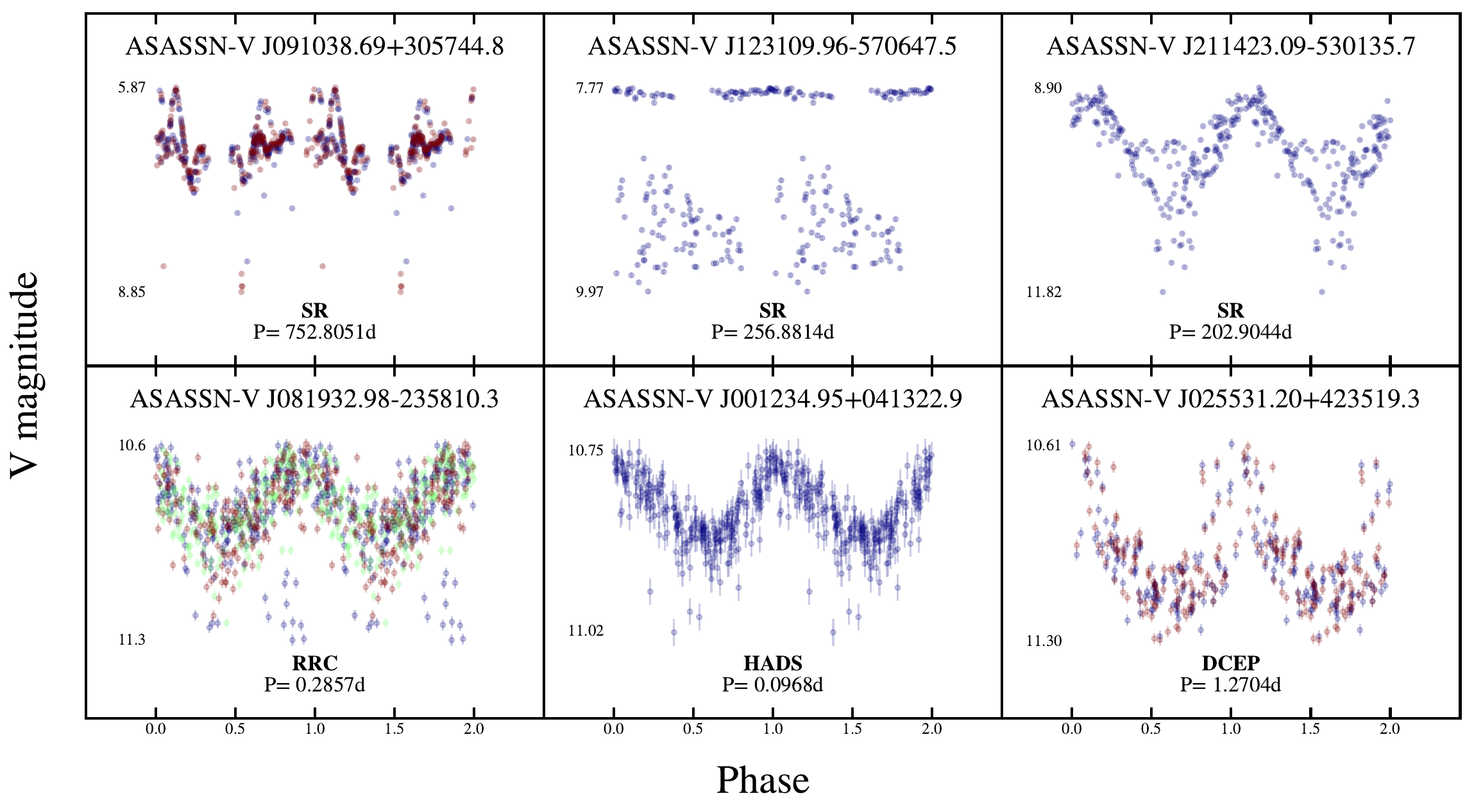}
    \caption{Examples of periodic variables in the saturated regime ($V<11$ mag). The format is the same as Figure \ref{fig:fig8}.}    
    \label{fig:fig44}
\end{figure*}

ASAS-SN uses an algorithm derived from the ASAS survey to correct the flux from the bleed trails of bright stars that saturate the detectors \citep{2017PASP..129j4502K}. In many cases, this leads to surprisingly good light curves for saturated sources. Figure \ref{fig:fig44} shows examples of periodic variables with $V<11$ mag. Generally, acceptable light curves are obtained for periodic variables with mean magnitudes in the range $10 \leq V\leq 11$ mag. However, saturation artifacts are prominent in the sources with mean magnitudes $V<9$ mag. We catalog ${\sim} 34,000$ variables with $V<11$ mag. 

\begin{figure*}
	\includegraphics[width=\textwidth]{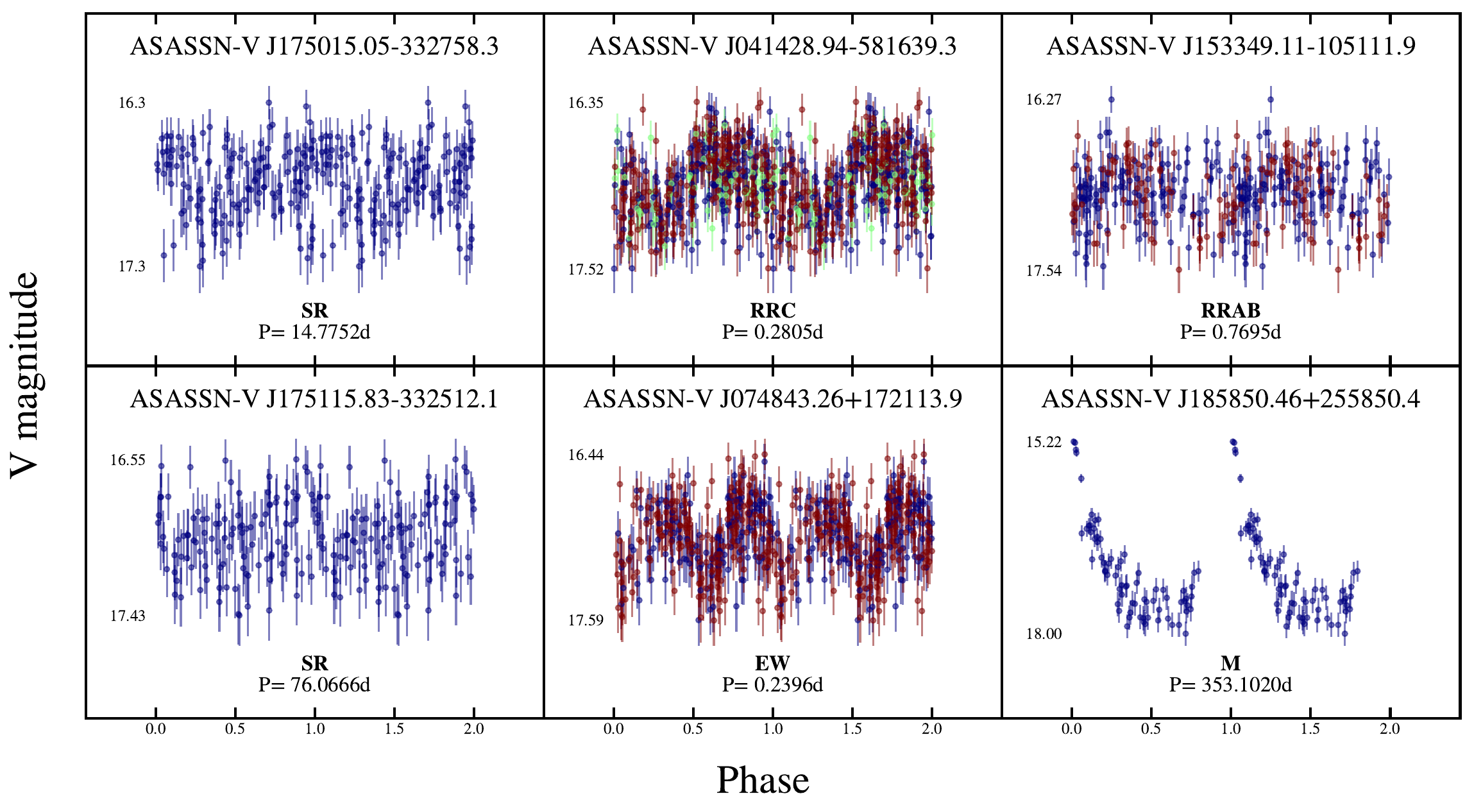}
    \caption{Examples of periodic variables with mean magnitude $V>17$ mag. The format is the same as Figure \ref{fig:fig8}.}    
    \label{fig:fig45}
\end{figure*}

ASAS-SN has a transient detection limit of $V\lesssim17$ mag (see for e.g, \citealt{2017MNRAS.471.4966H}). In some cases, sources with mean V-band magnitudes $V>17$ mag provide useful information, particularly in the cases of high amplitude Mira variables and semi-regular variables (Figure \ref{fig:fig45}). The detection of lower amplitude variables is increasingly difficult, and the light curves are noisy. We catalog ${\sim} 22,000$ variables with $V>17$ mag. 

Only ${\sim}30\%$ (${\sim}26\%$) of these variables have classification probabilities of >0.9 (>0.75), while a majority (${\sim}64\%$) of these sources have uncertain classifications. We derive periods for ${\sim} 6,000$ saturated/faint variables that do not have a period in the VSX catalog. 


%%%%%%%%%%%%%%%%%%%%%%%%%%%%%%%%%%%%%%%%%%%%%%%%%%

% Don't change these lines
\bsp	% typesetting comment
\label{lastpage}
\end{document}